\documentclass[11pt]{article}
\usepackage{cite}
\usepackage{amsmath,amsfonts,amssymb}
\usepackage[small,bf,hang]{caption}
\usepackage{slashed}
\usepackage{mathabx}
\usepackage{latexsym,epsfig}

\def\hybrid{
        \topmargin -20pt
        \oddsidemargin 0pt
        \headheight 0pt \headsep 0pt
        \textwidth 6.25in 
        \textheight 9.5in 
        \marginparwidth .875in
        \parskip 5pt plus 1pt \jot = 1.5ex}

\hybrid

 \csname
@addtoreset\endcsname{equation}{section}

\def\moth{\mathsurround=0pt}
\newdimen\zo \zo=0pt

\def\tick{\leaders\hrule height 0.5ex depth 0pt \hskip 0.5pt}
\def\upboxfill{$\moth \setbox\zo\hbox{\tick}%
  \hskip 3pt\hbox to 0pt{$\tick$\hss}\hrulefill \hbox to 7.5pt{$\tick$\hss}$}

\def\dtick{\leaders\hrule height .34pt depth 0.5ex \hskip 0.5pt}
\def\downboxfill{$\moth \setbox\zo\hbox{\dtick}%
  \hskip 2pt\hbox to 0pt{$\dtick$\hss}\hrulefill \hbox to 2pt{$\dtick$\hss}$}

\def\bec{\begin{center}}
\def\ec{\end{center}}

 \def\det{{\rm det\,}}
\def\be{\begin{equation}}
\def\ee{\end{equation}}
\def\bea{\begin{eqnarray}}
\def\eea{\end{eqnarray}}
\def\ba{\begin{array}}
\def\ea{\end{array}}

\newcommand{\kkappa}{\gamma}

\begin{document}

\begin{titlepage}
\rightline{}
\rightline{\tt  MIT-CTP 4519}
\rightline{December 2013}
\begin{center}
\vskip .6cm
{\Large \bf {Exceptional Field Theory I:\\[1.5ex]
E$_{6(6)}$ covariant Form of M-Theory and Type IIB }}\\
\vskip 2cm
{\large {Olaf Hohm${}^1$ and Henning Samtleben${}^2$}}
\vskip 1cm
{\it {${}^1$Center for Theoretical Physics}}\\
{\it {Massachusetts Institute of Technology}}\\
{\it {Cambridge, MA 02139, USA}}\\
ohohm@mit.edu
\vskip 0.2cm
{\it {${}^2$Universit\'e de Lyon, Laboratoire de Physique, UMR 5672, CNRS}}\\
{\it {\'Ecole Normale Sup\'erieure de Lyon}}\\
{\it {46, all\'ee d'Italie, F-69364 Lyon cedex 07, France}}\\
henning.samtleben@ens-lyon.fr

\vskip 1.5cm
{\bf Abstract}
\end{center}

\vskip 0.2cm

\noindent
\begin{narrower}
We present the details of the recently constructed E$_{6(6)}$ covariant extension of 
11-dimensional supergravity. This theory requires a $5+27$ dimensional spacetime 
in which the `internal' coordinates transform in the $\bar{\bf 27}$ of E$_{6(6)}$. 
All fields are E$_{6(6)}$ tensors and transform under (gauged) internal generalized 
diffeomorphisms. The `Kaluza-Klein' vector field acts as a gauge field 
for the E$_{6(6)}$ covariant `E-bracket' rather than a Lie bracket, requiring the 
presence of two-forms akin to the tensor hierarchy of gauged supergravity.
We construct the complete and unique action that is gauge invariant under 
generalized diffeomorphisms in the internal and external coordinates.
The theory is subject to covariant section constraints on the derivatives, 
implying that only a subset of the extra $27$ coordinates is physical. 
We give two solutions of the section constraints: the first preserves ${\rm GL}(6)$ and
embeds the action of the complete (i.e.~untruncated) 
11-dimensional supergravity; the second preserves ${\rm GL}(5)\times {\rm SL}(2)$
and embeds complete type IIB supergravity.  As a by-product, we thus obtain an off-shell action 
for type IIB supergravity.

\end{narrower}

\end{titlepage}

\newpage

\tableofcontents

\newpage

\section{Introduction}
For more than three decades, since the seminal work of Cremmer and Julia \cite{Cremmer:1979up}, it has been known 
that toroidal compatification of 11-dimensional supergravity~\cite{Cremmer:1978km} gives rise to the exceptional 
symmetries E$_{n(n)}(\mathbb{R})$, $n=6,7,8$,  in dimensions $D=11-n$. 
Later, in the mid 1990's, the discrete subgroups E$_{n(n)}(\mathbb{Z})$ were interpreted 
as part of the U-duality symmetries of M-theory~\cite{Hull:1994ys}, but ever since 
it has remained a mystery why 11-dimensional supergravity knows about the exceptional groups
and to which extent they are already present in the full theory. 
This fact has inspired various authors to speculate about a hidden new geometry in 
higher dimensions that transcends the Riemannian geometry underlying Einstein's theory
\cite{Julia:1982gx,deWit:1986mz,Nicolai:1986jk,Obers:1998fb,Koepsell:2000xg,deWit:2000wu,West:2001as,
HenryLabordere:2002dk,Damour:2002cu,Damour:2002et,West:2003fc,West:2004iz,Hull:2007zu,DallAgata:2007sr,
Pacheco:2008ps,Hillmann:2009ci,Hillmann:2009pp,Gunaydin:2009zz,Aldazabal:2010ef,Berman:2010is,Berman:2011pe,
Berman:2011jh,Coimbra:2011ky,Coimbra:2012af,Berman:2012uy,Berman:2012vc,Park:2013gaj,Aldazabal:2013mya,Godazgar:2013dma}, 
but it is fair to say that so far there was no scheme that casts the full 11-dimensional supergravity
into  a truly E$_{n(n)}$ covariant form. 
In this paper, we present in detail the construction announced recently in \cite{Hohm:2013pua}, 
which gives an extension of 
11-dimensional supergravity that makes the exceptional group E$_{6(6)}$ manifest 
prior to any toroidal compactification, while also hosting the type IIB theory~\cite{Schwarz:1983wa,Howe:1983sra}.  
The details for the remaining finite dimensional groups 
E$_{7(7)}$ and E$_{8(8)}$ will be presented in separate publications~\cite{HSEFT7}.

Our construction is a continuation and generalization of `double field theory' (DFT), which 
is an approach to make the $O(d,d)$ T-duality group of string theory manifest by 
introducing a generalized spacetime with doubled coordinates, subject to a `section constraint'
or `strong constraint',  and reorganizing the 
fields into $O(d,d)$ tensors \cite{Siegel:1993th,Hull:2009mi,Hull:2009zb,Hohm:2010jy,Hohm:2010pp}. 
(For earlier results see \cite{Tseytlin:1990va,Duff:1989tf,Siegel:1993xq,Kugo:1992md}.)
Remarkably, DFT is applicable not only to (the low-energy spacetime action of) 
bosonic string theory, but also to the heterotic string  \cite{Hohm:2011ex},  including their supersymmetric formulations 
\cite{Hohm:2011nu,Coimbra:2011nw,Jeon:2011sq}, as well as massless and massive type II theories 
\cite{Hohm:2011zr,Hohm:2011dv,Hohm:2011cp,Jeon:2012kd}.  
DFT also yields an intriguing generalization of Riemannian geometry 
\cite{Siegel:1993th,Hohm:2010xe,Hohm:2011si,Hohm:2012gk,Hohm:2012mf,Jeon:2010rw,Jeon:2011cn}, 
which in turn extends results in the `generalized geometry' developed in pure mathematics 
\cite{Hitchin:2004ut,Gualtieri:2003dx,Gualtieri:2007bq}.  
Moreover, it provides a natural framework for non-geometric fluxes 
\cite{Aldazabal:2011nj,Geissbuhler:2011mx,Andriot:2012wx,Andriot:2012an,Geissbuhler:2013uka}. 
Finally, 
an extension of DFT to higher-derivative $\alpha'$ corrections has recently been given \cite{Hohm:2013jaa}.
(For a more exhaustive list of references see the recent reviews \cite{Hohm:2013bwa,Aldazabal:2013sca,Berman:2013eva}.)

In contrast to $D=10$ string theory and DFT, where the fields naturally combine into 
tensors under $O(10,10)$, the fields of $D=11$ supergravity do not organize  directly 
into tensors under any of the exceptional groups. For instance, in order to realize
the E$_{n(n)}$ symmetry in dimensional reduction, some field components have to be 
dualized into forms of lower rank. As such transformations are specific to a given dimension, 
it is not obvious how to build complete E$_{n(n)}$ multiplets in $D=11$ prior to any reduction. 
We have recently shown how to overcome these obstacles by gauge fixing the local 
Lorentz group and decomposing 
the fields and coordinates as in Kaluza-Klein compactifications, 
but \textit{without} truncation \cite{Hohm:2013jma}.
The resulting formulation therefore captures all of the original 11-dimensional supergravity, 
at the cost of abandoning some of the Lorentz gauge freedom. The various field components, 
necessarily including some of their duals, can then be reorganized into E$_{n(n)}$ tensors. Extending the 
`internal' derivatives to transform in some fundamental representation of E$_{n(n)}$, 
subject to a generalization of the DFT section constraint proposed in~\cite{Coimbra:2012af,Berman:2012vc}, 
we arrive at a manifestly E$_{n(n)}$ covariant 
extension of 11-dimensional supergravity. The resulting theory, which we refer to 
in the following as `exceptional field theory' (EFT),  closely resembles DFT 
when subjected to an analogous Kaluza-Klein type gauge fixing of the local Lorentz group  
\cite{Hohm:2013nja}.

Already the early work of de~Wit and Nicolai~\cite{deWit:1986mz,Nicolai:1986jk} has identified directly in eleven dimensions
some of the structures found in dimensional reduction, following a Kaluza-Klein decomposition without 
truncation similar to the present construction. 
Manifest 11-dimensional covariance is abandoned, in favor of 
an enhanced local Lorentz symmetry in accordance with the 
(composite) gauge symmetries appearing in the $D=4$ or $D=3$ 
coset models.
However, these constructions do not yet manifest the exceptional groups, and further work in \cite{Koepsell:2000xg} suggested that 
additional coordinates should be introduced in order to achieve this, an idea that also features 
prominently in the proposal of \cite{West:2003fc}.  
Later work \cite{Hillmann:2009ci,Hillmann:2009pp} gave a manifestly ${\rm E}_{7(7)}$ invariant action functional for a certain 
7-dimensional truncation of $D=11$ supergravity by introducing coordinates in the ${\bf 56}$ of ${\rm E}_{7(7)}$. 
Recently, other subsectors of $D=11$ supergravity have been reformulated in terms of a generalized metric, 
see e.g., \cite{Berman:2010is,Berman:2011pe,Berman:2011jh}, together with a duality-covariant 
formulation of part of the gauge symmetries in form of generalized Lie derivatives.  
These constructions are also related to extensions of generalized 
geometry to the exceptional groups~\cite{Hull:2007zu,Coimbra:2011ky}.
In all these truncations the match to 11-dimensional supergravity
requires a Kaluza-Klein-type decomposition of the latter in which one 
sets to zero all off-diagonal components of the metric and the 3-form, sets to 
zero the external components of the 3-form  
and freezes the external metric to the Minkowski metric, possibly up to a warp factor. 
Finally, one truncates the coordinate dependence of all fields to the internal coordinates.  
We will explain in the appendix the embedding of these theories into 
the full EFT formulation, constructed in this paper.

This formulation to be constructed requires various new 
mathematical tools  \cite{Hohm:2013jma}, analogous to the Lorentz gauge fixed DFT \cite{Hohm:2013nja}.
 Most importantly, the off-diagonal 
vector field components of the Kaluza-Klein-like decomposition yield a generalization 
of a Yang-Mills gauge field. More precisely, these fields transform in the same way as a
Yang-Mills connection, but with a bracket, in the following referred to as the 
`E-bracket', that does not satisfy all axioms of a Lie bracket. This, in turn, requires the introduction of 
forms of higher rank in order to maintain gauge covariance of the field strengths, in precise 
analogy to the `tensor hierarchy' of gauged supergravity~\cite{deWit:2005hv,deWit:2008ta}. 
Moreover, these higher forms play a vital role as the duals of some physical fields,  which is implemented at the level of an 
off-shell action by means of topological Chern-Simons-like terms, as in gauged supergravity~\cite{Nicolai:2000sc,deWit:2004nw}.
Finally, the `internal' field components organize into a `generalized metric' ${\cal M}_{MN}$
that is a covariant tensor under E$_{n(n)}$, while the `external' metric $g_{\mu\nu}$ 
is an E$_{n(n)}$ singlet that, however, transforms as a scalar density under the (internal) 
generalized Lie derivatives.

In this paper, we present in detail the construction of the E$_{6(6)}$ EFT. 
Dimensional reduction from eleven dimensions on a torus $T^6$ is known to give rise to maximal
$D=5$ supergravity with global E$_{6(6)}$ symmetry~\cite{Cremmer:1980gs}.
It becomes manifest in five dimensions after proper dualization of all $p$-form tensors
to lowest possible degree. In particular, the three-form descending from eleven dimensions
is dualized into a scalar and joins the coordinates of the scalar target space
described by the coset space ${\rm E}_{6(6)}/{\rm USp}(8)$.
The E$_{6(6)}$ EFT keeps the field and multiplet structure of the five-dimensional theory,
but elevates all fields to functions of $5+27$ coordinates $(x^{\mu},Y^M)$, where the $Y^M$, 
with dual derivatives $\partial_M$, live in the fundamental representation $\bar{\bf 27}$ of E$_{6(6)}$.
The theory is subject to 
covariant section constraints, which can be written in terms of the E$_{6(6)}$ invariant $d$-symbols 
$d^{MNK}$ and $d_{MNK}$ as follows~\cite{Coimbra:2011ky,Berman:2012vc}
 \be
  d^{MNK}\,\partial_N \partial_K A \ = \ 0\;, \quad  d^{MNK}\,\partial_NA\, \partial_K B \ = \ 0 \,, 
 \label{sectioncondition}
 \ee  
where $A,B$ denote any fields or gauge parameters. This constraint is the analogue of
the `strong constraint' in DFT and implies that only a subset of the $27$ coordinates  is physical. 
While in DFT the strong constraint is motivated from string theory, as implementing  
a strong version of the level-matching constraint, eq.~(\ref{sectioncondition}) has been postulated 
by analogy. However, we will discuss below that for the SO$(5,5)$ T-duality subgroup of E$_{6(6)}$
it actually reduces to the strong constraint of DFT. 
The E$_{6(6)}$ covariant field content is 
given by 
\bea
\left\{e_\mu{}^a, {\cal M}_{MN}, A_\mu{}^M, B_{\mu\nu\,M} \right\}
\;, 
\label{fieldcontent}
\eea
where $e_{\mu}{}^{a}$ denotes the f\"unfbein corresponding to the external metric, while 
$A_\mu{}^M$ and $B_{\mu\nu\,M}$ are the tensor gauge fields relevant for the E$_{6(6)}$ EFT.
The symmetric matrix ${\cal M}_{MN}$ parametrizes the coset space ${\rm E}_{6(6)}/{\rm USp}(8)$ 
whose 42 coordinates describe the `scalar' fields of the theory.
The full action is given by 
\bea
\label{finalaction}
\begin{split}
 S_{\rm EFT} \ = \  \int d^5x\, d^{27}Y\,e\, \Big(& \widehat{R}
 +\frac{1}{24}\,g^{\mu\nu}{\cal D}_{\mu}{\cal M}^{MN}\,{\cal D}_{\nu}{\cal M}_{MN}\\
 &{}-\frac{1}{4}\,{\cal M}_{MN}{\cal F}^{\mu\nu M}{\cal F}_{\mu\nu}{}^N
 +e^{-1}{\cal L}_{\rm top}-V({\cal M}_{MN},g_{\mu\nu})\Big) \,.
\end{split}
\eea
This action takes the same structural form as $D=5$ gauged 
supergravity~\cite{deWit:2004nw}, with 
a (covariantized) Einstein-Hilbert term for $e_{\mu}{}^{a}$, a `scalar' kinetic term for 
${\cal M}_{MN}$ and a Yang-Mills term based on the field strength ${\cal F}_{\mu\nu}{}^{M}$, 
the latter also depending on the two-form $B_{\mu\nu\,M}$ in accordance with the tensor hierarchy. 
All fields depend on the `internal' coordinates, corresponding 
to the non-abelian structure of covariant derivatives and field strengths involving the derivatives 
$\partial_M$. In addition, the `potential' $V({\cal M},g)$ is the manifestly E$_{6(6)}$ covariant 
expression (built using only the $\partial_M$ derivatives) given by  
\be\label{fullpotentialIntro}
 \begin{split}
  V \ = \ &-\frac{1}{24}{\cal M}^{MN}\partial_M{\cal M}^{KL}\,\partial_N{\cal M}_{KL}+\frac{1}{2} {\cal M}^{MN}\partial_M{\cal M}^{KL}\partial_L{\cal M}_{NK}\\
  &-\frac{1}{2}g^{-1}\partial_Mg\,\partial_N{\cal M}^{MN}-\frac{1}{4}  {\cal M}^{MN}g^{-1}\partial_Mg\,g^{-1}\partial_Ng
  -\frac{1}{4}\,{\cal M}^{MN}\partial_Mg^{\mu\nu}\partial_N g_{\mu\nu}\;. 
 \end{split} 
 \ee 
All terms in the action (\ref{finalaction}) are separately gauge invariant under 
the internal (generalized) diffeomorphisms of the $Y^M$, generated by a parameter  
$\Lambda^M(x,Y)$, with the $A_\mu{}^M$ taking the role of a gauge connection for this symmetry. 
The action is further gauge invariant under ($A_\mu$-covariantized) `external' diffeomorphisms 
generated by $\xi^{\mu}(x,Y)$, but this symmetry is not manifest for $Y$-dependent parameter $\xi^\mu$. 
In fact, it is  this symmetry that relates the various terms in (\ref{finalaction}) and fixes all relative coefficients.

\begin{figure}[bt]
   \centering
   \includegraphics[width=11.6cm]{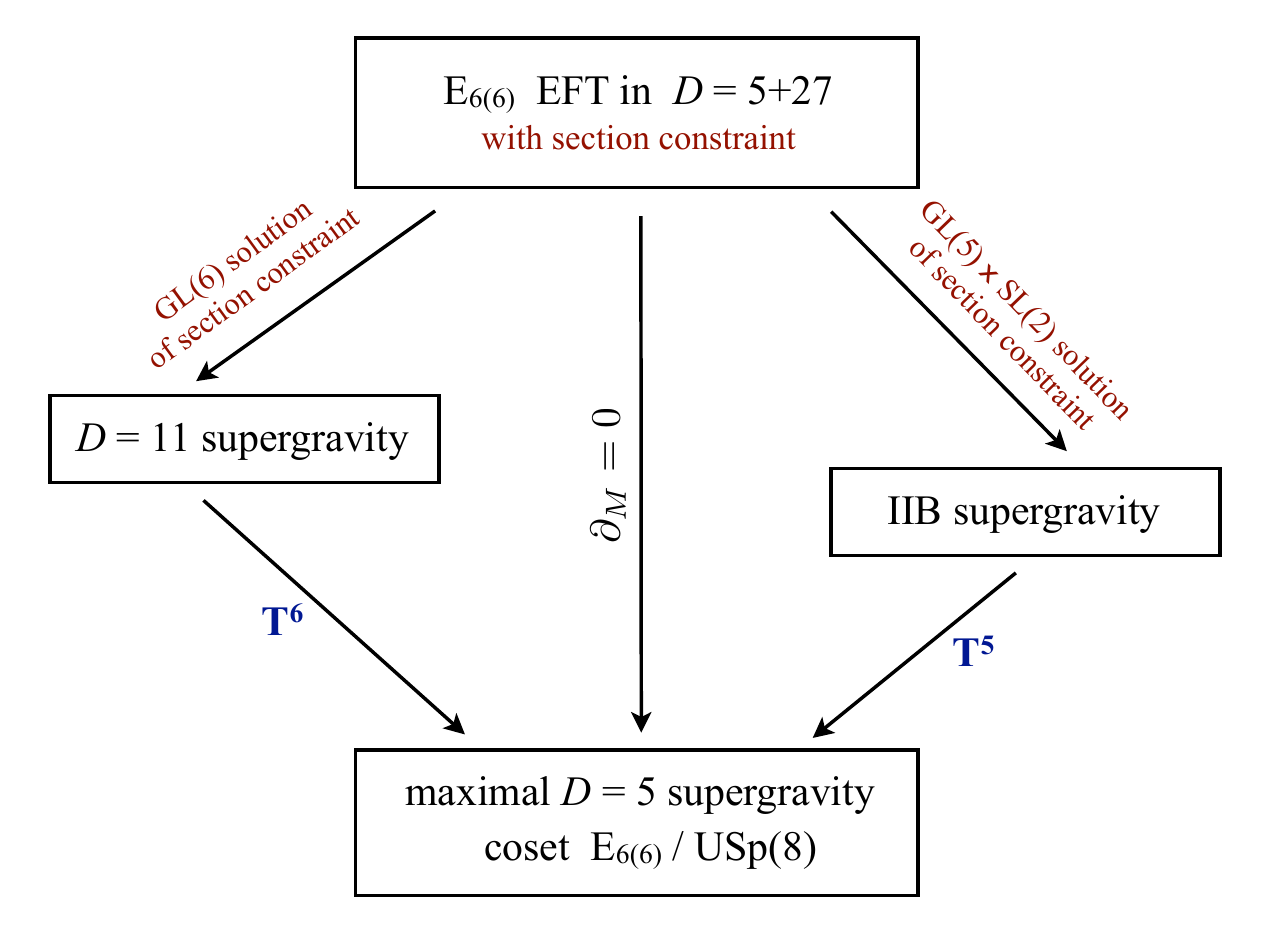}
     \caption{{\small ${\rm E}_{6(6)}$ EFT} embedding of $D=11$ supergravity, IIB supergravity,
   and $D=5$ supergravity.}
   \label{Figure:E6}
\end{figure}

Apart from the construction of the action (\ref{finalaction}), a central result of this paper is 
to show that this action after putting an appropriate solution
of the section condition (\ref{sectioncondition}) reduces to full (i.e.\ untruncated) 11-dimensional
supergravity after rearrangement of the fields according a 5+6 Kaluza-Klein split but keeping
the dependence on all eleven coordinates.
We work this out in full detail and reproduce from (\ref{finalaction}) the action of eleven-dimensional supergravity.
Moreover, it has been noted in~\cite{Hohm:2013pua} that the section condition (\ref{sectioncondition})
allows for (at least) two inequivalent solutions, the second of which reduces the theory (\ref{finalaction})
to the full ten-dimensional IIB theory.
To this end we first break E$_{6(6)}$ under SL$(6)\,\times\, $SL$(2)$ such that the fundamental representation
decomposes as
\be
{\bf 27} ~\rightarrow~ (15,1) + (6,2)
\;. 
\label{27}
\ee
If we let the fields depend on six coordinates from the  SL$(2)$ doublet, 
the section constraints are satisfied. We are left with an unbroken GL$(6)$ symmetry and fields 
depending on $5+6$ coordinates. 
For this choice, the action (\ref{finalaction}) reduces to an action that is 
on-shell equivalent to 11-dimensional supergravity. 
Alternatively, the section constraint is solved by letting fields depend on 
$5$ coordinates from the $15$ in (\ref{27}), which in turn breaks the symmetry to 
GL$(5)\,\times\, $SL$(2)$. For this choice, (\ref{finalaction}) reduces to a 
10-dimensional action with a global SL$(2)$ symmetry and we obtain an on-shell equivalent
formulation of type IIB supergravity. 
As a by-product, this yields an off-shell action for 
type IIB supergravity, at the cost of sacrificing manifest 10-dimensional spacetime covariance. 
In the sense just explained, the EFT defined by (\ref{finalaction}) unifies type IIB and 
M-theory (and thus type IIA), a feature shared with the type II DFT 
constructed in \cite{Hohm:2011zr,Hohm:2011dv}.
Instead, dropping all derivatives w.r.t.\ to the extra internal coordinates, i.e.\ setting $\partial_M =0$,
the theory (\ref{finalaction}) directly reduces to $D=5$ maximal supergravity in the form in 
which the exceptional symmetry ${\rm E}_{6(6)}$ is manifest without further dualization~\cite{Cremmer:1980gs}.
The various links are depicted in figure~\ref{Figure:E6}, which can be thought of as a commutative 
diagram that explains the emergence of ${\rm E}_{6(6)}$ from M-theory or type IIB. 

This paper is organized as follows. In sec.~\ref{sec2} we introduce the required 
E$_{6(6)}$ structures: the generalized Lie derivatives, the E-bracket, 
and the associated tensor hierarchy. Employing these techniques, 
we define in sec.~\ref{sec3} the various terms of the E$_{6(6)}$ EFT action 
and discuss the (non-manifest) gauge invariance under the external, $5$-dimensional 
diffeomorphisms. In sec.~\ref{sec4} we prove that 11-dimensional supergravity 
can be embedded in EFT, upon solving the section constraint as above 
and re-writing 11-dimensional supergravity appropriate for the Kaluza-Klein inspired 
gauge fixing of the Lorentz group.  In sec.~\ref{sec5} we discuss the 
embedding and decomposition of type IIB supergravity along the same lines. 
We close with a summary and outlook in sec.~\ref{sec6}. In the appendix we discuss 
truncations of our theory, in order to relate it to some of the 
duality-covariant truncations previously obtained in the literature.

\section{E$_{6(6)}$ Generalized Diffeomorphisms and the Tensor Hierarchy}\label{sec2}
We start by introducing the mathematical background needed for the definition of the 
theory (\ref{finalaction}), including the E$_{6(6)}$ generalized Lie derivatives that 
generate the internal (generalized) diffeomorphisms and the `E-bracket'. 
Then we introduce the gauge fields
$A_{\mu}{}^{M}$ which gauge this symmetry in the sense of making it local w.r.t.~the 
`external' $x$-space. Due to the non-trivial Jacobiator of the E-bracket, gauge covariance 
requires the introduction of the two-form $B_{\mu\nu M}$  in accordance with the 
general tensor hierarchy of non-abelian $p$-forms~\cite{deWit:2005hv,deWit:2008ta}.

\subsection{Generalized Lie derivatives and the E-bracket}
We begin by collecting the relevant facts about the exceptional Lie group E$_{6(6)}$. 
Its Lie algebra is of dimension 78, with generators that we denote by  
$t_{\alpha}$ with the adjoint index $\alpha=1,\ldots, 78$. 
In addition, E$_{6(6)}$ has two 
inequivalent fundamental representations of dimension $27$, 
which we denote by ${\bf 27}$, and $\bar{\bf 27}$ for its contragredient. 
These representations will be indicated by lower indices $M,N=1,\ldots,27$ for ${\bf 27}$ 
and upper indices for $\bar{\bf 27}$.
Note, in particular, that there is no invariant metric to raise and lower fundamental indices. 
In contrast, we raise and lower adjoint indices by the (rescaled) Cartan-Killing form
$\kappa_{\alpha\beta}\equiv (t_{\alpha})_M{}^N (t_{\beta})_N{}^M$\,.

In the fundamental representation,
there are two cubic E$_{6(6)}$-invariant tensors, the fully symmetric $d$-symbols $d^{MNK}$ and $d_{MNK}$, 
which we normalize as $d_{MPQ}d^{NPQ} = \delta_M^N$.
Below we will need the projector onto the adjoint representation 
\bea\label{adjproj}
\mathbb{P}^M{}_N{}^K{}_L
&\equiv& (t_\alpha)_N{}^M (t^\alpha)_L{}^K ~=~
\frac1{18}\,\delta_N^M\delta^K_L + \frac16\,\delta_N^K\delta^M_L
-\frac53\,d_{NLR}d^{MKR}\;, 
\eea
which satisfies 
\bea
\mathbb{P}^M{}_N{}^N{}_M &=& 78\;.
\eea
We note the useful cubic relations for the $d$-symbols 
 \be\label{cubicidentity}
  \begin{split}
   d_{S(MN} \,d_{PQ)T}\, d^{STR} \ &= \ \frac{2}{15}\delta^{R}{}_{(M} \,d_{NPQ)}\;, \\
   d_{STR}\, d^{S(MN} \, d^{PQ)T}  \ &= \ \frac{2}{15}\delta_{R}{}^{(M} \,d^{NPQ)}\;. 
  \end{split}
 \ee  
 
Next, we introduce the generalized Lie derivative w.r.t.~the vector parameter $\Lambda^M$  
acting on E$_{6(6)}$ tensors in the 
fundamental representation with an arbitrary number of upper and lower indices.  
Moreover, the tensors can carry an arbitrary {\em density weight} $\lambda$. On a vector 
$V^M$ of  weight $\lambda$ it acts as \cite{Coimbra:2011ky,Berman:2012vc}
\bea\label{genLie}
\delta V^M \ = \ \mathbb{L}_{\Lambda} V^M 
\ \equiv \ \Lambda^K \partial_K V^M - 6\, \mathbb{P}^M{}_N{}^K{}_L\,\partial_K \Lambda^L\,V^N
+\lambda\,\partial_P \Lambda^P\,V^M
\;. 
\eea
Similarly, it acts on a co-vector $W_M$ of weight $\lambda'$ as
 \bea\label{coaction}
\delta W_M \ = \ \mathbb{L}_{\Lambda} W_M 
\ \equiv \ \Lambda^K \partial_K W_M + 6\, \mathbb{P}^N{}_M{}^K{}_L\,\partial_K \Lambda^L\,W_N
+\lambda'\,\partial_P \Lambda^P\,W_M
\;,
\eea
and accordingly on an E$_{6(6)}$ tensor 
with an arbitrary number of covariant and contravariant fundamental indices.
Because of the projector in (\ref{genLie}), the generalized Lie derivative is compatible with the ${\rm E}_{6(6)}$ algebra
structure: the $d$-symbols are invariant tensors of weight $\lambda=0$
\bea
\mathbb{L}_{\Lambda}\,d_{MNK} &=& 0
\;,
\eea
and its action on the E$_{6(6)}$ valued generalized metric ${\cal M}_{MN}$ to be introduced below 
(carrying weight $\lambda=0$) preserves the group property. 
Moreover, the above definition is such that the E$_{6(6)}$ invariant contraction between a vector and 
a co-vector transforms as 
 \be\label{scalartrans}
  \delta_{\Lambda}\big(V^M W_M\big) \ = \ \Lambda^K \partial_K\big(V^M W_M\big)
  +\big(\lambda+\lambda'\big)\partial_P \Lambda^P\,V^M W_M\;.
 \ee 
In particular, the contraction transforms as a genuine scalar if the vectors have opposite weights, 
$\lambda=-\lambda'$. Writing out the projector (\ref{adjproj}), the Lie derivative 
on, say, a vector reads explicitly 
\be\label{explicitLie}
   \delta _{\Lambda}V^M \ = \   \Lambda^K \partial_K V^M-\partial_K\Lambda^M V^K
  +\Big(\lambda-\frac{1}{3}\Big)\,\partial_P\Lambda^P\, V^M
  +10\, d_{NLR} d^{MKR}\partial_K\Lambda^L V^N\;. 
\ee
We observe that the projector contributes an additional density-type term, leading to 
an `effective weight' of $(\lambda-\frac{1}{3})$ in the action (\ref{explicitLie}),  
which singles out the value $\lambda=\frac{1}{3}$. 
In fact, we will see below that the vector gauge parameter itself has to be thought 
of as a vector of weight $\lambda=\frac{1}{3}$, such that (\ref{explicitLie}) carries no explicit weight term. 
We stress that by referring to the weight $\lambda$ of a tensor $V$, 
sometimes denoted by $\lambda(V)$,  
we always denote the weight in (\ref{genLie}), as opposed to the effective weight of (\ref{explicitLie}). 
In the following, a careful treatment of 
the emerging weights will be crucial.   
A remarkable observation is the following:
if $V_M$ is a covariant vector of weight $\lambda(V)=\frac23$, then the following combination 
\be\label{covariantdoperation}
   W^M \ \equiv \ d^{MNK}\,\partial_K V_N\;, 
\ee 
is a contravariant vector of weight $\lambda(W)  =  \frac{1}{3}$\,.
This can be viewed as an ${\rm E}_{6(6)}$ analogue of the fact that
for standard diffeomorphisms the exterior derivative $\partial_{[m_0} C_{m_1\dots m_p]}$ 
of an antisymmetric $p$-form 
is a covariant tensor (note, however, that the tensor $d^{MNK}$ in (\ref{covariantdoperation}) is totally
symmetric). Indeed, embedding the structures of ten- and eleven-dimensional space-time
diffeomorphisms, the tensor structure of (\ref{covariantdoperation}) precisely encodes those exterior derivatives,
as we will
find from the  explicit decompositions of the $d$-symbol in (\ref{dbreak}) and (\ref{dbreakB}) below.
The tensorial nature of (\ref{covariantdoperation}) will prove crucial for the structure of the tensor hierarchy of non-abelian $p$-forms.  
For a general study of connections and connection-free covariant derivatives in such `exceptional geometries'  
see~\cite{Coimbra:2011ky,Aldazabal:2013mya,Cederwall:2013naa}.

Let us now discuss a few properties of the generalized Lie derivatives, which 
all require the section constraints (\ref{sectioncondition}). First, we note that 
there are `trivial'  gauge parameters, i.e., gauge parameters that 
do not generate a gauge transformation via (\ref{genLie}). These are of the form 
 \be\label{trivialLambda}
  \Lambda^M \ = \ d^{MNK}\partial_N\chi_K\;, 
 \ee  
for an arbitrary covariant vector $\chi_K$.  
To prove this claim we compute from (\ref{explicitLie}) 
 \be\label{trivVar}
   \delta _{\Lambda}V^M 
    \ = \ \big(-d^{MPQ}\partial_N\partial_P\chi_Q+10d_{NLR} d^{MKR} d^{LPQ}\partial_K\partial_P\chi_Q\big)V^N\;. 
 \ee 
Here we have set to zero the transport term and the density term, since for the above parameter 
they vanish by the section constraints (\ref{sectioncondition}). 
Next we apply the cubic identity (\ref{cubicidentity}), noticing that
 \be\label{cubicStep}
 \begin{split}
   d_{RLN}d^{R(MK} d^{PQ)L}\partial_K\partial_P\chi_Q \ &= \ \frac{1}{6}\,d_{RLN}
   \big(2 d^{RMK} d^{PQL}+2 d^{RQK} d^{PML}\big)\partial_K\partial_P\chi_Q\\
   \ &= \ \frac{2}{3}\,d_{RLN}d^{RMK} d^{PQL}\partial_K\partial_P\chi_Q\;, 
 \end{split}
 \ee  
where we used the symmetry in $K,P$ and the section constraint. 
The cubic identity thus implies 
 \be\label{CUBicStep}
 \begin{split}
  10\, d_{RLN}\, d^{RMK}\, d^{PQL}\partial_K\partial_P\chi_Q
  \ = \ 2\, \delta_N{}^{(M}\,d^{KPQ)} \partial_K\partial_P\chi_Q
  \ = \  d^{PMQ}\partial_N\partial_P\chi_Q\;, 
 \end{split} 
 \ee 
where,  in the last equality, we used again the section constraint. 
Inserting this in (\ref{trivVar}) we observe that this cancels the first term, thus proving 
$\delta_{\Lambda}V^M=0$ and so triviality of the action of this gauge parameter. 
In the above proof we have given the detailed steps that will recur in similar form 
in many of the computations below, making repeated use of the section constraints 
 (\ref{sectioncondition}) and the cubic identity  (\ref{cubicidentity}). As such, 
 in the following derivations we will not repeat all intermediate steps in similar detail.

Next, we turn to the gauge algebra. 
A direct computation as above shows that, modulo the section constraints (\ref{sectioncondition}), 
the gauge transformations close
 \bea\label{gaugeclosure}
{}\big[ \delta_{\Lambda_1},\delta_{\Lambda_2}\big] \ = \ \delta_{[\Lambda_2,\Lambda_1]_{\rm E}}\;,
\eea
according to   the `E-bracket' 
\bea\label{Ebracket}
\big[\Lambda_2,\Lambda_1\big]^M_{\rm E} \ = \ 
2\Lambda_{[2}^K \partial_K \Lambda_{1]}^M
-10 \,d^{MNP}d_{KLP}\,\Lambda_{[2}^K \partial_N \Lambda_{1]}^L
\;.
\eea
Put differently, the generalized Lie derivatives satisfy 
the algebra~\cite{Coimbra:2011ky,Berman:2012vc}\footnote{Note that the seeming
sign difference between (\ref{gaugeclosure}) and (\ref{Liealgebra}) originates from the difference 
between a field variation, acting on fields first, and an abstract operator like the Lie derivative.} 
 \be\label{Liealgebra}
  \big[\,\mathbb{L}_{\Lambda_1},\mathbb{L}_{\Lambda_2}\,\big] \ = \ \mathbb{L}_{[\Lambda_1,\Lambda_2]_{\rm E}}\;.
 \ee 
The E-bracket is the M-theory or EFT analogue of the C-bracket in DFT. 
Like the C-bracket, the E-bracket does not define a Lie algebra in that it has 
a non-trivial `Jacobiator' 
 \be\label{Jacobiator}
  J(U,V,W) \ \equiv \ \big[\big[U,V\big]_{\rm E},W\big]_{\rm E}
  + \big[\big[V,W\big]_{\rm E},U\big]_{\rm E}
  + \big[\big[W,U\big]_{\rm E},V\big]_{\rm E}\;.
 \ee 
As in DFT, however, the Jacobiator takes the form of a trivial parameter 
(\ref{trivialLambda}) and is therefore consistent with the Jacobi identity 
for the symmetry variations, $[[\delta_{\Lambda_1},\delta_{\Lambda_2}],\delta_{\Lambda_3}]+{\rm cycl.}=0$.  
The proof is formally identical to that for the Courant bracket in generalized geometry  \cite{Gualtieri:2003dx}
or for the C-bracket in DFT \cite{Hull:2009zb} and proceeds as follows.\footnote{See also the analysis in the 
context of exceptional generalized geometry \cite{Coimbra:2011ky}, to which our discussion reduces for one 
solution of the section constraint.} First, we define the 
Dorfman-type product (or bracket) between vectors of weight $\frac{1}{3}$,
 \be\label{Dorfman}
  (V\circ W)^M \ \equiv \ (\mathbb{L}_{V}W)^M
  \ = \ V^N\partial_NW^M-W^N\partial_NV^M
  +10\,d^{MKR} d_{PLR} \,\partial_KV^L \,W^P\;. 
 \ee 
Comparison with (\ref{Ebracket}) then shows that the product differs from the E-bracket by a term 
symmetric in the two arguments, 
 \be\label{symmePart}
   (V\circ W)^M \ = \ \big[V,W\big]^M_{\rm E}+5\,d^{MKR}\partial_K\big(d_{RPL}V^P W^L\big)\;.
 \ee  
Note that the symmetric contribution takes the trivial form (\ref{trivialLambda}) and so 
$(V\circ W)$ and $[V,W]_{\rm E}$
generate the same generalized Lie derivative. Using this and the algebra  (\ref{Liealgebra}) it is 
straightforward to verify that the product satisfies the Jacobi-like identity
 \be\label{Jacobilike}
  U\circ (V\circ W)-V\circ(U\circ W)-(U\circ V)\circ W \ = \ 0\;.
 \ee 
In fact, with (\ref{Dorfman}) we compute 
 \be\label{JacobiProof}
  \begin{split} 
    U\circ (V\circ W)-V\circ(U\circ W) \ &= \ U\circ (\mathbb{L}_{V}W)-V\circ(\mathbb{L}_UW)\\
    \ &= \ \mathbb{L}_U\mathbb{L}_VW-\mathbb{L}_V\mathbb{L}_UW\\
    \ &= \ \mathbb{L}_{[U,V]_{\rm E}}W\\
    \ &= \ \mathbb{L}_{(U\circ V)}W \ = \ (U\circ V)\circ W\;, 
   \end{split}
  \ee   
thus proving (\ref{Jacobilike}). Next  we use (\ref{symmePart}) to compute 
 \be\label{JacobiSTEP}
 \begin{split}
   \big[\big[U,V\big]_{\rm E},W\big]_{\rm E} \ &= \ \big(\big[U,V\big]_{\rm E}\circ W\big)^M
   -5\,d^{MKR}\partial_K\big(\,d_{RPL}\,[U,V]_{\rm E}^P\, W^L\big)\\
   \ &= \ \big((U\circ V)\circ W\big)^M
  -5\,d^{MKR}\partial_K\big(\,d_{RPL}\,[U,V]_{\rm E}^P\, W^L\big) \;. 
  \end{split}
  \ee 
Using that as a consequence of (\ref{symmePart}) the E-bracket Jacobiator is proportional to 
the `Jacobiator' for the Dorfman product, one computes with the identity (\ref{Jacobilike}) 
 \be\label{JacobiatorFinal}
  J^M(U,V,W) \ = \ \frac{5}{3}\,d^{MKR}\partial_K
  \Big(d_{RPL}\big(\,[U,V]_{\rm E}^P\, W^L+[W,U]_{\rm E}^P\, V^L+[V,W]_{\rm E}^P \,U^L\,\big)\Big)\;.
 \ee 
This completes the proof that the Jacobiator is of the trivial form (\ref{trivVar}).

\subsection{E$_{6(6)}$ Tensor Hierarchy}

We now turn to a discussion of external covariant derivatives, gauge connections, and 
covariant curvatures.  
These are necessary because in the above gauge transformations we will take the gauge 
parameters  $\Lambda^M$ to be functions of the (internal) E$_{6(6)}$ coordinates $Y^M$ but also of 
the (external) 5-dimensional coordinates $x^{\mu}$. Thus, the gauge transformations are local 
w.r.t.~the $x$-space and the corresponding partial derivatives $\partial_{\mu}$ need to be 
covariantized. We thus introduce a gauge connection $A_{\mu}{}^{M}$ 
and define the covariant derivative 
 \be\label{GeneralCOV}
  {\cal D}_{\mu} \ \equiv \ \partial_{\mu}-\mathbb{L}_{A_{\mu}}\;.
 \ee 
For instance, the covariant derivative of a vector (of weight $\lambda$) is given by 
 \bea\label{specificcov}
{\cal D}_{\mu}V^M &=& \partial_{\mu} V^M -A_{\mu}{}^K \partial_K V^M + 6\, \mathbb{P}^M{}_N{}^K{}_L\,\partial_K 
A_{\mu}{}^L\,V^N
- \lambda \,\partial_P A_{\mu}{}^P\,V^M\;. 
\eea
Sometimes, we will explicitly split off the density term and write
\bea\label{Dderivative}
{\cal D}_{\mu}V^M &=& {D}_{\mu}V^M - \lambda \,\partial_P A_{\mu}{}^P\,V^M
\;,
\eea
for a vector $V^M$ of weight $\lambda$\,.
The transformation of the gauge connection 
is obtained by requiring gauge covariance of the covariant derivatives.
An explicit computation shows that with 
\bea\label{GaugeVar}
\delta A_\mu{}^M &=& \partial_\mu\Lambda^M - A_\mu{}^K\partial_K \Lambda^M + \Lambda^K \partial_K A_\mu{}^M 
-10\,d^{MNP}d_{KLP}\,\Lambda^L\, \partial_N A_\mu{}^K
\nonumber\\
&=& D_\mu \Lambda^M -\frac13\,(\partial_K A_\mu{}^K)\,\Lambda^M 
\nonumber\\
&\equiv & {\cal D}_\mu \Lambda^M
\;,
\eea
the covariant derivatives are indeed covariant. 
This confirms that the gauge parameter $\Lambda^M$ is a contravariant tensor
of weight $\lambda=\frac13$.

Next, we introduce a non-abelian field strength for the above gauge connection. 
The naive non-abelian Yang-Mills field strength reads 
\bea\label{NaiveYM}
\begin{split}
F_{\mu\nu}{}^M \ &= \  2\, \partial_{[\mu} A_{\nu]}{}^M  -\big[A_{\mu},A_{\nu}\big]^M_{\rm E}  \\
 \ &= \  2\, \partial_{[\mu} A_{\nu]}{}^M 
-2\,A_{[\mu}{}^K \partial_K A_{\nu]}{}^M 
+10\, d^{MKR}d_{NLR}\,A_{[\mu}{}^N\,\partial_K A_{\nu]}{}^L
\;.
\end{split}
\eea
Since the E-bracket does not satisfy the Jacobi identity, however, this field strength does not transform 
fully covariantly. We first compute its variation w.r.t.~an arbitrary $\delta A_{\mu}{}^{M}$,  which 
is a contravariant vector of weight $\lambda=\frac13$, 
\bea\label{deltaF}
\delta F_{\mu\nu}{}^M &=& 2\,  {\cal D}_{[\mu}\, \delta A_{\nu]}{}^M 
+10\, d^{MKR} d_{NLR}\,\partial_K \left( A_{[\mu}{}^N\, \delta A_{\nu]}{}^L \right)
\;.
\eea
The final term here is non-covariant, but of the
`trivial' form (\ref{trivialLambda}).  In the spirit of the tensor hierarchy~\cite{deWit:2005hv,deWit:2008ta}, this 
suggests to introduce two-form potentials $B_{\mu\nu\,M}$ and define the full covariant field strength by
\bea
{\cal F}_{\mu\nu}{}^M &\equiv&
F_{\mu\nu}{}^M + 10 \, d^{MNK}\,\partial_K B_{\mu\nu\,N}
\;,
\label{modF}
\eea
such that its general variation is given by
\bea\label{covFVar}
\delta {\cal F}_{\mu\nu}{}^M &=&
 2  {\cal D}_{[\mu}\, \delta A_{\nu]}{}^M  + 10 \, d^{MNK}\,\partial_K \Delta B_{\mu\nu\,N}
\;,
\eea
with
\bea\label{DELTAB}
\Delta B_{\mu\nu\,N} &\equiv& \delta B_{\mu\nu\,N} + d_{NKL}\,A_{[\mu}{}^K\, \delta A_{\nu]}{}^L
\;.
\eea
The covariant field strength also appears in the commutator of covariant derivatives, 
 \be
  \big[ {\cal D}_{\mu},{\cal D}_{\nu}\big] \ = \ -\mathbb{L}_{F_{\mu\nu}} \ = \  -\mathbb{L}_{{\cal F}_{\mu\nu}}\;, 
 \ee
where the last equality uses the triviality of  (\ref{trivialLambda}). 
With these results at hand we can now verify the gauge covariance of the curvature. In addition to the 
gauge symmetry parameterized by $\Lambda^M$, the newly introduced gauge potential $B_{\mu\nu\, M}$
comes with its own tensor gauge symmetry, whose parameter we denote by $\Xi_{\mu\, M}$. 
Explicitly, the complete gauge variations are given by
 \be
 \begin{split}
   \delta A_\mu{}^M \ &= \ D_\mu \Lambda^M -\frac13\,(\partial_K A_\mu{}^K)\,\Lambda^M -10\, d^{MNK} \partial_K\Xi_{\mu N}\;, \\
   \Delta B_{\mu\nu M} \ &= \ 2D_{[\mu}\Xi_{\nu]\,M} -\frac{4}{3}\left(\partial_K A_{[\mu}{}^K\right) \Xi_{\nu] M}
   +d_{MKL}\Lambda^K{\cal F}_{\mu\nu}{}^{L}+ {\cal O}_{\mu\nu M}\;, 
 \end{split}  
 \label{deltaAB}
 \ee
up to yet unspecified terms ${\cal O}_{\mu\nu M}$ satisfying 
 \be
  d^{MNK} \partial_K{\cal O}_{\mu\nu N} \ = \ 0\;,
 \ee 
which do not contribute to (\ref{covFVar}). 
It is a straightforward calculation to show that under (\ref{deltaAB}), the field strength (\ref{modF}) 
transforms as a contravariant vector (\ref{explicitLie}) of weight $\lambda=\frac13$\,.
Moreover, the form of (\ref{deltaAB}) shows that
the two-form gauge parameter $\Xi_{\mu\,M}$ is a covariant vector of weight $\lambda=\frac23$\,.

After having introduced a gauge covariant field strength, we will now discuss the Bianchi identities, 
which is also a convenient trick in order to define the covariant 
field strength of the two-form $B_{\mu\nu\,M}$. 
To this end we note the following useful relation, which follows from the observation in (\ref{covariantdoperation}), 
\bea\label{calDcommute}
{\cal D}_\mu \left(d^{MNK}\,\partial_K V_N\right) &=&
d^{MNK}\,\partial_K {\cal D}_\mu V_N
\;,
\eea
valid for any covariant vector $V_N$ of weight $\lambda=\frac23$\,.
Explicit computation shows that 
the field strength (\ref{modF}) satisfies the Bianchi identities
 \be
  3 \,{\cal D}_{[\mu}{\cal F}_{\nu\rho]}{}^M \ = \ 10\, d^{MNK}\partial_K{\cal H}_{\mu\nu\rho\,N}\;, 
  \label{Bianchi}
 \ee 
with the 3-form field strength ${\cal H}_{\mu\nu\rho \,M}$ defined by this equations
(up to terms that vanish under the projection with $d^{MNK} \partial_K$):
\bea\label{HCurvature}
\ {\cal H}_{\mu\nu\rho\,M} &=&
3\,{\cal D}_{[\mu} B_{\nu\rho]\,M}
-3\,d_{MKL}\, A_{[\mu}{}^K\,\partial_{\vphantom{[}\nu} A_{\rho]}{}^L 
+ 2\,d_{MKL}\, A_{[\mu}{}^K A_{\vphantom{[}\nu}{}^P \partial_P A_{\rho]}{}^L 
\nonumber\\
&&{}
-10\, d_{MKL}d^{LPR}d_{RNQ}\, A_{[\mu}{}^K A_{\vphantom{[}\nu}{}^N\,\partial_P A_{\rho]}{}^Q+\cdots 
\;.
\eea
W.r.t.~the generalized Lie derivative, this is a covariant vector of weight $\lambda=\frac23$\,.
Next, we determine the Bianchi identity for ${\cal H}_M$. 
From the derivative of~(\ref{Bianchi}) 
\bea\label{BianchiCOMP}
20\, d^{MNK}\partial_K {\cal D}_{[\mu} {\cal H}_{\nu\rho\sigma]N}
&=&
6{\cal D}_{[\mu}{\cal D}_{\vphantom{[}\nu} {\cal F}_{\rho\sigma]}{}^M 
\nonumber\\
&=&
 -15\,d^{MNK} \partial_K \left( d_{NPQ} {\cal F}_{[\mu\nu}{}^P {\cal F}_{\rho\sigma]}{}^Q \right) 
 \;, 
\eea
we conclude the Bianchi identity
\bea\label{calHBianchi}
4\, {\cal D}_{[\mu} {\cal H}_{\nu\rho\sigma]M} &=& 
-3 \,d_{MPQ} {\cal F}_{[\mu\nu}{}^P {\cal F}_{\rho\sigma]}{}^Q
+\dots\;,
\eea
again up to terms annihilated by the projection with $d^{MNK}\partial_K$.

\section{Covariant E$_{6(6)}$ Theory}\label{sec3}
We are now in the position to define all terms in the 
E$_{6(6)}$ EFT action (\ref{finalaction}), specifically the kinetic terms for 
the propagating fields $e_{\mu}{}^{a}$, ${\cal M}_{MN}$ and $A_{\mu}{}^{M}$. 
The dynamics of the two-form tensors $B_{\mu\nu M}$ is governed by a 
topological Chern-Simons-type term that implies the required duality relations between 
$A_{\mu}{}^{M}$ and $B_{\mu\nu M}$.  We 
define the `potential' term as a function of the generalized metric ${\cal M}_{MN}$
and the external metric $g_{\mu\nu}$, and
prove its gauge invariance under 
the internal generalized diffeomorphisms. Finally, 
we discuss the non-manifest invariance of the action
under the (covariantized) 5-dimensional external diffeomorphisms, 
which in turn fixes all relative coefficients of the action.

\subsection{Kinetic and Topological Terms}
Let us start by recalling the field content as given in 
(\ref{fieldcontent}) above:
\bea
\left\{e_\mu{}^a, {\cal M}_{MN}, A_\mu{}^M, B_{\mu\nu\,M} \right\}
\;.
\label{fieldcontent2}
\eea
In the following we define the kinetic terms for the first three fields. 
The 5-dimensional vielbein (`f\"unfbein') $e_{\mu}{}^{a}$ is a scalar-density 
under $\Lambda^M$ gauge transformations, with weight $\lambda=\frac{1}{3}$. 
In order to write a gauge invariant action we thus have to employ the 
covariant derivatives 
\bea
{\cal D}_\mu e_\nu{}^a &\equiv& \partial_\mu e_\nu{}^a - A_\mu{}^M\partial_M e_\nu{}^a
-\frac13\, \partial_MA_\mu{}^M e_\nu{}^a
\,, 
\label{covderE}
\eea
in the usual definition of the spin connection coefficients $\omega_{\mu}{}^{ab}$, 
which then become $\Lambda^M$ scalars (i.e.\ carry weight $\lambda=0$). 
The correspondingly covariantized Riemann tensor $R_{\mu\nu}{}^{ab}$ defined in the usual 
fashion then also transforms as a $\Lambda^M$ scalar.   However, because of 
the non-commutativity of the covariant derivatives ${\cal D}_{\mu}$, the covariantized 
Riemann tensor does not transform tensorially under local Lorentz transformations 
$\delta_{\lambda} \omega_{\mu}{}^{ab}=-{\cal D}_{\mu}\lambda^{ab}$. 
This can be repaired by defining the improved Riemann tensor \cite{Hohm:2013nja}
  \be
  \widehat{R}_{\mu\nu}{}^{ab} \ \equiv \  R_{\mu\nu}{}^{ab}+{\cal F}_{\mu\nu}{}^{M}
  e^{a}{}^{\rho}\partial_M e_{\rho}{}^{b}\,,
  \label{improvedRE6}
 \ee
which transforms covariantly under internal generalized diffeomorphisms 
and local Lorentz transformations.\footnote{One could also write an $A$-covariantized Einstein-Hilbert term  
in terms of the metric $g_{\mu\nu}$, in which case there is no such extra term present, 
Lorentz symmetry being already manifest.} The covariantized Einstein-Hilbert term 
 \be
  S_{\rm EH} \ = \ \int d^5x \,d^{27}Y\,e\,\widehat{R} \ = \ 
  \int d^5x \,d^{27}Y\,e\,e_{a}{}^{\mu}e_{b}{}^{\nu} \widehat{R}_{\mu\nu}{}^{ab}\;, 
 \ee
then is gauge invariant under  these symmetries. In particular, 
the weight $\lambda=\frac{5}{3}$ carried by the f\"unfbein determinant $e$ according to (\ref{covderE}),
combines with the weights of the inverse f\"unfbeins to a total weight of $1$,
as required in order for the Lagrangian to vary under $\Lambda^M$ transformations into a 
total derivative. 

Next, we turn to the kinetic term for ${\cal M}_{MN}$. 
This matrix parametrizes the scalar coset space ${\rm E}_{6(6)}/{\rm USp}(8)$ 
whose 42 coordinates describe the scalar fields of the theory. Under 
generalized diffeomorphisms (\ref{coaction}) it 
transforms as a symmetric 2-tensor of weight $\lambda'=0$. Note in particular, that this transformation
is compatible with the group property ${\rm det}\,{\cal M}=1$\,.
Introducing its covariant 
derivative according to (\ref{GeneralCOV}),
we can define the gauge invariant kinetic term  
\bea
{\cal L}_{\rm sc} &=& \frac1{24}\,e\,g^{\mu\nu}\,{\cal D}_{\mu}{\cal M}_{MN}\,{\cal D}_{\nu}{\cal M}^{MN}
\;,
\eea
with the inverse matrix ${\cal M}^{MN}$.
In particular, with the inverse metric $g^{\mu\nu}$ 
carrying weight $\lambda=-\frac{2}{3}$ and the f\"unfbein determinant 
carrying weight $\lambda=\frac{5}{3}$, the total weight of this term in the Lagrangian is $1$, 
as required for $\Lambda^M$ gauge invariance. 
Similarly, the Yang-Mills kinetic term $-\frac{1}{4}\,e\,{\cal M}_{MN}\,{\cal F}^{\mu\nu M}{\cal F}_{\mu\nu}{}^{N}$
in  (\ref{finalaction}) carries the correct weight of 1 and is hence gauge invariant.  
Indeed, we saw above that the field strengths ${\cal F}_{\mu\nu}{}^{M}$ are gauge covariant and 
carry  a weight of $\lambda=\frac{1}{3}$, which is precisely the correct weight 
given the presence of two inverse metrics $g^{\mu\nu}$. 
 
After having discussed the kinetic terms, we now turn to the topological 
Chern-Simons-like term. By this we mean a term that is written without use 
of the metric (i.e., only through exterior products of forms) and that contains 
bare gauge potentials such that it is only gauge invariant up to boundary 
terms. Its structure is analogous to the topological term in general $D=5$
gauged supergravity \cite{deWit:2004nw}, such that its field equations yield the
desired first order duality equations relating $A_\mu{}^M$ and $B_{\mu\nu\,M}$\,.
Such a term may be written more conveniently as a total derivative in 
one higher dimension, which has the advantage of making the gauge invariance manifest. 
Using form notation for the invariant curvatures introduced in (\ref{modF}) and (\ref{HCurvature}), 
 \be
  {\cal F}^M \ \equiv \ \frac{1}{2}\,{\cal F}_{\mu\nu}{}^{M}\,dx^{\mu}\wedge dx^{\nu}\;, \qquad
  {\cal H}_{M} \ \equiv \ \frac{1}{3!}\,{\cal H}_{\mu\nu\rho M}\,dx^{\mu}\wedge dx^{\nu}\wedge dx^{\rho}\;, 
 \ee
the topological term can be written as  an integral of an exact 6-form over a 6-dimensional space ${\cal M}_6$,   
\bea
S_{\rm top} &=& 
\int d^5 x\,d^{27}Y\,{\cal L}_{\rm top} 
\nonumber\\
&=&
\kappa \!\int d^{27}Y \int_{{\cal M}_6}\,\left(
d_{MNK}\,{\cal F}^M \wedge  {\cal F}^N \wedge  {\cal F}^K
-40\, d^{MNK}{\cal H}_M\,  \wedge \partial_N{\cal H}_K
\right)
\,,
\label{CSlike}
\eea
whose overall constant $\kappa$ will be determined below. 
From this we may determine the non-manifestly gauge invariant 5-dimensional form,
but it is not very illuminating and will also not be needed in the following. What will be needed in the 
following is the general variation of the topological term, which is derived from (\ref{CSlike})
and takes the form
 \bea
\delta{\cal L}_{\rm top} &=&
\kappa\,\varepsilon^{\mu\nu\rho\sigma\tau}
\Big(\,\frac34\,d_{MNK}\,
{\cal F}_{\mu\nu}{}^M {\cal F}_{\rho\sigma}{}^N  \delta A_\tau{}^K
+5\,d^{MNK} d_{KPQ}\,\partial_N{\cal H}_{\mu\nu\rho \,M} \,
A_{\sigma}{}^P\delta A_{\tau}{}^Q
\nonumber\\
&&{}\qquad\qquad \;\,
+5\,d^{MNK}\,\partial_N{\cal H}_{\mu\nu\rho \,M} \,
\delta B_{\sigma\tau\,K} \Big)
\;. 
\label{vartopo0}
\eea
In terms of the covariant variation (\ref{DELTAB}) it takes the even simpler form 
  \bea
\delta{\cal L}_{\rm top} &=&
\kappa\,\varepsilon^{\mu\nu\rho\sigma\tau}
\Big(\,\frac34\,d_{MNK}\,
{\cal F}_{\mu\nu}{}^M {\cal F}_{\rho\sigma}{}^N  \delta A_\tau{}^K
+5\,d^{MNK}\,\partial_N{\cal H}_{\mu\nu\rho \,M} \,
\Delta B_{\sigma\tau\,K} \Big)
\;.
\label{vartopo}
\eea
With this form it is straightforward to explicitly verify gauge invariance under 
$\Lambda$ and $\Xi$ transformations (\ref{deltaAB}), integrating by parts and 
using the Bianchi identities (\ref{Bianchi}) and (\ref{calHBianchi}). 
Note that due to (\ref{calDcommute}) in this computation 
we can exchange the relevant $\partial_M$ and ${\cal D}_{\mu}$
derivatives. 

We close this subsection by giving the field equations of 
the topological fields $B_{\mu\nu M}$, which enter the topological term  and 
the Yang-Mills term via the covariant field strength 
${\cal F}_{\mu\nu}{}^{M}$. The field equations obtained 
by varying $B_{\mu\nu P}$ in these terms read 
 \be\label{dualityrel}
d^{PML}\partial_L  \left(e{\cal M}_{MN} {\cal F}^{\mu\nu N}
 +\kappa  \varepsilon^{\mu\nu\rho\sigma\tau}\,
  {\cal H}_{\rho\sigma\tau M}\right) \ = \ 0
  \,. 
  \ee
We will see in the following sections that upon taking 
appropriate solutions of the constraints (\ref{sectioncondition}), 
these relations reduce to the required first-order duality 
relations of either 11-dimensional supergravity or type IIB
supergravity.

\subsection{The Potential}

We now discuss the final term in the EFT action: the potential, which is 
a function of $g_{\mu\nu}$ and ${\cal M}_{MN}$ 
given by
\be\label{fullpotential}
 \begin{split}
  V \ = \ &-\frac{1}{24}{\cal M}^{MN}\partial_M{\cal M}^{KL}\,\partial_N{\cal M}_{KL}+\frac{1}{2} {\cal M}^{MN}\partial_M{\cal M}^{KL}\partial_L{\cal M}_{NK}\\
  &-\frac{1}{2}g^{-1}\partial_Mg\,\partial_N{\cal M}^{MN}-\frac{1}{4}  {\cal M}^{MN}g^{-1}\partial_Mg\,g^{-1}\partial_Ng
  -\frac{1}{4}{\cal M}^{MN}\partial_Mg^{\mu\nu}\partial_N g_{\mu\nu}\;. 
 \end{split} 
 \ee 
The relative coefficients in here are determined by $\Lambda^M$ gauge invariance, 
and in the following we will verify this gauge symmetry.
As the potential is an E$_{6(6)}$ singlet, with all indices being properly contracted, 
it is sufficient to verify cancellation of all terms that are `non-covariant' in the following sense.   
For a generic object with an arbitrary number of upper and lower E$_{6(6)}$ fundamental indices, 
we define 
 \be
  \Delta_{\Lambda} \ \equiv \ \delta_{\Lambda}-\mathbb{L}_{\Lambda}\;.
 \ee 
Put differently, by $\Delta$ we denote all terms in its variation that differ from the covariant ones (in turn given by 
the generalized Lie derivative). As the covariant generalized Lie derivative terms automatically 
combine into the Lie derivative of a scalar, it is sufficient to verify cancellation of 
the non-covariant terms. 
The only terms that lead to a non-trivial $\Delta$ are those 
involving a partial derivative, so we have to compute those terms for $\partial {\cal M}$ and 
$\partial g$. First, we  compare  
 \be\label{firstgaugevar}
  \delta_{\Lambda}\big(\partial_M{\cal M}^{KL}\big) \ = \ 
  \partial_M\big(\Lambda^P\partial_P {\cal M}^{KL}
   -12\,\mathbb{P}^{(K}{}_{R}{}^{|P}{}_{Q}\partial_P\Lambda^{Q|} {\cal M}^{L)R}\big)\;, 
 \ee 
with the covariant 
 \be\label{genLieparM}
 \begin{split}
  \mathbb{L}_{\Lambda} \big(\partial_M{\cal M}^{KL}\big) 
  \ = \ &\Lambda^{P}\partial_P\big(\partial_M{\cal M}^{KL}\big) 
   -12\,\mathbb{P}^{(K}{}_{R}{}^{|P}{}_{Q}\,\partial_P\Lambda^{Q|}\, \partial_M{\cal M}^{L)R} \\
   &+6\, \mathbb{P}^R{}_{M}{}^{P}{}_{Q}\,\partial_P\Lambda^Q\,\partial_R{\cal M}^{KL}
   +\lambda\, \partial_P\Lambda^P\,\partial_M{\cal M}^{KL}\;. 
 \end{split}
 \ee
Here we introduced $\lambda$ in order to allow for a possible weight of $\partial{\cal M}$. 
In fact, we will show momentarily that although ${\cal M}$ has weight zero, its derivative 
has a non-trivial weight. To see this we note that the first term in the second line of 
(\ref{genLieparM}) simplifies by the section constraint, so that writing out 
the projector according to (\ref{adjproj}) we obtain 
 \be\label{genLieparM2}
 \begin{split}
  \mathbb{L}_{\Lambda} \big(\partial_M{\cal M}^{KL}\big) 
  \ = \ &\Lambda^{P}\partial_P\big(\partial_M{\cal M}^{KL}\big) 
   -12\,\mathbb{P}^{(K}{}_{R}{}^{|P}{}_{Q}\,\partial_P\Lambda^{Q|}\, \partial_M{\cal M}^{L)R} \\
   &+\frac{1}{3}\,\partial_P\Lambda^P\,\partial_M{\cal M}^{KL}+\partial_M\Lambda^P \,\partial_P{\cal M}^{KL}
   +\lambda\, \partial_P\Lambda^P\,\partial_M{\cal M}^{KL}\;. 
 \end{split}
 \ee
In (\ref{firstgaugevar}) there are no density-type terms, so in order to match this as closely 
as possible with (\ref{genLieparM2}) we have to cancel the density term by setting $\lambda=-\frac{1}{3}$. 
We then infer that (\ref{firstgaugevar})  agrees with (\ref{genLieparM2}), up to terms that 
involve second derivatives of the gauge parameter. In total, we have shown that 
$\partial{\cal M}$ comes with weight $\lambda=-\frac{1}{3}$ while its non-covariant variation is given by 
  \be\label{noncovdelM}
   \Delta_{\Lambda}\big(\partial_M{\cal M}^{KL}\big) \ = \ 
   -12\,\mathbb{P}^{(K}{}_{R}{}^{|P}{}_{Q}\,\partial_M\partial_P\Lambda^{Q|} \,{\cal M}^{L)R}\;. 
 \ee
Similarly, we have 
 \be
    \Delta_{\Lambda}(\partial_M{\cal M}_{KL}) \ = \ 
   +12\, \mathbb{P}^{R}{}_{(K}{}^{P}{}_{|Q}\partial_M\partial_{P|} \Lambda^{Q} {\cal M}_{L)R}\;,  
 \ee
again taking  $\partial{\cal M}$ to have weight $\lambda=-\frac{1}{3}$. 
Taking the trace of (\ref{noncovdelM}) we obtain in particular
 \be
  \Delta_{\Lambda}\big(\partial_N{\cal M}^{MN}\big) \ = \ 
  -\frac{5}{3}\partial_N \partial_P\Lambda^P  {\cal M}^{MN}
  -\partial_N\partial_P\Lambda^M {\cal M}^{PN}+\cdots\;,
 \ee 
up to terms that vanish upon contraction with $\partial_M$ by the section constraint.  
Finally we need to determine $\Delta_{\Lambda}$ for $\partial g$. 
By an exactly analogous computation we find that $g^{-1}\partial g$
has weight $\lambda=-\frac{1}{3}$. Moreover, derivatives $\partial_M$ acting on $g^{\mu\nu}$ and 
$g_{\mu\nu}$ induce additional weights of $-\frac{1}{3}$, such that we find the total weights to be
 \be
  \lambda\big(g^{-1}\partial_M g\big) \ = \ -\frac{1}{3}\;, \qquad 
  \lambda\big(\partial_M g^{\mu\nu}\big) \ = \ -1\;, \qquad
  \lambda\big(\partial_M g_{\mu\nu}\big) \ = \ \frac{1}{3}\;,
 \ee 
with the non-covariant gauge variations given by    
 \be\label{noncovdg}
 \begin{split}
  \Delta_{\Lambda}(g^{-1}\partial_Mg) \ &= \ \frac{10}{3}\partial_M\partial_P\Lambda^P
  \;, \\
  \Delta_{\Lambda}(\partial_Mg^{\mu\nu})  \ &= \ -\frac{2}{3}\partial_M\partial_P\Lambda^P g^{\mu\nu}\;, \\  
   \Delta_{\Lambda}(\partial_Mg_{\mu\nu})  \ &= \ 
   \frac{2}{3}\partial_M\partial_P\Lambda^P g_{\mu\nu}\;.
 \end{split}
 \ee  
 
Let us now verify gauge invariance of the potential. First, we note that 
the weights of the partial derivatives of the fields are as required in order to 
combine to a total weight of $1$ with the weight $\lambda=\frac{5}{3}$ of the f\"unfbein determinant $e$
multiplying the potential term in the action. 
Thus, complete $\Lambda$ invariance of the action is proven once we checked  
that all $\Delta_{\Lambda}$ variations above cancel, which we will now show.  
We compute for the first term of (\ref{fullpotential})
 \be\label{noncovcomp}
 \begin{split}
  \delta_{\Lambda}\Big(-\frac{1}{24}\,e{\cal M}^{MN}\partial_M{\cal M}^{KL}\,\partial_N{\cal M}_{KL}\Big)
  \ &= \ \frac{1}{6} \,e\partial_M\partial_P\Lambda^K \,{\cal M}^{MN}{\cal M}^{PL}\partial_N{\cal M}_{KL}\\
  &\quad -\frac{5}{3} \,e d_{RQS} d^{KPS} \partial_M\partial_P\Lambda^Q{\cal M}^{MN}{\cal M}^{RL}\partial_N{\cal M}_{KL} \\
  \ &= \  e\partial_M\partial_P\Lambda^K \,{\cal M}^{MN}{\cal M}^{PL}\partial_N{\cal M}_{KL}\;. 
 \end{split}
 \ee 
Here, in the second equality, we used that ${\cal M}$ is E$_{6(6)}$ valued 
with determinant 1, 
which allows for simplifications. In order to explain this  we first note that 
the current 
\be
 (J_N)^K{}_{L}\ \equiv \ {\cal M}^{KP}\partial_N{\cal M}_{PL}\;,  
\ee 
lives in the adjoint representation and is traceless.  
Therefore it satisfies 
 \be
  \mathbb{P}^{M}{}_{N}{}^{K}{}_{L} (J_P)^{L}{}_{K} \ = \ (J_P)^{M}{}_{N}\;. 
 \ee  
Spelling out the projector with (\ref{adjproj}), this condition implies:
 \be
  d_{NLS}\,d^{MKS}\, J^L{}_{K} \ = \ -\frac{1}{2}J^{M}{}_{N}\;.
 \ee  
Using this in the second term on the right-hand side of the first equality in (\ref{noncovcomp})
then reproduces the final equality.  
For the second term in the potential (\ref{fullpotential})
we compute  
 \be\label{secondMcube}
 \begin{split}
  \delta_{\Lambda}\Big(\,\frac{1}{2}\, e {\cal M}^{MN}\partial_M{\cal M}^{KL}&\partial_L{\cal M}_{NK}\Big) 
  \ = \ \frac{2}{3} e\partial_M\partial_P\Lambda^P\,\partial_N{\cal M}^{MN}\\
  &- e\partial_M\partial_P\Lambda^K\,{\cal M}^{MN}{\cal M}^{PL}\partial_N{\cal M}_{LK}
  +e \partial_M\partial_P\Lambda^L\,\partial_L{\cal M}^{MP}\;. 
 \end{split}
 \ee 
Here we used again that the current $J$ is Lie algebra valued, so that the invariance of 
the $d$-symbol implies
 \be
 0 \ = \  3 d^{K(SP} J^{M)}{}_{K}  \ = \ d^{KPM} J^{S}{}_{K}
 +2 d^{SK(P} J^{M)}{}_{K}\;.
 \ee
The last term in here appears in the above variation, and by this relation has been rewritten 
in terms of the first term, which then in turn gives zero by the section constraint. 
We observe that the cubic term in ${\cal M}$ in (\ref{secondMcube}) precisely 
cancels the same term in (\ref{noncovcomp}), which in turn determined the 
relative coefficient between these terms.  
Using (\ref{noncovdg}) it is straightforward to verify that the remaining terms linear in 
$\partial{\cal M}$ are cancelled by the $\Delta_{\Lambda}$ variation of the terms in the second line 
of (\ref{fullpotential}). This proves the full $\Lambda^M$ gauge invariance of the 
potential.

\subsection{$(4+1)$-dimensional Diffeomorphisms}

In the previous subsections, we have determined the various terms of the EFT action (\ref{finalaction})
by invariance under generalized internal $\Lambda^M$ diffeomorphisms. While this has uniquely
fixed the form of the five different terms in (\ref{finalaction}), they could in principle have appeared
with arbitrary relative coefficients. In this section we show that all relative factors are determined by
invariance of the full action under the remaining gauge symmetries, which are a covariantized 
version of the $(4+1)$-dimensional diffeomorphisms with parameters $\xi^{\mu}(x,Y)$. 
If $\xi^{\mu}$ is independent of $Y$ these 
are manifest symmetries for each term in the action separately. 
For general $\xi^{\mu}$, however, this gauge invariance is far from manifest
and in particular relates all terms in the action. As a result, the action (\ref{finalaction})
is the unique action (with no free parameter left up to an overall rescaling) 
that is not only invariant under generalized internal
diffeomorphisms $\Lambda^{M}(x,Y)$ but also under the appropriate version of the
external diffeomorphisms $\xi^{\mu}(x,Y)$.
The action of these diffeomorphisms on the various fields are given by 
 \bea
 \delta e_{\mu}{}^{a} &=& \xi^{\nu}{\cal D}_{\nu}e_{\mu}{}^{a}
 + {\cal D}_{\mu}\xi^{\nu} e_{\nu}{}^{a}\;, \nonumber\\
\delta {\cal M}_{MN} &=& \xi^\mu \,{\cal D}_\mu {\cal M}_{MN}\;,\nonumber\\
\delta A_{\mu}{}^M &=& \xi^\nu\,{\cal F}_{\nu\mu}{}^M + {\cal M}^{MN}\,g_{\mu\nu} \,\partial_N \xi^\nu
\;,\nonumber\\
\Delta B_{\mu\nu\,M} &=& \frac1{16\kappa}\,\xi^\rho\,
 e\varepsilon_{\mu\nu\rho\sigma\tau}\, {\cal F}^{\sigma\tau\,N} {\cal M}_{MN} 
 \;, 
 \label{skewD}
\eea
written for $B_{\mu\nu\,M}$ in terms of the covariant variation (\ref{DELTAB}). 
They take the form of conventional diffeomorphisms, but `covariantized' with 
respect to the connection $A$ of the separate $\Lambda$ gauge symmetry, except 
for an additional ${\cal M}$-dependent term in $\delta A_{\mu}{}^M$ and 
an on-shell modification in $\Delta B_{\mu\nu\,M}$. More precisely, 
the naive covariant variation of $B_{\mu\nu\,M}$ would take the form 
$\Delta_{\xi}B_{\mu\nu\,M}=\xi^{\rho}\,{\cal H}_{\mu\nu\rho \,M}$, with the covariant 
field strength defined in (\ref{HCurvature}), but it turns out that 
off-shell gauge invariance of the action requires to replace this field strength 
according to the duality relation (\ref{dualityrel}). Thus, the gauge variations 
(\ref{skewD}) are only on-shell equivalent to the conventional form of 
(covariantized) diffeomorphisms. 

Next, we discuss the gauge invariance of the action under (\ref{skewD})
in some detail. The explicit verification of this gauge invariance is quite  
tedious and so we focus on a subset of terms that provide a very strong consistency 
check and that are 
sufficient in order to determine all relative coefficients in the action. 
Specifically, for various structures the cancellation proceeds completely parallel to the calculation
that ensures standard diffeomorphism invariance in eleven-dimensional supergravity
in a $5+6$ splitting of fields and coordinates. They can therefore be omitted. In particular, 
as explained in \cite{Hohm:2013jma}, terms linear in ${\cal M}$ that are of the structural form 
${\cal M}^{MN}\partial_M(\cdots)\partial_{N}(\dots)$ have to cancel separately, and 
this computation is formally identical to the corresponding one for standard diffeomorphisms. 
In the following we focus on those terms for which cancellation 
involves the novel features of the EFT action.

We start by computing the variation of the sum of Yang-Mills and the topological term, 
denoted in the following by ${\cal L}_{\rm VT}$, 
 \bea
{\cal L}_{\rm VT}&\equiv &
-\frac14\,e\,{\cal F}_{\mu\nu}{}^M{\cal F}^{\mu\nu\,N}\,{\cal M}_{MN}
+\kappa\,{\cal L}_{\rm CS}
\;.
\label{LVT}
\eea
Using (\ref{vartopo})  one easily sees that its general variation is given by  
\bea
\delta {\cal L}_{\rm VT} &=&{}
\left( \kappa\,\varepsilon^{\mu\nu\rho\sigma\tau}\,
d_{MNK}\,{\cal F}_{\nu\rho}{}^K {\cal F}_{\sigma\tau}{}^N
- {\cal D}_{\nu}\left( e{\cal M}_{MN}\,{\cal F}^{\mu\nu\,N}\right)
\right) \delta A_{\mu}{}^M
\nonumber\\
&&{}
+5 d^{MKN}\, \partial_K\left(   e\, {\cal F}^{\mu\nu\,N}\,{\cal M}_{MN} 
+\frac{4\kappa}3\,\varepsilon^{\mu\nu\rho\sigma\tau}\,
  {\cal H}_{\rho\sigma\tau\,M}\right) \Delta B_{\mu\nu\,N} 
  \nonumber\\
&&{}
+ {\cal O}(\delta g_{\mu\nu}) + {\cal O}(\delta {\cal M}_{MN})
  \;.
  \label{varLVT}
\eea
Next, we insert the gauge variations (\ref{skewD}) and first focus on the 
${\cal F}\wedge{\cal F}$ terms in the variation: 
 \bea\label{VTvariation}
\delta {\cal L}_{\rm VT}\Big|_{{\cal F}\wedge{\cal F}} &=&{}
 \kappa\,\varepsilon^{\mu\nu\rho\sigma\tau}\,
d_{MNK}\,{\cal F}_{\nu\rho}{}^K {\cal F}_{\sigma\tau}{}^N
 {\cal M}^{ML}\,g_{\mu\lambda} \,\partial_L \xi^\lambda
\nonumber\\
&&{}
+\frac5{16\kappa} d^{MKN}\, \partial_K\left(   e\, {\cal F}^{\mu\nu\,Q}\,{\cal M}_{MQ} 
\right) 
\xi^\rho\,
 e\varepsilon_{\mu\nu\rho\sigma\tau}\, {\cal F}^{\sigma\tau\,P} {\cal M}_{NP}  
 \nonumber\\[2ex]
 &=&
 \kappa\,\varepsilon^{\mu\nu\rho\sigma\tau}\,
d_{MNK}\, {\cal M}^{ML}\,{\cal F}_{\nu\rho}{}^K {\cal F}_{\sigma\tau}{}^N\,
g_{\mu\lambda} \,\partial_L \xi^\lambda
\nonumber\\
&&{}
-\frac5{32\kappa} \,\varepsilon^{\mu\nu\sigma\tau\rho}\,
d^{MKN}\,{\cal M}_{MQ}  {\cal M}_{NP}\,  
    {\cal F}_{\mu\nu}{}^{P}{\cal F}_{\sigma\tau}{}^{Q}\, 
g_{\rho\lambda}\,\partial_K \xi^\lambda
\;.
\eea
We can simplify this variation by 
using that ${\cal M}$ is E$_{6(6)}$ valued, so that the 
invariance of the $d$-symbol implies  
$d^{MKN}{\cal M}_{MQ}  {\cal M}_{NP}=d_{PQM}{\cal M}^{MN}$. 
Using this in (\ref{VTvariation}) we infer that this variation vanishes for
\bea
\kappa^2 &=& \frac5{32}
\;. 
\eea
Let us now return to (\ref{varLVT}) 
and focus on the variation coming from the second term in the first line, 
restricted to the covariant, ${\cal M}$-independent term of $\delta A_{\mu}{}^{M}$ in (\ref{skewD}).
Integrating by parts we compute 
\bea
e\,  {\cal F}^{\mu\nu\,N}\,{\cal M}_{MN}
\, {\cal D}_{\nu}\left( \xi^\rho\,{\cal F}_{\rho\mu}{}^M \right)
&=&
e\,  {\cal F}^{\mu\nu\,N}\,{\cal M}_{MN}
\, {\cal D}_{\nu} \xi^\rho \,{\cal F}_{\rho\mu}{}^M 
-\frac12\,e\,  {\cal F}^{\mu\nu\,N}\,{\cal M}_{MN}
\,  \xi^\rho\, {\cal D}_{\rho} {\cal F}_{\mu\nu}{}^M 
\nonumber\\
&&{}
+5\,e\, d^{MPQ}  \xi^\rho\,{\cal F}^{\mu\nu\,N}\,{\cal M}_{MN}
\, \partial_P {\cal H}_{\mu\nu\rho\,Q} 
\;,
\label{testX}
\eea
where we rewrote the ${\cal D}{\cal F}$ term as a total curl and then 
used the Bianchi identity (\ref{Bianchi}) 
in the last term in the second line. Let us note that the first two terms 
of (\ref{testX}) occur already in completely analogous form in the usual 
diffeomorphism variation, and so their cancellation against 
the variation of $g_{\mu\nu}$ and ${\cal M}_{MN}$ from (\ref{varLVT}) is standard.
The term in the last line originating from the novel Bianchi 
identity, however, needs to be cancelled separately. 
This is achieved by the variation originating from the second term in 
the second line of (\ref{varLVT}). In fact, inserting $\Delta B$ 
from (\ref{skewD}) we compute for this term 
\be\nonumber
\frac{5}{12}\, e\varepsilon_{\mu\nu\lambda\sigma'\tau'}\,
\varepsilon^{\mu\nu\rho\sigma\tau}\,
d^{MKN}\,  
\partial_K  {\cal H}_{\rho\sigma\tau\,M} 
\xi^\lambda\,
  {\cal F}^{\sigma'\tau'\,Q} {\cal M}_{NQ} 
  \ = \ 
  -5\,e\,
  d^{MKN}\,  
\partial_K  {\cal H}_{\rho\sigma\tau\,M} 
\xi^\rho\,
  {\cal F}^{\sigma\tau\,Q} {\cal M}_{NQ} 
  \;, 
\ee
which cancels precisely the final term in (\ref{testX}). 

Let us next inspect the variation of the second term in the first line of (\ref{varLVT}), 
but now under the non-covariant, ${\cal M}$-dependent term of $\delta A_{\mu}{}^{M}$ in (\ref{skewD}). 
Upon integration by parts we obtain
\bea
- {\cal D}_{\nu}\left( e{\cal M}_{MN}\,{\cal F}^{\mu\nu\,N}\right)
 {\cal M}^{MK}\,g_{\mu\rho}\,\partial_K\xi^\rho
 &=&
 e{\cal F}_{\mu\nu}{}^M\,{\cal M}_{MN}\,
  {\cal D}^{\nu} {\cal M}^{NK}\,\partial_K\xi^\mu
  \nonumber\\
  &&{} -
{\cal F}^{\mu\nu\,M}
 \, {\cal D}_{\mu} \left(g_{\nu\rho}\,\partial_M\xi^\rho\right)\;. 
 \label{vartermF}
\eea
The second term precisely cancels against the main contribution from variation of the 
Einstein-Hilbert term. This computation is formally identical to that 
presented in   \cite{Hohm:2013jma}, c.f.~eq.~(4.16) in that paper. 
The first term in (\ref{vartermF}) will cancel against the variation of the 
scalar kinetic term. In order to show this, let us first compute the 
variation of the `scalar current':   
\bea
\delta_{\xi}\big( {\cal D}_\mu {\cal M}_{MN}\big) &=&
{\cal D}_\mu \left(\xi^\nu {\cal D}_\nu {\cal M}_{MN} \right)
-\mathbb{L}_{\delta A_\mu}  {\cal M}_{MN} 
\nonumber\\
&=&
  {\cal L}_\xi \left({\cal D}_\mu {\cal M}_{MN}\right)
  - \xi^\nu\,\mathbb{L}_{{\cal F}_{\mu\nu}}{\cal M}_{MN}
+\mathbb{L}_{\xi^\nu {\cal F}_{\mu\nu}}  {\cal M}_{MN}
-\mathbb{L}_{{\cal M}^{\bullet K} g_{\mu\nu}\partial_K\xi^\nu}  {\cal M}_{MN}
\nonumber\\
&=&
{\cal L}_\xi \left({\cal D}_\mu {\cal M}_{MN}\right)
+12\,\mathbb{P}^P{}_Q{}^K{}_{(M} {\cal M}_{N)K}\,{\cal F}_{\mu\nu}{}^Q\,\partial_P \xi^\nu 
\nonumber\\
&&
-{\cal M}^{KL}  \partial_L {\cal M}_{MN}\,g_{\mu\nu}\partial_K\xi^\nu
+12\,\mathbb{P}^P{}_Q{}^K{}_{(M}{\cal M}_{N)K}\,
\partial_P \left({\cal M}^{LQ} g_{\mu\nu}\partial_L\xi^\nu \right)
\nonumber\\
&=&
{\cal L}_\xi \left({\cal D}_\mu {\cal M}_{MN}\right)
+\frac23\,{\cal M}_{MN}\,{\cal F}_{\mu\nu}{}^P\,\partial_P \xi^\nu 
+2\,{\cal F}_{\mu\nu}{}^K\,{\cal M}_{K(M}\partial_{N)} \xi^\nu 
\nonumber\\
&&
-20\,d^{PKL}d_{QL(M} {\cal M}_{N)K}\,{\cal F}_{\mu\nu}{}^Q\,\partial_P \xi^\nu 
-{\cal M}^{KL}  \partial_L {\cal M}_{MN}\,g_{\mu\nu}\partial_K\xi^\nu
\nonumber\\
&&
+\frac23\,{\cal M}_{MN}\,
\partial_P \left({\cal M}^{LP} g_{\mu\nu}\partial_L\xi^\nu \right)
+2\,{\cal M}_{K(M}\,
\partial_{N)} \left({\cal M}^{KL} g_{\mu\nu}\partial_L\xi^\nu \right)
\nonumber\\
&&
-20\,d^{PKL}d_{QL(M}{\cal M}_{N)K}\,
\partial_P \left({\cal M}^{RQ} g_{\mu\nu}\partial_R\xi^\nu \right)\;. 
\eea
After some tedious algebra, using in particular that 
$({\cal D}^\mu{\cal M}^{-1} {\cal M})^M{}_N$ is
an $\mathfrak{e}_{6(6)}$ algebra-valued matrix on which the projector $\mathbb{P}^P{}_Q{}^N{}_{M}$
acts as the identity, 
one then computes for the 
variation of the scalar kinetic term 
\bea
\delta {\cal L}_{\rm kin}&=&
{\cal D}^\mu{\cal M}^{MN} {\cal M}_{NK}\,
{\cal F}_{\mu\nu}{}^K\,\partial_M \xi^\nu 
+{\cal D}^\mu{\cal M}^{MN} \,
\partial_M \left( g_{\mu\nu}\partial_N\xi^\nu \right)
\nonumber\\
&&
+ \Big(
{\cal M}_{NL}\,
\partial_M {\cal M}^{LK} \,
-\frac1{12}\,{\cal M}^{KL} \partial_L {\cal M}_{MN}
\Big)\,
{\cal D}_\mu{\cal M}^{MN} \partial_K\xi^\mu 
\;.
\label{varkinS}
\eea
The first term in here precisely cancels the first term in (\ref{vartermF}). 
The second term is of the form ${\cal M}^{MN}\partial_M\partial_N$, which 
we consistently omitted, c.f.~the discussion above and ref.~\cite{Hohm:2013jma}. 
Finally, the last line will be cancelled against part of the variation of the potential 
(thereby determining the overall coefficient of the potential). In fact, it is not 
difficult to see, using the analogue of the first of the eqs.~(4.22) in \cite{Hohm:2013jma}, 
that the variation of the leading terms in the potential read 
\bea
\delta V &=&
\delta\Big(\frac12\,
{\cal M}_{NL}\,
\partial_M {\cal M}^{LK} \,
-\frac1{24}\,{\cal M}^{KL} \partial_L {\cal M}_{MN}
\Big)\,
\partial_K {\cal M}^{MN}  + \cdots
\nonumber\\
&=&
\Big(
{\cal M}_{NL}\,
\partial_M {\cal M}^{LK} \,
-\frac1{12}\,{\cal M}^{KL} \partial_L {\cal M}_{MN}
\Big)\,
{\cal D}_\mu{\cal M}^{MN} \partial_K\xi^\mu 
+ \cdots \;.
\eea
As claimed, in the combination ${\cal L}_{\rm kin}-V$ they 
cancel the terms in (\ref{varkinS}). 
We have thus succeeded in determining all relative coefficients in the 
action (\ref{finalaction}) from $\xi^{\mu}$ gauge invariance and have shown how the non-standard 
diffeomorphism symmetry is realized in the EFT action. 
This concludes our discussion 
of the $(4+1)$-dimensional diffeomorphisms.

\section{Embedding of $D=11$ Supergravity}\label{sec4}
In this section we show explicitly how to embed 11-dimensional supergravity 
into the EFT constructed above. To this end, in the first subsection we rewrite $D=11$ supergravity 
in a Lorentz gauge fixed form that would be appropriate for Kaluza-Klein compactification  
to $D=5$, but keeping the dependence on all 11 coordinates. In the second subsection 
we reduce the EFT (\ref{finalaction}) by choosing a specific solution for the section constraint (\ref{sectioncondition})
that breaks  E$_{6(6)}$ to GL$(6)$, with all fields depending on $5+6$ coordinates. After explicit
dualization of some fields, we establish complete equivalence 
with $D=11$ supergravity.

\subsection{Decomposition of $D=11$ Supergravity}
\label{subsec:decomposition}

We start by briefly recalling the bosonic sector of $D=11$ supergravity~\cite{Cremmer:1978km}, 
whose fields consist of the elfbein $E_{\hat{\mu}}{}^{\hat{a}}$ and 
the 3-form potential $C_{\hat{\mu}\hat{\nu}\hat{\rho}}$, where $\hat{\mu}, \hat{\nu}=0,\ldots,10$, 
and $\hat{a}, \hat{b}=0,\ldots, 10$, denote $D=11$ curved and flat indices, respectively. 
The action reads 
 \be\label{D11sugra}
   S_{11} \ = \ \int d^{11}x\,E\Big(R-\frac{1}{12}F^{\hat{\mu}\hat{\nu}\hat{\rho}\hat{\sigma}}F_{\hat{\mu}\hat{\nu}\hat{\rho}\hat{\sigma}}
  +\frac{1}{12\cdot 216}E^{-1}\varepsilon^{\hat{\mu}_1\cdots \hat{\mu}_{11}}F_{\hat{\mu}_1\cdots \hat{\mu}_4}
  F_{\hat{\mu}_5\cdots \hat{\mu}_8}C_{\hat{\mu}_9\hat{\mu}_{10}\hat{\mu}_{11}}\Big)\; , 
 \ee
with the abelian field strength
 \be
   F_{\hat{\mu}\hat{\nu}\hat{\rho}\hat{\sigma}} \ = \ 4\partial_{[\hat{\mu}}C_{\hat{\nu}\hat{\rho}\hat{\sigma}]}\,.
 \ee  
This theory is invariant under 3-form gauge transformations 
$\delta C_{\hat{\mu}\hat{\nu}\hat{\rho}}=3\partial_{[\hat{\mu}}\Lambda_{\hat{\nu}\hat{\rho}]}$
and under 11-dimensional diffeomorphisms as well as local Lorentz transformations. 
Next we reduce the Lorentz gauge symmetry from ${\rm SO}(1,10)$ to ${\rm SO}(1,4)\times {\rm SO}(6)$,  
choosing an upper-triangular gauge for the elfbein, and 
accordingly split the indices and field components in the above three terms of the action.

\subsubsection*{Einstein-Hilbert Term}
First we consider the decomposition of the Einstein-Hilbert term, following \cite{Aulakh:1985un,Hohm:2005sc}. 
For future application it is convenient to keep the decomposition general, 
so for the moment we consider a $D$-dimensional Einstein-Hilbert term and 
split the indices as $D=n+d$, 
 \be
  \hat{\mu} \ = \ (\mu,m)\;, \qquad \hat{a} \ = \ (a,\alpha)\;,
 \ee 
where $\mu=1,\ldots n$, and $m=1,\dots,d$, and similarly for the flat indices.  
The Lorentz gauge symmetry is partially fixed by choosing the upper-triangular 
 form of the $D$-dimensional vielbein as follows 
 \be\label{KKgauge}
  E_{\hat{\mu}}{}^{\hat{a}} \ = \ \left(\begin{array}{cc} \phi^{\kkappa}e_{\mu}{}^{a} &
  A_{\mu}{}^{m} \phi_{m}{}^{\alpha} \\ 0 & \phi_{m}{}^{\alpha}
  \end{array}\right)\,, 
 \ee 
where $\phi=\det (\phi_{m}{}^{\alpha})$. The inverse is then given by  
 \be
  E_{\hat{a}}{}^{\hat{\mu}} \ = \ \left(\begin{array}{cc} \phi^{-\kkappa}e_a{}^{\mu} &
  -\phi^{-\kkappa}e_a{}^{\nu}A_{\nu}{}^{m}  \\ 0 & \phi_{\alpha}{}^{m}
  \end{array}\right)\,. 
 \ee 
The constant parameter $\kkappa$ depends on the `external' dimension $n$ and
is determined as
 \be\label{kappa}
  \kkappa \ = \ -\frac{1}{n-2}\;,
 \ee 
by requiring an Einstein-frame metric in the $n$-dimensional theory.

Before we compute the form of the Einstein-Hilbert term in the gauge (\ref{KKgauge})
it is convenient to investigate the form of the gauge symmetries after this splitting. 
The original Einstein-Hilbert term is invariant under 
$D$-dimensional diffeomorphisms $x^{\hat{\mu}}\rightarrow x^{\hat{\mu}}-\xi^{\hat{\mu}}$ 
and local Lorentz transformations parametrized by $\lambda^{\hat{a}}{}_{\hat{b}}$, which act on the elfbein as 
 \be\label{fullDiff}
  \delta E_{\hat{\mu}}{}^{\hat{a}} \ = \ \xi^{\hat{\nu}}\partial_{\hat{\nu}}E_{\hat{\mu}}{}^{\hat{a}}
  +\partial_{\hat{\mu}}\xi^{\hat{\nu}} E_{\hat{\nu}}{}^{\hat{a}}+\lambda^{\hat{a}}{}_{\hat{b}} E_{\hat{\mu}}{}^{\hat{b}}\;. 
 \ee 
After the splitting of indices, the diffeomorphisms give rise to two type of gauge symmetries according to 
 \be
  \xi^{\hat{\mu}} \ = \ (\xi^{\mu}\,,\;\Lambda^m)\;. 
 \ee 
We will refer to the gauge transformations parametrized by $\Lambda^m$ as 
`internal' diffeomorphisms. From (\ref{fullDiff}) we compute 
 \be\label{Lambdaguge}
  \begin{split}
   \delta_{\Lambda}e_{\mu}{}^{a} \ &= \ \Lambda^m\partial_m e_{\mu}{}^{a}-\kkappa\, \partial_m\Lambda^m\,e_{\mu}{}^{a}\;, \\
   \delta_{\Lambda}\phi_m{}^{\alpha} \ &= \ \Lambda^n\partial_n \phi_m{}^{\alpha}+\partial_m\Lambda^n\,\phi_n{}^{\alpha}\;, \\
   \delta_{\Lambda}\phi \ &= \ \Lambda^n\partial_n\phi+\partial_n\Lambda^n \,\phi\;, \\
   \delta_{\Lambda} A_{\mu}{}^m \ &= \ \partial_{\mu}\Lambda^m-A_{\mu}{}^{n}\partial_n\Lambda^m
   +\Lambda^n\partial_n A_{\mu}{}^{m}\;.
  \end{split}
 \ee  
We infer that $e$ and $\phi$ transform as tensor(-densities) under the symmetry of $\Lambda^m$ transformations, 
for which $A_{\mu}{}^{m}$ provides a gauge connection. In fact, we can define 
covariant derivatives and field strengths as follows 
  \be\label{covderfieldstr}
  \begin{split}
   D_{\mu}e_{\nu}{}^{a} \ &= \ \partial_{\mu} e_{\nu}{}^{a}-A_{\mu}{}^{m}\partial_m e_{\nu}{}^{a}+\kkappa\, \partial_n A_{\mu}{}^{n} \,e_{\nu}{}^{a}\;, \\
   D_{\mu}\phi_m{}^{\alpha} \ &= \ \partial_{\mu}\phi_m{}^{\alpha}-A_{\mu}{}^{n}\partial_n\phi_m{}^{\alpha}-\partial_mA_{\mu}{}^{n} \phi_n{}^{\alpha}\;, \\
   F_{\mu\nu}{}^{m} \ &= \ \partial_{\mu}A_{\nu}{}^{m}-\partial_{\nu}A_{\mu}{}^{m}-A_{\mu}{}^{n}\partial_n A_{\nu}{}^m
   +A_{\nu}{}^{n}\partial_n A_{\mu}{}^m\;, 
  \end{split}
 \ee  
and it is straightforward to verify that they transform covariantly under (\ref{Lambdaguge}). 
In order to compute the form of the gauge transformations parametrized by $ \xi^{{\mu}}$, 
to which we refer as  `external' diffeomorphisms in the following,  
we have to add a compensating local Lorentz transformation in order to preserve the gauge choice 
in (\ref{KKgauge}). The Lorentz parameter is found to be 
 \be
  \lambda^{a}{}_{\beta} \ = \ -\phi^{\kkappa} \phi_{\beta}{}^{m}\partial_m\xi^{\nu}\,e_{\nu}{}^{a}\;. 
 \ee 
Moreover, it turns out to be convenient to present these `external' diffeomorphisms in the form 
of covariant or `improved' diffeomorphisms, for which we add a field-dependent gauge transformation 
with parameter  $\Lambda^m=-\xi^{\nu} A_{\nu}{}^{m}$. The full transformation rules 
can then be written directly in terms of the covariant objects from (\ref{covderfieldstr}): 
 \be
  \begin{split}
  \delta_{\xi}e_{\mu}{}^{a} \ &= \ \xi^{\nu}D_{\nu} e_{\mu}{}^{a}+D_{\mu}\xi^{\nu}\,e_{\nu}{}^{a}\;, \\
  \delta_{\xi}\phi_m{}^{\alpha} \ &= \ \xi^{\nu}D_{\nu}\phi_m{}^{\alpha}\;, \\
  \delta_{\xi} A_{\mu}{}^m \ &= \ \xi^{\nu} F_{\nu\mu}{}^m + \phi^{2\kkappa}\phi^{mn} g_{\mu\nu}\partial_n\xi^{\nu}\;,
 \end{split}
 \ee 
 with $\phi^{mn}=\phi_\alpha{}^m\phi^{\alpha\,n}$\,.

After having discussed the form of the gauge symmetries, we are now ready to 
decompose the Einstein-Hilbert term. To this end it is convenient to use the following 
formula: 
  \be\label{EHeasy}
  S_{\rm EH} \ = \ \int d^Dx\,E\,E_{\hat{a}}{}^{\hat{\mu}} E_{\hat{b}}{}^{\hat{\nu}} R_{\hat{\mu}\hat{\nu}}{}^{\hat{a}\hat{b}}
  \ = \ \int d^nx\, d^dy\,E\Big( -\frac{1}{4} \widehat{\Omega}^{\hat{a}\hat{b}\hat{c}}\, \widehat{\Omega}_{\hat{a}\hat{b}\hat{c}}
  +\frac{1}{2}\widehat{\Omega}^{\hat{a}\hat{b}\hat{c}}\, \widehat{\Omega}_{\hat{b}\hat{c}\hat{a}}
  +\widehat{\Omega}_{\hat{c}\hat{b}}{}^{\hat{b}}\,\widehat{\Omega}^{\hat{c}}{}_{\hat{a}}{}^{\hat{a}}\Big)\;, 
 \ee 
where  we introduced the coefficients of anholonomy, 
 \be
  \widehat{\Omega}_{\hat{a}\hat{b}\hat{c}} \ = \ 
  E_{\hat{a}}{}^{\hat{\mu}} E_{\hat{b}}{}^{\hat{\nu}}\big(\partial_{\hat{\mu}} E_{\hat{\nu}\hat{c}}
  -\partial_{\hat{\nu}} E_{\hat{\mu}\hat{c}}\big) \;. 
 \ee 
Inserting the elfbein (\ref{KKgauge}) and its inverse in here we find for the 
various components 
 \be\label{allomegas}
  \begin{split}
   \widehat{\Omega}_{abc} \ &= \ \phi^{-\kkappa}\Omega_{abc}+2\kkappa \phi^{-\kkappa-1} e_{[a}{}^{\mu}\eta_{b]c} D_{\mu}\phi\;, \\
   \widehat{\Omega}_{ab\gamma} \ &= \ \phi^{-2\kkappa} e_{a}{}^{\mu} e_b{}^{\nu} F_{\mu\nu}{}^{m} \phi_{m\gamma}\;, \\
   \widehat{\Omega}_{a\beta\gamma} \ &= \ \phi^{-\kkappa} \phi_{\beta}{}^{m} e_{a}{}^{\mu} D_{\mu}\phi_{m\gamma}\;, \\
   \widehat{\Omega}_{\alpha bc} \ &= \ e_{b}{}^{\nu}\phi_{\alpha}{}^{m}D_me_{\nu c}\;, \\
   \widehat{\Omega}_{\alpha \beta c} \ &= \ 0\;, \\
   \widehat{\Omega}_{\alpha \beta \gamma} \ &= \ \Omega_{\alpha \beta \gamma}\;, 
  \end{split}
 \ee  
where we introduced the `external' and `internal' coefficients of anholonomy 
 \be
 \begin{split}
  \Omega_{abc} \ &= \ 2e_{[a}{}^{\mu} e_{b]}{}^{\nu} D_{\mu} e_{\nu c}\;, \\
  \Omega_{\alpha\beta\gamma} \ &= \ 2 \phi_{[\alpha}{}^{m}\phi_{\beta]}{}^{n} \partial_m\phi_{n\gamma }\;, 
 \end{split}
\ee  
and defined 
 \be\label{defDe}
  D_me_{\nu c} \ \equiv  \ \partial_m e_{\nu c}+\kkappa\, 
    \phi^{-1}\partial_m\phi\, e_{\nu c}\;. 
 \ee   
This latter derivative is covariant under the internal diffeomorphisms (\ref{Lambdaguge}) in that 
$D_me_{\nu c}$ transforms as a vector-density (with the same weight $-\kkappa$ as $e_{\nu c}$). 
Moreover, we see that in (\ref{allomegas}) all components organized already into the 
covariant objects (\ref{covderfieldstr}), so that the $\Lambda$ gauge invariance 
of the action will be manifest. 

Next we determine the form of the Einstein-Hilbert term by inserting  the components (\ref{allomegas})
into (\ref{EHeasy}) and using  
 \be
  E \ \equiv \ \det E_{\hat{\mu}}{}^{\hat{a}} \ = \ \phi^{n\kkappa+1}\,e\;. 
 \ee 
We find 
 \be\label{firstEH}
  \begin{split}
    S_{\rm EH} \ = \  \int &d^nx\, d^dy\,e\Big[-\frac{1}{4}\Omega^{abc}\,\Omega_{abc}+\frac{1}{2}\Omega^{abc}\,\Omega_{bca}+\Omega^a\,\Omega_a
    -e^{a\mu} e^{b\nu} F_{\mu\nu}{}^m\left(e_{b}{}^{\rho}\partial_m e_{\rho a}\right) \\
    &-\frac{1}{2}\phi^{mn}g^{\mu\nu} D_{\mu}\phi_m{}^{\alpha}\,D_{\nu}\phi_{n\alpha}
    -\kkappa^2\left(n-2\right)\phi^{-2}g^{\mu\nu} D_{\mu}\phi D_{\nu}\phi\\
    &-\frac{1}{2}g^{\mu\nu}(\phi^{\alpha m}D_{\mu} \phi_m{}^{\gamma})(\phi_{\gamma}{}^{n} D_{\nu}\phi_{n\alpha})
    -\frac{1}{4}\phi^{-2\kkappa}\phi_{mn} F^{\mu\nu m} F_{\mu\nu}{}^{n}\\
    &+\phi^{2\kkappa}\Big(-\frac{1}{2}\phi^{mn}g^{\mu\nu}D_me_{\mu}{}^{a}\,D_n e_{\nu a}-\frac{1}{2}\phi^{mn}(e^{b\mu}D_m e_{\mu}{}^{c})(e_{c}{}^{\nu}D_n e_{\nu b})
    \\
    &\qquad\qquad +\phi^{mn}(e^{-1}D_me)(e^{-1}D_ne)\\
    &\qquad\qquad -\frac{1}{4}\Omega^{\alpha\beta\gamma}\,\Omega_{\alpha\beta\gamma}
    +\frac{1}{2}\Omega^{\alpha\beta\gamma}\,\Omega_{\beta\gamma\alpha}+\Omega^{\alpha}\,\Omega_{\alpha} 
         +2\phi^{\alpha m}\,\Omega_{\alpha} \, e^{-1}D_m e \Big)\Big]\;. 
  \end{split}
 \ee   
Let us now write the various terms more geometrically.  The terms in the first line combine into 
the $n$-dimensional Einstein-Hilbert term for $e_{\mu}{}^{a}$, but with the additional 
covariantization that all derivatives are covariant according to (\ref{covderfieldstr}) 
and the Ricci scalar is based on the `improved' Riemann tensor  
 \be
  \widehat{R}_{\mu\nu}{}^{ab}  =  R_{\mu\nu}{}^{ab}+F_{\mu\nu}{}^{m}
  e^{a}{}^{\rho}\partial_m e_{\rho}{}^{b}\,,
  \label{improvedR}
 \ee
which is necessary in order to preserve local ${\rm SO}(1,4)$ Lorentz invariance, as discussed above for 
the full EFT. 
Next, the terms in the last line in the potential can also be written more geometrically, using 
  \be
  \begin{split}
   e\phi^{2\kkappa}\,R(\phi_m{}^{\alpha}) \ = \ e\phi^{2\kkappa}\Big(&-\frac{1}{4}\Omega^{\alpha\beta\gamma}\,\Omega_{\alpha\beta\gamma}
    +\frac{1}{2}\Omega^{\alpha\beta\gamma}\,\Omega_{\beta\gamma\alpha}+\Omega^{\alpha}\,\Omega_{\alpha}\\
      &+2\phi_{\alpha}{}^m e^{-1}\partial_m e\,\Omega^{\alpha}+2(2\kkappa-1)\phi_{\alpha}{}^m\phi^{-1}\partial_m\phi\,\Omega^{\alpha}\Big)
    +\text{total~der.}\;, 
   \end{split}
   \ee 
which for $\kkappa$ as determined in (\ref{kappa}) reproduces the last line of (\ref{firstEH}). Finally, 
we can reorganize 
the $De$ terms into $Dg$ terms in order to make the local Lorentz invariance manifest. In total we obtain 
 \be
  \begin{split}
   S_{\rm EH} \ = \  \int d^nx\, d^dy\,e\Big[ &\widehat{R}-\frac{1}{4}\phi^{-2\kkappa}\phi_{mn} F^{\mu\nu m} F_{\mu\nu}{}^{n}\\
   &-\frac{1}{2}\phi^{mn}g^{\mu\nu} D_{\mu}\phi_m{}^{\alpha}\,D_{\nu}\phi_{n\alpha}
    -\kkappa^2\left(n-2\right)\phi^{-2}g^{\mu\nu} D_{\mu}\phi D_{\nu}\phi\\
    &-\frac{1}{2}g^{\mu\nu}(\phi^{\alpha m}D_{\mu} \phi_m{}^{\gamma})(\phi_{\gamma}{}^{n} D_{\nu}\phi_{n\alpha})-V(\phi,e)\Big]\;, 
  \end{split}
  \label{EHkin}
 \ee   
with the `Einstein-Hilbert  potential' 
 \be\label{VEH}
  V_{\rm EH} (\phi,e) 
  \ = \ -\phi^{2\kkappa}\Big(R(\phi)+\frac{1}{4}\phi^{mn}\left(D_mg^{\mu\nu}\,D_{n}g_{\mu\nu}+g^{-1}D_mg\,g^{-1}D_ng\right)\Big)\;.
 \ee 
Below we will also need the form of the potential in terms of the symmetric tensor $\phi_{mn}=\phi_{m}{}^\alpha \phi_{n\,\alpha}$, as opposed 
to the vielbein. Integrating by parts, and setting $\kkappa=-\frac{1}{3}$, the term involving the internal Ricci scalar can be written as  
\be\label{alterscalar}
 \begin{split}
  e\phi^{-\frac{2}{3}} R(\phi) \ = \ e\phi^{-\frac{2}{3}}\Big[\,&\frac{1}{2}\phi^{mn}\phi^{kl}\phi^{pq}
  \partial_k\phi_{mq} \partial_p\phi_{nl}-\frac{1}{4}\phi^{mn}\phi^{kl}\phi^{pq}
  \partial_p\phi_{mk} \partial_q\phi_{nl}\\
  &-\frac{2}{3}\partial_m\phi^{mn}\,\phi^{-1}\partial_n\phi
  -\frac{21}{9}\phi^{mn}(\phi^{-1}\partial_m\phi)(\phi^{-1}\partial_n\phi)\\
  &+\partial_m\phi^{mn}\,e^{-1}\partial_ne+2\phi^{mn}(e^{-1}\partial_me)(\phi^{-1}\partial_n\phi)
  \,\Big]\;, 
 \end{split}
 \ee 
 which is the form convenient for the comparison with the E$_{6(6)}$ covariant theory.

\subsubsection*{3-Form Kinetic and Topological Term}
We now turn to the decomposition of the kinetic term for the 3-form. 
First, we have to perform field redefinitions of the various components of $C_{\hat{\mu}\hat{\nu}\hat{\rho}}$ 
in terms of the Kaluza-Klein vector 
in order to obtain forms that transform covariantly under the gauge symmetries. The general 
prescription for Kaluza-Klein reductions is to `flatten' all $D=11$ curved indices with 
$E_{a}{}^{\hat{\mu}}$ and then to `un-flatten' with the external $n$-bein components 
$E_{\mu}{}^{a}$. For instance, the vectors originating from the 3-form are redefined according to 
 \be
  A_{\mu\,mn} \ \equiv \ E_{\mu}{}^{a} E_{a}{}^{\hat{\nu}} C_{\hat{\nu}mn}\;.
 \ee 
Performing the analogous field redefinition for the other components we obtain 
the following field variables originating from the 3-form $C_{\hat{\mu}\hat{\nu}\hat{\rho}}$, 
denoted by $A$:
 \begin{equation}
  \begin{split}
   A_{mnk} \ &= \ C_{mnk}\;, \\
   A_{\mu\,mn} \ &= \ C_{\mu mn}-A_{\mu}{}^k\,C_{kmn}\;, \\
   A_{\mu\nu\,m} \ &= \ C_{\mu\nu m}-2 A_{[\mu}{}^n\,C_{\nu]mn}+A_{\mu}{}^n A_{\nu}{}^k\,C_{mnk}\;, \\
   A_{\mu\nu\rho} \ &= \ C_{\mu\nu\rho}-3 A_{[\mu}{}^m\,C_{\nu\rho]m}+3 A_{[\mu}{}^m A_{\nu}{}^n\, C_{\rho]mn}
   -A_{\mu}{}^m A_{\nu}{}^n A_{\rho}{}^k\,C_{mnk}\;. 
  \end{split}
  \label{comp3form}
 \ee  
This definition is such that the fields transform covariantly under internal diffeomorphisms, i.e., 
simply according to their `internal' index structure.  
In order to display the transformation under the components of 3-form gauge parameter $\Lambda_{\hat{\mu}\hat{\nu}}$, 
we also have to perform  redefinitions of the parameters with the 
Kaluza-Klein vector, following exactly the same prescription as for the fields. 
Thus, we define the new parameters
 \be
  \Lambda_{\mu m}' \ = \ \Lambda_{\mu m}-A_{\mu}{}^{n} \Lambda_{nm}\;, \qquad \text{etc.}
 \ee 
Dropping the prime on the parameters in the following, we obtain the gauge transformations 
under  $(\Lambda_{mn}, \Lambda_{\mu m}, \Lambda_{\mu\nu})$ which act on the fields as 
 \be\label{gaugevar}
  \begin{split}
   \delta A_{mnk} \ &= \ 3\partial_{[m}\Lambda_{nk]}\;, \\
   \delta A_{\mu mn} \ &= \ D_{\mu}\Lambda_{mn}-2\partial_{[m}\Lambda_{|\mu|n]}\;, \\
   \delta A_{\mu\nu\, m} \ &= \ 2 D_{[\mu}\Lambda_{\nu]m}-F_{\mu\nu}{}^n\Lambda_{mn}+\partial_m\Lambda_{\mu\nu}\;, \\
   \delta A_{\mu\nu\rho} \ &= \ 3 D_{[\mu}\Lambda_{\nu\rho]}-3 F_{[\mu\nu}{}^m \Lambda_{\rho]m}\;. 
  \end{split}
 \ee  
As usual, all derivatives are covariant w.r.t.~the internal diffeomorphisms. 
We observe that after the decomposition the formerly 
abelian 3-form gauge transformations of $D=11$ supergravity take a non-trivial form
with non-commuting covariant derivatives and extra St\"uckelberg-type 
transformations, reminiscent of the 
tensor hierarchy introduced above.  Moreover, the 
Kaluza-Klein Yang-Mills field strength $F_{\mu\nu}{}^n$ explicitly appears in the transformation rules.

Let us now turn to the form of the field strength components. 
As for the fields, redefinitions are required, in order to arrive at field strengths 
that are covariant under internal diffeomorphisms and invariant under (\ref{gaugevar}). We define 
 \be\label{newfieldstrength}
   F_{\mu\, mnk}' \ \equiv \ E_{\mu}{}^{a} E_{a}{}^{\hat{\nu}} F_{\hat{\nu}\, mnk}\;, \quad \text{etc.}\;, 
 \ee
which are manifestly invariant under the 3-form gauge transformations as a consequence 
of the invariance of the original field strength $F_{\hat{\mu}\hat{\nu}\hat{\rho}\hat{\sigma}}$.     
Dropping the primes in the following, one finds for the redefined field strength 
in terms of the redefined fields    
 \be\label{invfieldstr}
  \begin{split}
   F_{mnkl}  \ &= \ 4\partial_{[m}A_{nkl]}\;, \\
   F_{\mu\, nkl} \ &= \ D_{\mu}A_{nkl}-3\partial_{[n}A_{|\mu| kl]}\;, \\
   F_{\mu\nu\, mn} \ &= \ 2 D_{[\mu}A_{\nu]mn}+F_{\mu\nu}{}^k A_{kmn}+2\partial_{[m}A_{|\mu\nu| n]}\;, \\
   F_{\mu\nu\rho\, m} \ &= \  3D_{[\mu}A_{\nu\rho]m}+3 F_{[\mu\nu}{}^n A_{\rho]mn}
   -\partial_m A_{\mu\nu\rho}\;, \\
   F_{\mu\nu\rho\sigma} \ &= \ 4 D_{[\mu}A_{\nu\rho\sigma]}+6 F_{[\mu\nu}{}^m A_{\rho\sigma]m}\;.  
  \end{split}
 \ee   
These field strengths are manifestly covariant w.r.t.~internal diffeomorphisms. 
Moreover, one may verify by an explicit computation that the field strengths are 
gauge invariant under (\ref{gaugevar}).  
Due to the non-abelian gauge connections entering the fields strengths, 
the latter satisfy non-standard Bianchi identities:  
 \be\label{Bianchiss}
  \begin{split}
   D_{\mu}F_{mnkl} \ &= \ 4\partial_{[m}F_{|\mu| nkl]} \;, \\
   2D_{[\mu} F_{\nu]nkl} \ &= \ -3 \partial_{[n} F_{|\mu\nu|kl]}-F_{\mu\nu}{}^m F_{mnkl}\;, \\
   3 D_{[\mu} F_{\nu\rho]mn} \ &= \ 2\partial_{[m}F_{|\mu\nu\rho |n]}+3 F_{[\mu\nu}{}^k F_{\rho]kmn}\;, \\
   4 D_{[\mu}F_{\nu\rho\sigma]m} \ &= \ -\partial_m F_{\mu\nu\rho\sigma}+6 F_{[\mu\nu}{}^n F_{\rho\sigma]mn}\;, \\
   5 D_{[\mu}F_{\nu\rho\sigma\lambda]} \ &= \ 10 F_{[\mu\nu}{}^m F_{\rho\sigma\lambda]m}\;. 
  \end{split}
 \ee    
As for the tensor hierarchy, the Bianchi identities relate the exterior derivatives of 
a field strength to the `next higher' field strength in the hierarchy.   
 
We are now in a position to give the decomposition of the kinetic term for the 3-form. 
Due to the form of the redefinition (\ref{newfieldstrength}) of the field strengths,  
it is straightforward to rewrite the $F^2$ term, by simply going to flattened indices:
 \be\label{3formterm}
 \begin{split}
 {\cal L}_{\text{\,3-form}} 
 \ &= \  -\frac{1}{12}E\,
 F^{\hat{\mu}\hat{\nu}\hat{\rho}\hat{\sigma}}F_{\hat{\mu}\hat{\nu}\hat{\rho}\hat{\sigma}}
  \ = \  -\frac{1}{12}E\,F^{\hat{a}\hat{b}\hat{c}\hat{d}}F_{\hat{a}\hat{b}\hat{c}\hat{d}} \\
  \ &= \ -\frac{1}{12}\phi^{n\kkappa+1}e\,\Big(\phi^{-8\kkappa}
  F^{\mu\nu\rho\sigma}F_{\mu\nu\rho\sigma}+4\phi^{-6\kkappa}\phi^{mn}F^{\mu\nu\rho}{}_{m}
  F_{\mu\nu\rho\,n}
  +6\phi^{-4\kkappa}\phi^{mn}\phi^{kl}F^{\mu\nu}{}_{mk} F_{\mu\nu\, nl}
  \\ &\qquad\qquad \qquad \quad 
  +4\phi^{-2\kkappa}\phi^{mn}\phi^{kl}\phi^{pq} F^{\mu}{}_{mkp} F_{\mu\,nlq}
  +\phi^{mn}\phi^{kl}\phi^{pq}\phi^{rs}F_{mkpr} F_{nlqs}\Big)\\
  \ &= \ -\frac{1}{12}e\,\Big(\phi^{2}
  F^{\mu\nu\rho\sigma}F_{\mu\nu\rho\sigma}+4\phi^{\frac{4}{3}}\phi^{mn}F^{\mu\nu\rho}{}_{m}
  F_{\mu\nu\rho\,n}+6\phi^{\frac{2}{3}}\phi^{mn}\phi^{kl}F^{\mu\nu}{}_{mk} F_{\mu\nu\, nl}\\
  &\qquad\qquad \qquad \quad +4\phi^{mn}\phi^{kl}\phi^{pq} F^{\mu}{}_{mkp} F_{\mu\,nlq}
  +\phi^{-\frac{2}{3}}\phi^{mn}\phi^{kl}\phi^{pq}\phi^{rs}F_{mkpr} F_{nlqs}\Big) \;. 
\end{split}  
\ee 
Here we left the raising of spacetime indices with $g^{\mu\nu}$ implicit, and we inserted 
the value for $\kkappa$, see eq.~(\ref{kappa}), for $n=5$.

Next we have to decompose the topological Chern-Simons-like term in (\ref{D11sugra}) 
and write it in terms of the invariant field strengths defined in (\ref{invfieldstr}). One finds 
 \be\label{finaltop}
  \begin{split}
   {\cal L}_{\rm top} \ = \  -\frac{1}{108}\,&\varepsilon^{\mu\nu\rho\sigma\lambda}\varepsilon^{mnklpq}\Big(
   A_{\mu\nu\, m} F_{\rho\sigma\lambda\,n}F_{klpq}+6A_{\mu\nu\, m}F_{\rho\sigma\, nk} F_{\lambda\,lpq}
   -\frac{1}{2} A_{\mu\nu\rho} F_{\sigma\lambda\,mn} F_{klpq}\\
   &\; +\frac{2}{3}A_{\mu\nu\rho} F_{\sigma\, mnk} F_{\lambda\,lpq}-\frac{1}{4}A_{\mu\, mn}F_{klpq} F_{\nu\rho\sigma\lambda}
   +4A_{\mu\, mn} F_{\nu\, klp} F_{\rho\sigma\lambda\,q}\\
   &\; -\frac{9}{2}A_{\mu\, mn}F_{\nu\rho\, kl} F_{\sigma\lambda\, pq}
   +\frac{1}{3}A_{mnk} F_{\mu\,lpq} F_{\nu\rho\sigma\lambda}
   +2 A_{mnk} F_{\mu\nu\, lp} F_{\rho\sigma\lambda\, q}\Big)\;. 
  \end{split}
 \ee  
The validity of this expression can be checked explicitly by verifying gauge invariance under
(\ref{gaugevar}). As the field strengths are already gauge invariant by construction, we only have to 
vary the bare gauge potentials $A$. After this we may integrate by parts and 
show cancellation by use of the Bianchi identities (\ref{Bianchiss}). 
This computation requires repeated use of Schouten identities 
according to which terms with total antisymmetrization over seven internal indices $m,n,\ldots$ 
vanish identically. Let us note that up to total derivatives, 
the form of (\ref{finaltop}) is uniquely determined by gauge 
invariance under (\ref{gaugevar}), up to the overall coefficient that is determined by 
$D=11$ supergravity.

Finally, we can give the complete action of $D=11$ supergravity under the 
$5+6$ decomposition and the corresponding gauge fixing of the local Lorentz group:
\be
  \begin{split}
   S_{11} \ = \ \int d^5x\, d^6y\,e\,\Big[&\,\widehat{R}-\frac{1}{4}\tilde{\cal M}_{\frak{m}\frak{n}}{\cal F}^{\mu\nu\,\frak{m}}
   {\cal F}_{\mu\nu}{}^{\frak{n}}-\frac{1}{12}\phi^{2}F^{\mu\nu\rho\sigma}F_{\mu\nu\rho\sigma}
   -\frac{1}{3}\phi^{\frac{4}{3}}\phi^{mn}F^{\mu\nu\rho}{}_{m} F_{\mu\nu\rho\, n}\\
   &-\frac{1}{2}\phi^{mn}D^{\mu}\phi_{m}{}^{\alpha}\,D_{\mu}\phi_{n\alpha}-\frac{1}{3}\phi^{-2}D^{\mu}\phi\,D_{\mu}\phi\\
   &-\frac{1}{2}(\phi^{\alpha m}D^{\mu}\phi_{m}{}^{\gamma})(\phi_{\gamma}{}^{n}D_{\mu}\phi_{n\alpha})
   -\frac{1}{3} \phi^{mn}\phi^{kl}\phi^{pq}F^{\mu}{}_{mkp} F_{\mu\, nlq}\\
   &-V(e,\phi)+e^{-1}{\cal L}_{\rm top}\,\Big]\;. 
  \end{split}
  \label{finalKK}
 \ee  
Here we fixed $\kkappa=-\frac{1}{3}$ according to (\ref{kappa}). 
Moreover, we combined the two-form field strengths of the Kaluza-Klein gauge vector
and the vector originating from the 3-form,  
  \be
  {\cal F}_{\mu\nu}{}^{\frak{m}} \ = \  (   {\cal F}_{\mu\nu}{}^{m}\,,\; {\cal F}_{\mu\nu\,mn})
  \ \equiv \ (F_{\mu\nu}{}^{m}\,,\; F_{\mu\nu\,mn}-F_{\mu\nu}{}^{k}A_{kmn})\;, 
  \label{newFKK}
 \ee 
by introducing the scalar dependent kinetic metric 
 \be
  \begin{split}
   \tilde{\cal M}_{m,n} \ &= \ \phi^{\frac{2}{3}}\big(\phi_{mn}+2 \phi^{kl}\phi^{pq}A_{mkp} A_{nlq}\big)\;, \\
   \tilde{\cal M}_{m,}{}^{kl} \ &= \ 2\phi^{\frac{2}{3}}\phi^{kp}\phi^{lq} A_{mpq}\;, \\
   \tilde{\cal M}^{mn,kl} \ &= \ 2\phi^{\frac{2}{3}}\phi^{m[k}\phi^{l]n}\;, 
  \end{split}
 \ee
with the index ${}_{\frak{m}} \ = \ ({}_{m}\,,\;{}^{[mn]})$. The topological term 
is given by (\ref{finaltop}) and the full potential reads 
  \be
  \begin{split}
  eV \ = \ - e&\phi^{-\frac{2}{3}}\Big[\,\frac{1}{2}\phi^{mn}\phi^{kl}\phi^{pq}
  \partial_k\phi_{mq}\partial_p\phi_{nl}-\frac{1}{4}\phi^{mn}\phi^{kl}\phi^{pq}
  \partial_p\phi_{mk}\partial_q\phi_{nl}\\
  &\hspace{0.45cm}-\frac{2}{3}\partial_m\phi^{mn}\,\phi^{-1}\partial_n\phi
  -\frac{1}{9}\phi^{mn}(\phi^{-1}\partial_m\phi)(\phi^{-1}\partial_n\phi)\\
  &\hspace{0.45cm}+\partial_m\phi^{mn}\,e^{-1}\partial_ne-\frac{2}{3}\phi^{mn}(e^{-1}\partial_me)(\phi^{-1}\partial_n\phi)\\
  &\hspace{0.45cm}+\frac{1}{4}\phi^{mn}\big(\partial_mg^{\mu\nu}\,\partial_{n}g_{\mu\nu}
  +g^{-1}\partial_mg\,g^{-1}\partial_ng\big)-\frac{1}{12}\phi^{mn}\phi^{kl}\phi^{pq}\phi^{rs}
  F_{mkpr} F_{nlqs}
  \,\Big]\;. 
 \end{split}
 \label{finalpot}
 \ee 
It is obtained by combining (\ref{VEH}) with the purely internal $F^2$ term from (\ref{3formterm}). 
Moreover, we used (\ref{alterscalar}) and expanded the $Dg$ terms according to (\ref{defDe}). 
This is the final form of the action, still equivalent to the full $D=11$ supergravity.
In the following, we will compare and match this result with the action obtained by evaluating 
the EFT (\ref{finalaction}) for a particular solution of the section constraints.

\subsection{${\rm GL}(6)$ invariant reduction of EFT }
\label{subsec:gl6}

In this subsection, we will consider the E$_{6(6)}$ covariant EFT (\ref{finalaction}) 
upon specifying an explicit solution of the section condition, that breaks
E$_{6(6)}$ down to ${\rm GL}(6)$. We will show that the resulting theory upon further dualization
precisely coincides with eleven-dimensional supergravity in the form 
presented in the previous subsection.

\subsubsection{${\rm GL}(6)$ invariant solution of the section condition}

The relevant embedding of ${\rm GL}(6)$ into E$_{6(6)}$
is given by
\bea
{\rm GL}(6) &=& {\rm SL}(6) \times {\rm GL}(1) ~\subset~ {\rm SL}(6) \times {\rm SL}(2) ~\subset~ {\rm E}_{6(6)}
\;,
\eea
with the fundamental representation of ${\rm E}_{6(6)}$ breaking as
\bea
\bar{\bf 27} &\rightarrow& 6_{+1} + 15'_{0} + 6_{-1} 
\;,
\label{split27}
\eea
and the adjoint breaking into
\bea
{\bf 78} &\rightarrow& 1_{-2} + 20_{-1} + \left(1+35\right)_{0} + 20_{+1} + 1_{+2}
\;,
\label{split78}
\eea
with the subscripts referring to the ${\rm GL}(1)$ charges.
An explicit solution to the section condition (\ref{sectioncondition}) is given by 
restricting the $Y^M$ dependence of all fields to the six coordinates in the $6_{+1}$.
Explicitly, splitting the coordinates $Y^M$ according to (\ref{split27}) into
\bea
\left\{Y^M\right\} &\rightarrow&\left\{\, y^m\,,\; y_{mn}\,, \; y^{\bar{m}} \,\right\}
\;,
\label{Ybreak}
\eea
with indices $m, n = 1, \dots, 6$,
the non-vanishing components of the $d$-symbol are given by\footnote{
We use summation conventions $X^M Y_M = X^m Y_m + X_{mn} Y^{mn} + X^{\bar{m}}Y_{\bar{m}}$\,.} 
 \bea
d^{MNK} &:&
  d^{m\,\bar{n}}{}_{kl} \ = \  
  \frac{1}{\sqrt{5}}\,\delta_{[k}^m \delta_{l]}^n\;, \qquad \;\, d_{mn\,kl\,pq} \ = \ \frac{1}{4\sqrt{5}}\,\varepsilon_{mnklpq}\;, \nonumber\\
d_{MNK} &:&   d_{m\,\bar{n}}{}^{kl} \ = \ \frac{1}{\sqrt{5}}\,\delta_{[m}^k \delta_{n]}^l\;, \qquad d^{mn\,kl\,pq} \ = \ \frac{1}{4\sqrt{5}}\,\varepsilon^{mnklpq}\;, 
 \label{dbreak}
 \eea
and all those related by symmetry $d^{MNK}=d^{(MNK)}$\,.
In particular, the ${\rm GL}(1)$ grading guarantees that all components $d^{m\,n\,k}$ vanish,
such that the section condition (\ref{sectioncondition}) indeed is solved by restricting the coordinate dependence
of all fields according to
\bea
\left\{\partial_{\bar{m}} A \ = \ 0\;,\; \partial^{mn} A \ = \ 0\right\}\qquad
\Longleftrightarrow\qquad
A(x^\mu, Y^M) &\rightarrow& A(x^\mu, y^m)
\;.
\label{explicit_section}
\eea

Let us first revisit the resulting field content of the model. The ${\rm E}_{6(6)}$-covariant formulation presented above
carries all 27 vector fields $A_\mu{}^M$, now breaking according to (\ref{split27}), whereas the two-forms
appear only under projection $d^{MNK}\partial_N B_{\mu\nu\,K}$. With (\ref{dbreak}) we find, that only the
components $B_{\mu\nu\,\bar{n}}$ and $B_{\mu\nu}{}^{mn}$ enter the Lagrangian, moreover they enter
under $\partial_m$-derivatives according to
\bea
\partial_{m} B_{\mu\nu\,\bar{n}}-\partial_{n} B_{\mu\nu\,\bar{m}}\;,\quad\mbox{and} \qquad \partial_m B_{\mu\nu}{}^{mn}
\;.
\label{onlyB}
\eea
In other words, with this parametrization the Lagrangian comes with an additional local shift symmetry 
\bea
\delta B_{\mu\nu\,\bar{n}}~=~ \partial_n \Omega_{\mu\nu} \;,\qquad
\delta B_{\mu\nu}{}^{mn}~=~ \partial_k \Omega_{\mu\nu}{}^{[kmn]}
\;,
\label{shiftB}
\eea
for arbitrary $\Omega_{\mu\nu}$, $\Omega_{\mu\nu}{}^{[kmn]}$\,.
In total, the full $p$-form field content of the ${\rm E}_{6(6)}$ Lagrangian in this basis
is thus given by
\bea
\left\{ A_\mu{}^m\,,\; A_{\mu\,mn}\,, \; A_{\mu}{}^{\bar{m}} \right\}\;,\qquad
\left\{B_{\mu\nu\,\bar{m}}\,, \; B_{\mu\nu}{}^{mn} \right\}
\;,
\label{AB}
\eea
modulo (\ref{shiftB}).
Comparing (\ref{AB}) to the field content of the Kaluza-Klein reduction of $D=11$ supergravity in the split of 
section~\ref{subsec:decomposition} suggests to identify
the $A_\mu{}^m$ with the Kaluza-Klein vector fields sitting in the eleven-dimensional vielbein (\ref{KKgauge}), and to relate
the fields $\{ A_{\mu\,mn}, B_{\mu\nu\,\bar{m}}\}$ to the different components of the eleven-dimensional 3-form (\ref{comp3form}).
The index structure of the remaining fields $\{ B_{\mu\nu}{}^{mn}, A_{\mu}{}^{\bar{m}}\}$ suggests to relate them to the
corresponding components of the eleven-dimensional 6-form, i.e.\ to describe degrees of freedom on-shell dual 
to $\{ A_{\mu\,mn}, B_{\mu\nu\,\bar{m}}\}$. Finally the six two-form tensors $B_{\mu\nu\,m}$ that are absent in (\ref{AB}) 
represent the degrees of freedom that are on-shell dual to the Kaluza-Klein vector fields, i.e.\ descending from the eleven-dimensional
dual graviton. They do not figure in the action (\ref{finalaction}) and we comment on their role in the conclusions.
We recall that in the EFT formulation, all vector fields appear with a Yang-Mills kinetic term 
whereas the two-forms couple via a topological term. The latter do not represent additional degrees of freedom
but are on-shell dual to the vector fields. In order to match the structure of $D=11$ supergravity, we will thus have to trade the
YM vector field $A_{\mu}{}^{\bar{m}}$ for a propagating two-form $B_{\mu\nu\,\bar{m}}$ as we shall describe in 
detail in  section~\ref{subsubsec:dualization} below.

Let us now work out the details of this identification by 
evaluating the general EFT formulas 
in the basis (\ref{split27}) and imposing the explicit solution of the section condition (\ref{explicit_section}) on all fields.
We first consider the six vector fields $A_\mu{}^m$ transforming in the same 
representation as the surviving coordinates (\ref{explicit_section}).
Under general gauge transformations (\ref{GaugeVar}) they transform according to
\bea
\delta_\Lambda A_\mu{}^m & = & \partial_\mu \Lambda^m - A_\mu{}^n \partial_n \Lambda^m + 
\Lambda^n \partial_n A_\mu{}^m 
\;,
\eea
while they remain invariant under all higher tensor gauge transformations from (\ref{deltaAB}). The associated
gauge transformations close into the Lie algebra
 \bea
{}\big[ \delta_{\Lambda_1},\delta_{\Lambda_2}\big] \ = \ \delta_{\Lambda_{12}}
\;,
\qquad
\Lambda_{12}^m ~\equiv~ \Lambda_{2}^k \partial_k \Lambda_{1}^m-\Lambda_{1}^k \partial_k \Lambda_{2}^m
\;,
\eea
of standard six-dimensional diffeomorphisms, embedded into the E-bracket~(\ref{Ebracket}).
The six vector fields $A_\mu{}^m$ thus ensure that the theory is invariant under internal
diffeomorphisms with parameters $\Lambda^m$\,. As anticipated above, we will identify them with the 
Kaluza-Klein vector fields from the eleven-dimensional vielbein (\ref{KKgauge}).
For the following and just as in the previous section, c.f.~(\ref{covderfieldstr}), we thus define the covariant derivatives
\bea
D_\mu &=& \partial_\mu - {\cal L}_{A_\mu}
\;,
\eea
corresponding to the action of six-dimensional internal diffeomorphisms.
Accordingly, the covariant field strength as evaluated from the corresponding components
of the ${\rm E}_{6(6)}$ object ${\cal F}_{\mu\nu}{}^M$ 
coincides with the non-abelian field strength for the Kaluza-Klein vector field in (\ref{covderfieldstr})
\bea
{\cal F}_{\mu\nu}{}^m &=&  2\partial_{[\mu} A_{\nu]}{}^m 
- A_{\mu}{}^n\partial_n A_{\nu}{}^m+A_{\nu}{}^n\partial_n A_{\mu}{}^m~=~ F_{\mu\nu}{}^m
\;.
\eea
Evaluating the remaining components of the 
covariant field strengths (\ref{modF}) yields the
field strengths for the other gauge fields as 
\bea
{\cal F}_{\mu\nu\,mn}  &=& 2\,D_{[\mu}{A}_{\nu]\,mn}
  +\partial_{m}\tilde{B}_{\mu\nu\,\bar{n}}-\partial_{n}\tilde{B}_{\mu\nu\,\bar{m}}
  \;,\nonumber\\
{\cal F}_{\mu\nu}{}^{\bar{m}} &=& 2\,D_{[\mu}A_{\nu]}{}^{\bar{m}}
   -2\,(\partial_k A_{[\mu}{}^k) A_{\nu]}{}^{\bar{m}}
  -\frac{1}{2}\,\epsilon^{mnrskl}\,A_{[\mu|rs}\partial_{n|} A_{\nu]kl}
  +2\,\partial_n\tilde{B}_{\mu\nu}{}^{nm}
  \;,
\eea
where we have redefined the two-form tensors as
 \bea
  \tilde{B}_{\mu\nu \,\bar{m}} &=& \sqrt{5}\,B_{\mu\nu \bar{m}}+A_{[\mu}{}^n A_{\nu]\,nm}\;,
    \nonumber\\
  \tilde{B}_{\mu\nu}{}^{mn} &=& 
\sqrt{5} \, {B}_{\mu\nu}{}^{mn}
+\frac12 \left(A_{[\mu}{}^m A_{\nu]}{}^{\bar{n}}
- A_{[\mu}{}^n A_{\nu]}{}^{\bar{m}}\right)
\;.
\label{redB}
 \eea
In turn, we obtain the field strengths for these two-form tensors
by evaluating the corresponding components of the ${\rm E}_{6(6)}$ object ${\cal H}_{\mu\nu\rho\,M}$\,:
\bea
\tilde{\cal H}_{\mu\nu\rho\,\bar{m}}&\equiv& 
 \sqrt{5}\,{\cal H}_{\mu\nu\rho\,\bar{m}} - \partial_m {\cal O}_{\mu\nu\rho}
 ~=~ 3 D_{[\mu} \tilde{B}_{\nu\rho]\bar{m}}
+3 A_{[\mu|mn|}{}F_{\nu\rho]}{}^n \;, 
\nonumber\\[1ex]
\tilde {\cal H}_{\mu\nu\rho}{}^{mn} &\equiv& \sqrt{5}\,{\cal H}_{\mu\nu\rho}{}^{mn} 
-\partial_k {\cal O}_{\mu\nu\rho}{}^{[kmn]}
\nonumber\\
&=&
3  D_{[\mu} \tilde{B}_{\nu\rho]}{}^{mn} 
- 3  \,  \partial_k A_{[\mu}{}^k  \tilde{B}_{\nu\rho]}{}^{mn} 
+ \frac32 \left(A_{[\mu}{}^{\bar{m}} {F}_{\nu\rho]}{}^{n} -A_{[\mu}{}^{\bar{n}} {F}_{\nu\rho]}{}^{m}\right)
\nonumber\\
&&{}
- \frac34\, \epsilon^{mnklpq} \left( A_{[\mu}{}_{|kl|} D^{\vphantom{p}}_\nu A_{\rho]}{}_{pq} 
+2 \,  A_{[\mu|kl} \partial_{p|} \tilde B_{\nu\rho]\bar{q}}\right)
\;,
\label{defHHH}
\eea
where we have split off the additional contributions
\bea
{\cal O}_{\mu\nu\rho} &\equiv & - A_{[\mu}{}^k A_\nu{}^l A_{\rho]\,kl} 
\;,
\nonumber\\
{\cal O}_{\mu\nu\rho}{}^{[kmn]} &\equiv&  
A_{[\mu}{}^k A_\nu{}^m A_{\rho]}{}^{\bar{n}} + 
 A_{[\mu}{}^n A_\nu{}^k A_{\rho]}{}^{\bar{m}} + 
A_{[\mu}{}^m A_\nu{}^n  A_{\rho]}{}^{\bar{k}} 
\nonumber\\
&&{}
+\frac12\, \epsilon^{kmnlpq} \,
\left( 3 A_{[\mu|lp|} \tilde B_{\nu\rho]\bar{q}} -2 A_{[\mu}{}_{|lp|} A_{\nu\vphantom{[}}{}^r A_{\rho]rq}\right)
\;,
\eea
that are projected out from the Lagrangian, since just as the tensor fields
also their field strengths appear only under projection
$d^{MNK}\partial_N {\cal H}_{\mu\nu\rho\,K}$, cf.~(\ref{onlyB}).

For completeness, let us also give the vector and tensor gauge transformations of the 
various components as obtained from evaluating the general formulae (\ref{deltaAB})
 \bea
  \delta A_{\mu\,mn} &=&  D_\mu \Lambda_{mn}+{\cal L}_\Lambda A_{\mu\,mn}
  -2\,\partial_{[m}\tilde\Xi_{|\mu|\,n]}\;,
\nonumber\\
\delta A_\mu{}^{\bar{m}} &=&
D_\mu \Lambda^{\bar{m}} 
-\partial_n A_\mu{}^n \Lambda^{\bar{m}} +{\cal L}_\Lambda A_\mu{}^{\bar{m}} 
-2\, \partial_n \tilde\Xi_\mu{}^{nm}
\;,
\nonumber\\
\delta \tilde{B}_{\mu\nu\, \bar{m}} &=& 2D_{[\mu}\tilde{\Xi}_{\nu]m}
   +{\cal L}_\Lambda  \tilde{B}_{\mu\nu m}  +\Lambda_{km}F_{\mu\nu}{}^k  
   -\partial_m(\Lambda^k\tilde{B}_{\mu\nu\,{\bar k}})\;,
   \label{delAAB}
\eea
with tensor gauge parameters redefined in accordance with (\ref{redB})
\bea
  \tilde{\Xi}_{\mu\, m} &\equiv& \sqrt{5}\,{\Xi}_{\mu\, m}+ \Lambda^n A_{\mu\, nm}\;,
\nonumber\\
\tilde\Xi_\mu{}^{mn}&=& \sqrt{5}\,  \Xi_\mu{}^{mn}
+\frac12\left(\Lambda^{m} A_\mu{}^{\bar{n}} 
-\Lambda^{n} A_\mu{}^{\bar{m}}\right)
\;.
\eea

 \subsubsection{Scalar sector}
 
Let us now discuss the scalar field content of the theory. In the E$_{6(6)}$-covariant formulation
they parametrize the coset space ${\rm E}_{6(6)}/{\rm USp}(8)$ in terms of the
symmetric matrix ${\cal M}_{MN}$\,. To relate to $D=11$ supergravity,
we need to choose a parametrization of this matrix in accordance with the decomposition (\ref{split78}).
Following~\cite{Cremmer:1997ct}, we build the matrix as ${\cal M}={\cal V}{\cal V}^T$ from a `vielbein' ${\cal V}$
in triangular gauge
\bea
{\cal V}^T &\equiv& {\rm exp} \left[\Phi\, t_{(0)}\right]\,{\cal V}_6\;
{\rm exp}\left[c_{kmn}\,t_{(+1)}^{kmn}\right]\,{\rm exp}\left[\varphi \, t_{(+2)}\right]
\;.
\label{V27}
\eea
Here, $t_{(0)}$ is the E$_{6(6)}$ generator associated to the GL(1) grading, ${\cal V}_6$ denotes a general matrix in the SL(6) subgroup,
whereas the $t_{(+n)}$ refer to the E$_{6(6)}$ generators of positive grading in (\ref{split78}). All generators are evaluated in the
fundamental ${\bf 27}$ representation (\ref{split27}), such that the symmetric matrix ${\cal M}_{MN}$ takes the block form
\bea
{\cal M}_{KM} &=& \left(
\begin{array}{ccc}
{\cal M}_{km}&{\cal M}_k{}^{mn}&{\cal M}_{k\bar{m}}\\
{\cal M}^{kl}{}_m & {\cal M}^{kl,mn} & {\cal M}^{kl}{}_{\bar{m}}\\
{\cal M}_{\bar{k}m}& {\cal M}_{\bar{k}}{}^{mn}&{\cal M}_{\bar{k}\bar{m}}
\end{array}
\right)
\;.
\label{M27}
\eea
Explicit evaluation of (\ref{V27}) determines the various blocks in (\ref{M27}). E.g.\ its last line is given
by
\bea
{\cal M}_{\bar{m}n} &=& \frac1{24} \,e^\Phi  m_{mk} \,\epsilon^{klpqrs} \,c_{nlp}c_{qrs}-e^\Phi  m_{mn}\,\varphi
\;,
\nonumber\\
{\cal M}_{\bar{m}}{}^{kl} &=& -\frac1{6\,\sqrt{2}}\,m_{mn} \epsilon^{nklpqr}\,e^\Phi  c_{pqr}\;,
\qquad
{\cal M}_{\bar{m}\bar{n}} ~=~ e^\Phi m_{mn}\;,
\label{compM1}
\eea
parametrized by $\Phi$, $\varphi$, $c_{kmn}$. The symmetric matrix
$m_{mn}\equiv ({\nu}{\nu}{}^T)_{mn}$ is built from the ${\rm SL}(6)$ vielbein $\nu$ that parametrizes 
the standard embedding of this subgroup via ${\cal V}_6$ in (\ref{V27})  as
\bea
({\cal V}_{6})_{M}{}^A &=& \left(
\begin{array}{ccc}
{\nu}_{m}{}^a&0&0\\
0 & ({\nu}^{-1}){}^{[m}{}_a({\nu}^{-1}){}^{n]}{}_b & 0\\
0&0&{\nu}_{\bar{m}}{}^{\bar{a}}
\end{array}
\right)
\;.
\label{sl6}
\eea
The remaining blocks of (\ref{M27})
yield more lengthy expressions, but can be expressed in compact form via the corresponding 
blocks of the matrix
\bea
\tilde{\cal M}_{MN} &\equiv& {\cal M}_{MN}- {\cal M}_{M\bar{m}} ({\cal M}_{\bar{m}\bar{n}})^{-1} {\cal M}_{\bar{n}N}
\;,
\label{tildeM27}
\eea
which take the form
\bea
\tilde{\cal M}_{mn}  &=& e^{-\Phi}\,m_{mn} + \frac12\,c_{mkp}c_{nlq} \,m^{kl} m^{pq}
\;,\nonumber\\
\tilde{\cal M}_{m}{}^{kl}  &=& -\frac1{\sqrt{2}}\,c_{mpq} \,m^{pk} m^{ql}
\;,\qquad
\tilde{\cal M}^{kl,mn}  ~=~ m^{m[k}m^{l]n}
\;.
\label{compM2}
\eea
The matrix (\ref{tildeM27}) will play a central role in the following after re-dualizing 
some of the vector fields. From the inverse matrix ${\cal M}^{MN}$ we will need only 
the particular block
\bea
{\cal M}^{mn} &=& e^\Phi m^{mn}
\;.
\label{MI}
\eea

Now, that we have specified the field content according to the explicit 
solution~(\ref{explicit_section}), we can work out the E$_{6(6)}$ covariant Lagrangian
in this parametrization.
Let us start with the scalar kinetic term. First, we should evaluate the
covariant derivatives ${\cal D}_\mu {\cal M}_{MN}$ in the split (\ref{split27}).
With (\ref{dbreak}) we find for the covariant derivatives of the components
of a general vector~$V^M$
\bea
{\cal D}_\mu V^m &=& 
D_\mu V^m+ \frac13\,(\partial_k A_\mu{}^k)\,V^m 
\;,
\nonumber\\
{\cal D}_\mu V_{mn} &=&D_\mu V_{mn} 
+ \frac13\,(\partial_kA_\mu{}^k)\,V_{mn} 
+V^k \partial_{k}A_{\mu\,mn}+V^k \partial_{m}A_{\mu\,nk}+V^k \partial_{n}A_{\mu\,km}
\;,
\nonumber\\
{\cal D}_\mu V^{\bar{m}} &=& D_\mu V^{\bar{m}}- \frac23\,(\partial_k\Lambda^k)\,V^{\bar{m}}
+\frac12\,\epsilon^{mnklpq}\,\partial_n A_{\mu\,kl}\,V_{pq}
+(\partial_k A_\mu{}^{\bar{k}})\,V^m
\;,
\label{dVsplit}
\eea
where as above the derivatives $D_\mu$ are only covariantized with respect to the 
Kaluza-Klein gauge transformations, i.e.~$D_\mu \equiv \partial_\mu - {\cal L}_{A_\mu}$\,.
Comparing this to the parametrization (\ref{compM1}) of the matrix ${\cal M}_{MN}$, we derive
the covariant derivatives on the parameters of this matrix as
\bea
{\cal D}_\mu \,m_{mn} &=& D_\mu \, m_{mn} + \frac13\,(\partial_k A_\mu{}^k)\,m_{mn}
\;,
\nonumber\\
{\cal D}_\mu \Phi &=& D_\mu \Phi +  (\partial_n A_\mu{}^n)
\;,
\nonumber\\
{\cal D}_\mu c_{klm} &=& D_\mu \, c_{klm} {+} 3\,\sqrt{2}\, \partial_{[k} A_{|\mu|lm]}
\;,
\nonumber\\
{\cal D}_\mu \varphi &=& D_\mu \varphi -  (\partial_n A_\mu{}^n) \, \varphi
+\partial_n A_\mu{}^{\bar{n}} 
+\frac{\sqrt{2}}{24} \,\epsilon^{klmnpq} \,c_{klm} \, \partial_n A_{\mu\,pq}
\;.
\label{covcomp}
\eea
From the first two lines we infer that the combination
\bea
\phi_{mn} &\equiv& e^{-\Phi/3}\,m_{mn}
\;,
\eea
transforms as a genuine tensor (of vanishing weight) under six-dimensional diffeomorphisms.
As anticipated by the notation, we will identify it with the internal part 
$\phi_{mn}=\phi_{m\,\alpha} \phi_n{}^{\alpha}$ 
of the metric of eleven-dimensional supergravity (\ref{KKgauge}).

Putting all this together,
we obtain after some calculation the explicit form of the scalar kinetic term from (\ref{finalaction})
\bea
e^{-1}\,{\cal L}_{\rm kin,0} ~\equiv~
\frac1{24}\,{\cal D}_\mu {\cal M}_{MN}\,{\cal D}^\mu {\cal M}^{MN}
&=&
\frac14\, {D}_\mu \phi_{mn}\,{D}^\mu \phi^{mn}
-\frac1{3}\,\phi^{-2}\,D_\mu \phi   D^\mu \phi
\nonumber\\
&&{}
-\frac1{12}\,\phi^{-2/3}\,\phi^{kn}\phi^{lp}\phi^{mq}\,{\cal D}_{\mu} c_{klm} {\cal D}^{\mu} c_{npq}
\nonumber\\
&&{}
-\frac12\,\phi^{-2}
\left({\cal D}_\mu\varphi + \frac1{72} \,\epsilon^{klmnpq}\, c_{klm} {\cal D}_{\mu} c_{npq}\right)^2 \;
\label{kinfromE6}
\eea
with $\phi\equiv e^{-\Phi}= ({\rm det}\, \phi_{mn})^{1/2}$ as above.
Next, we can evaluate the E$_{6(6)}$ covariant potential (\ref{fullpotential}) in the parametrization 
(\ref{compM1}), (\ref{compM2}) and obtain
\bea
  V &= &
  -\frac13\,\phi^{-2/3} \partial_{m}\phi_{nk} \,\partial_{l} \phi_{pq} \,\phi^{mn}\phi^{kl}\phi^{pq}
 +\frac1{36}\,\phi^{-2/3} \partial_{m}\phi_{nk} \,\partial_{l} \phi_{pq} \,\phi^{ml}\phi^{nk}\phi^{pq}
 \nonumber\\
  &&
  +\frac1{4}\,\phi^{-2/3} \partial_{m}\phi_{nk} \,\partial_{l} \phi_{pq} \,\phi^{ml}\phi^{np}\phi^{kq}
  -\frac1{2}\,\phi^{-2/3} \partial_{m}\phi_{nk} \,\partial_{l} \phi_{pq} \,\phi^{mp}\phi^{nq}\phi^{kl}
  \nonumber\\
  &&
  +\frac23\,\phi^{-5/3} \phi^{mn}\,e^{-1}\partial_me\,\partial_n \phi
  -e^{-1}\phi^{-2/3} \partial_me\, \partial_n \phi^{mn}
  +\frac{1}{3}\, \phi^{-2/3} \,\partial_{[k} c_{lmn]}\, \partial_{[p} c_{qrs]}  \,\phi^{kp}\phi^{lq}\phi^{mr}\phi^{ns}
 \nonumber\\
  &&
    -  \phi^{-2/3} \phi^{mn}\,e^{-1}\partial_me\,e^{-1}\partial_ne
    -\frac{1}{4}\phi^{-2/3} \phi^{mn} \partial_m g^{\mu\nu}\partial_n g_{\mu\nu}
\;.
\label{potfromE6}
 \eea
In particular, the second line of the potential (\ref{fullpotential}) is straightforwardly evaluated with (\ref{MI}).

\subsubsection{Dualization}
\label{subsubsec:dualization}

Before explicitly evaluating the remaining parts of the E$_{6(6)}$ covariant Lagrangian, let us recall the 
field content. From (\ref{AB}) and the subsequent discussion, we have vectors and two-forms given by
\bea
\left\{ A_\mu{}^m\,,\; A_{\mu\,mn}\,, \;A_{\mu}{}^{\bar{m}} \right\}\;,\qquad
\left\{\tilde{B}_{\mu\nu\,\bar{m}}\,, \;\tilde{B}_{\mu\nu}{}^{mn} \right\}
\;, 
\eea
of which only the vectors represent propagating degrees of freedom.
In the previous subsection we have introduced the parametrization of the scalar fields of the model as
\bea
\left\{ \phi_{mn}\,, \;c_{kmn}\,, \; \varphi \,\right\}
\;.
\eea
Comparing this to the form of eleven-dimensional supergravity in the 5+6 split 
presented in section~\ref{subsec:decomposition}, we see that we will have to dualize the 
singlet scalar field $\varphi$ into a three-form tensor field 
and eliminate the fields $A_{\mu}{}^{\bar{m}}$ and $\tilde{B}_{\mu\nu}{}^{mn}$. 
In particular, the latter step should introduce a kinetic term for the two-form 
tensor fields $\tilde{B}_{\mu\nu\,\bar{m}}$, promoting these fields to
propagating degrees of freedom. 

For the dimensionally reduced theory this is precisely the pattern of dualizations of
$p$-forms into $(3-p)$-forms that is required to make the E$_{6(6)}$ symmetry apparent~\cite{Cremmer:1997ct}.
In the following, we give a version of that dualization which applies even for the fully $y$-dependent fields despite the
non-abelian structure of the internal diffeomorphisms that may put a seeming obstacle to the possibility of dualization.
It is rather similar to the mechanisms of non-abelian dualizations appearing in gauged 
supergravity~\cite{Nicolai:2003bp,deWit:2003ja} empowered by the compensating fields of the tensor hierarchy.
As a result, we will show in this section that upon this dualization, the Lagrangian evaluated from (\ref{finalaction})
precisely coincides with $D=11$ supergravity.

We start by dualizing the singlet scalar field $\varphi$ into a three-form.
To this end, we first note that the Lagrangian (\ref{finalaction})
after resolution of the section condition according to (\ref{explicit_section})
has a global symmetry that acts by shift on $\varphi$. 
Its origin is the E$_{6(6)}$ generator $t_{(+2)}$ in the basis of (\ref{V27}) with action
\bea
\delta_\lambda \varphi &=& \lambda
\;,
\qquad
\delta_\lambda A_\mu{}^{\bar{m}} ~=~ \lambda A_\mu{}^m
\;,
\label{shifts}
\eea
on scalar and vector fields.
This symmetry is compatible with the solution of the section constraint (\ref{explicit_section}) due to
\bea
\delta_\lambda \,\partial_{\bar{m}} &=& 0
\;,\qquad
\delta_\lambda\, \partial^{mn} ~=~ 0
\;,
\label{compcon}
\eea
as an immediate consequence of the grading (\ref{split27}), (\ref{split78}).
As a result, this symmetry survives after imposing
the explicit solution of the section constraint.
Moreover, due to our field redefinitions (\ref{redB}), the same generator has a non-trivial action on
the two-forms as
\bea
\delta_\lambda \tilde{B}_{\mu\nu}{}^{mn} &=& \lambda A_{[\mu}{}^m A_{\nu]}{}^n
\;.
\label{shifts2}
\eea
For dualizing the scalar fields $\varphi$ we will now follow a standard routine:
gauge the shift symmetry (\ref{shifts}) by introduction of an auxiliary vector field
and eliminate the latter by its field equations.
Specifically, in the scalar sector we introduce covariant derivatives
\bea
{\cal D}_\mu &\longrightarrow&\widehat{\cal D}_\mu \ \equiv \ {\cal D}_\mu - a_\mu \, t_{(+2)}
\;,
\label{covnew}
\eea
such that the kinetic term (\ref{kinfromE6}) remains invariant under the local form of (\ref{shifts}) provided
the auxiliary vector $a_\mu$ transforms as
\bea
\delta_\lambda a_\mu &=& \partial_\mu \lambda
\;,\qquad
\delta_\Lambda a_\mu ~=~  
 {\cal L}_\Lambda a_\mu + (\partial_k\Lambda^k)\,  a_\mu
\;.
\label{deltaa}
\eea
In the vector sector, gauging of (\ref{shifts}) is more intricate, since the new gauge symmetry interferes
with the existing non-abelian structure~(\ref{delAAB}) of the vector fields. As a result, this further deformation 
necessitates the introduction of additional St\"uckelberg type couplings
on the level of the field strengths according to
 \bea
{{\cal F}}_{\mu\nu}{}^{\bar{m}} ~\rightarrow~ \widehat{{\cal F}}_{\mu\nu}{}^{\bar{m}} &\equiv& 
2\widehat{D}_{[\mu}A_{\nu]}{}^{\bar{m}}
  -2(\partial_k A_{[\mu}{}^k) A_{\nu]}{}^{\bar{m}}
  -\frac{1}{2}\epsilon^{mnrskl}A_{[\mu|rs}\partial_{n|} A_{\nu]kl}
  \nonumber\\
  &&{}
  +2\partial_n\tilde{B}_{\mu\nu}{}^{nm} + b_{\mu\nu}{}^m
  \;,
  \label{Fmod}
 \eea
with the new auxiliary two-form $b_{\mu\nu}$ transforming as
\bea
\delta_\lambda b_{\mu\nu}{}^m~=~0\;,\qquad
\delta_\Lambda b_{\mu\nu}{}^m &=& 
{\cal L}_\Lambda b_{\mu\nu}{}^m+(\partial_k\Lambda^k)\,b_{\mu\nu}{}^m  
+2 a_{[\mu} \partial_{\nu]}  \Lambda^{{m}}
\;,
\eea
in order to guarantee covariant transformation behaviour of the field strength.
With these extra fields and modified transformations, the kinetic part of the 
Lagrangian is thus invariant under $\lambda$ and $\Lambda^M$
transformations. 
Moreover, the auxiliary two-form $b_{\mu\nu}{}^m$ comes with its own tensor gauge invariance 
\bea
\delta_\xi b_{\mu\nu}{}^m &=& 
2\partial_{[\mu} \xi_{\nu]}{}^{\bar{m}}
\;,\qquad
\delta_\xi a_\mu ~=~ -\partial_n \xi_\mu{}^{\bar{n}}
\;,\nonumber\\
\delta_\xi A_\mu{}^{\bar{m}} &=& -\xi_\mu{}^{\bar{m}}\;,
\quad\;\,
\delta_\xi \tilde B_{\mu\nu}{}^{mn}~=~ -A_{[\mu}{}^m\xi_{\nu]}{}^{\bar{n}}
+A_{[\mu}{}^n\xi_{\nu]}{}^{\bar{m}}
\;,
\label{newgauge}
\eea
which separately leaves the kinetic part of the Lagrangian invariant.

Let us now turn to the topological term (\ref{CSlike}) in order to render it invariant under the
new gauge symmetries (\ref{shifts}), (\ref{shifts2}), (\ref{newgauge}). After evaluating this term 
with the solution of the section condition (\ref{explicit_section}), it is invariant under the global 
symmetry (\ref{shifts}), (\ref{shifts2}) but acquires a non-trivial
variation for a local gauge parameter $\lambda$ according to
\bea
  \delta_\lambda {\cal L}_{\rm top,0} 
&=&
-\frac1{\sqrt{2}} \varepsilon^{\mu\nu\rho\sigma\tau}
\,\partial_\mu \lambda
\left(
  { F}_{\nu\rho}{}^m   A_{\sigma\,mn} A_\tau{}^{n}
+ \partial_m \tilde{B}_{\nu\rho\,\bar{n}}\, A_\sigma{}^m A_\tau{}^n\right)
\;.
\label{vartopoL}
\eea
In view of (\ref{deltaa}), this variation can be cancelled by 
adding the additional topological term
\bea
{\cal L}_{\rm top,1}&\equiv&
\frac1{\sqrt{2}}\, \varepsilon^{\mu\nu\rho\sigma\tau}
\,a_\mu 
\left(
  { F}_{\nu\rho}{}^m   A_{\sigma\,mn} A_\tau{}^{n}
+ \partial_m \tilde{B}_{\nu\rho\,\bar{n}}\, A_\sigma{}^m A_\tau{}^n\right)
\;,
\label{newtop}
\eea
such that the sum ${\cal L}_{\rm top,0} + {\cal L}_{\rm top,1}$ is invariant under local $\lambda$
transformations. 
In turn, the variation of this combined topological term under the local tensor gauge symmetry~(\ref{newgauge}) is given by
\bea
 \delta_\xi {\cal L}_{\rm top,0+1} &=&
-\frac1{\sqrt{2}}\, \varepsilon^{\mu\nu\rho\sigma\tau}
\left(
2 \partial_\mu A_\nu{}^k \partial_{[k} \tilde{B}_{\rho\sigma\,\bar{m}]}
-2 A_\mu{}^k   \partial_\nu \partial_{[k}\tilde{B}_{|\rho\sigma|\,\bar{m}]}
- \partial_\mu(A_{\nu\,mn} F_{\rho\sigma}^n)
\right) \xi_\tau{}^{\bar{m}}
\nonumber\\
&=&
\frac1{3\sqrt{2}}\, \varepsilon^{\mu\nu\rho\sigma\tau}
\left(
\tilde{\cal H}_{\mu\nu\rho\,\bar{m}}+3\,\partial_m(A_\mu{}^n\tilde{B}_{\nu\rho\,\bar{n}})\right)\partial_\sigma \xi_\tau{}^{\bar{m}}
\;,
\label{vario}
\eea
and thus can be cancelled by introduction of a second addition to the topological term
\bea
{\cal L}_{\rm top,2} &=&
-\frac1{6\sqrt{2}}\, \varepsilon^{\mu\nu\rho\sigma\tau}
\left(
\tilde{\cal H}_{\mu\nu\rho\,\bar{m}}+3\,\partial_m(A_\mu{}^n\tilde{B}_{\nu\rho\,\bar{n}})\right)b_{\sigma\tau}{}^m
\;.
\label{topo2}
\eea

Finally, we have to ensure that the combined topological term ${\cal L}_{\rm top,0+1+2}$
remains invariant under the original
$\Lambda^M$ and $\Xi_{\mu\,M}$ gauge transformations of (\ref{deltaAB}).
After some lengthy but straightforward calculation, 
we find for this variation 
\bea
\delta  {\cal L}_{\rm top,0+1+2}
&=&
\frac1{2\sqrt{2}}\,\varepsilon^{\mu\nu\rho\sigma\tau} \left(
2A_\mu{}^kA_\nu{}^n\partial_k \tilde\Xi_{\rho\,n}
-A_\mu{}^kF_{\nu\rho}{}^n \Lambda_{kn}
-\Lambda^n \partial_\mu \tilde{B}_{\nu\rho\,\bar{n}}
\right)\left( 2 \partial_\sigma a_\tau + \partial_m b_{\sigma\tau}{}^m\right)
\nonumber\\
&&{}
-\frac1{2\sqrt{2}}\,\varepsilon^{\mu\nu\rho\sigma\tau} \partial_m \left(
2  A_\mu{}^k\,\tilde\Xi_{\nu\,k}
-\Lambda^n \tilde{B}_{\mu\nu\,\bar{n}}
\right)  \partial_\rho b_{\sigma\tau}{}^m
\;.
\label{remain12}
\eea
This variation is cancelled by adding to the topological Lagrangian the final contribution
\bea
{\cal L}_{\rm top,3} &=& \frac1{4\sqrt{2}}\, \varepsilon^{\mu\nu\rho\sigma\tau}
\left( 
2 a_\mu \partial_\nu {\cal A}_{\rho\sigma\tau} + \partial_m b_{\mu\nu}{}^m {\cal A}_{\rho\sigma\tau} \right)
\;,
\label{Ldual}
\eea
with the new field ${\cal A}_{\rho\sigma\tau}$, 
transforming as
\bea
\delta {\cal A}_{\mu\nu\rho} &=& 
{\cal L}_\Lambda {\cal A}_{\mu\nu\rho} 
+2\Lambda^n  \partial_{[\mu} \tilde B_{\nu\rho]\,n}
+
2\,A_{[\mu}{}^{m} { F}_{\nu\rho]}{}^n     \Lambda_{mn} 
-4\,\partial_{m} \tilde\Xi_{[\mu\,|\bar{n}|}\, A_\nu{}^m A_{\rho]}{}^n
\nonumber\\
&&{}
+2\,\partial_{[\mu} (2  A_{\nu}{}^k\,\tilde\Xi_{\rho]\,k}
-\Lambda^n \tilde{B}_{\nu\rho]\,\bar{n}})
\;.
\label{varA3}
\eea
A short calculation shows that also the terms in the variation of (\ref{Ldual}) proportional to ${\cal A}_{\rho\sigma\tau}$  cancel.
Moreover, the term (\ref{Ldual}) is separately invariant under the new gauge symmetries (\ref{shifts}), (\ref{newgauge}),
so no further compensation is required.
To clean up the construction, we may eventually combine all new contributions to the topological term, which can
be put into the more compact form
\bea
{\cal L}_{\rm top,1+2+3}
&=&
\frac1{4\sqrt{2}}\,\varepsilon^{\mu\nu\rho\sigma\tau}
\,\left(
2 a_\mu 
\left(
{D}_\nu\tilde {\cal A}_{\rho\sigma\tau} -\tilde{B}_{\nu\rho\,\bar{m}} F_{\sigma\tau}{}^m
\right)
- 
\frac13\,\tilde{b}_{\mu\nu}{}^m
\left(
2\,\tilde{\cal H}_{\rho\sigma\tau\,\bar{m}}+3\partial_m \tilde{\cal A}_{\rho\sigma\tau} \right)
\right)
\;,
\nonumber\\
\label{fulltop}
\eea
with the auxiliary fields redefined as
\bea
\tilde{b}_{\mu\nu}{}^m &\equiv& b_{\mu\nu}{}^m - 2a_{[\mu}A_{\nu]}{}^m
\;,
\nonumber\\
\tilde{\cal A}_{\mu\nu\rho} &\equiv&
{\cal A}_{\mu\nu\rho}+2 A_\mu{}^n\tilde{B}_{\nu\rho\,\bar{n}}
\;.
\label{newaux}
\eea
After these redefinitions, the gauge transformations of ${\cal A}_{\mu\nu\rho}$ in (\ref{varA3})
take the fully covariant and more compact form
\bea
\delta \tilde{\cal A}_{\mu\nu\rho} &=& 
{\cal L}_\Lambda \tilde{\cal A}_{\mu\nu\rho} 
+2\, F_{[\mu\nu}{}^n \,\tilde{\Xi}_{\rho]n}
\;.
\eea
In the course of our construction,
something interesting has happened. Recall that the original Lagrangian 
carried the two-form $\tilde{B}_{\mu\nu\,\bar{n}}$ exclusively under $\partial_m$ derivative \`a la (\ref{onlyB}). 
This is still true for its variation~(\ref{vario}) (although not manifest in the final expression),
but no longer for the compensating term (\ref{topo2}).
Consequently, the new topological term (\ref{fulltop})
carries the longitudinal part of $\tilde{B}_{\mu\nu\,\bar{n}}$ as a new field.
Nevertheless, the shift symmetry (\ref{shiftB}) of the original Lagrangian can be preserved,
if the field $\tilde{\cal A}_{\mu\nu\rho}$ simultaneously transforms as
\bea
\delta \tilde{\cal A}_{\mu\nu\rho} &=& -2\, D_{[\mu} \Omega_{\nu\rho]}\;,
\qquad
\delta \tilde{B}_{\mu\nu\,\bar{m}} ~=~ 
\partial_m \Omega_{\mu\nu}
\;.
\label{gaugeA3}
\eea
I.e.\ this symmetry is identified with 
the tensor gauge symmetry of the new three-form $\tilde{\cal A}_{\mu\nu\rho}$\,.

Let us pause and summarize what we have achieved. Upon introducing new covariant derivatives and field strengths 
(\ref{covnew}) and (\ref{Fmod}) in the Lagrangian, as well as extending its topological term ${\cal L}_{\rm top,0}$
to ${\cal L}_{\rm top,0+1+2+3}$ from (\ref{fulltop}) we have modified the original
Lagrangian such that in addition to the former gauge symmetries it is also invariant under the new local gauge
symmetries (\ref{shifts}), (\ref{newgauge}), (\ref{gaugeA3}). The modification has introduced the auxiliary 
vector and tensor gauge fields $a_\mu$, $b_{\mu\nu}{}^m$, and ${\cal A}_{\mu\nu\rho}$\,.
The resulting Lagrangian provides an efficient tool to perform the dualization of the original theory.
We can show that depending on how we treat the auxiliary fields, the Lagrangian either reduces to the original one
or takes a different form, in which the former fields $\varphi$ and $A_\mu{}^{\bar{m}}$ disappear.  
Thereby we arrive at the dual version of the original Lagrangian.

Let us first show that the new Lagrangian is equivalent to the original theory obtained from 
the E$_{6(6)}$-covariant EFT after solving the section condition.
Recall that the only term in which $\tilde{B}_{\mu\nu\,\bar{m}}$ appears
without derivative, is (\ref{topo2}). It thus gives separate equations of motion 
(by variation of the type (\ref{shiftB}) under which all other terms are invariant) 
implying that
\bea
\partial_m \partial_{[\mu}  b_{\nu\rho]}{}^m  &=& 0
\;.
\eea
With the local gauge symmetry (\ref{newgauge}) we can thus set 
\bea
\partial_m b_{\mu\nu}{}^m \ = \ 0\quad\Rightarrow\quad
b_{\mu\nu}{}^m \ = \ \partial_n\Upsilon_{\mu\nu}{}^{[mn]}
\;,
\eea
for some locally defined $\Upsilon_{\mu\nu}{}^{[mn]}$. Upon making use
of yet another local symmetry of the full Lagrangian,\footnote{
This is not a novel gauge symmetry but simply illustrates some redundancy in the introduction 
of the auxiliary field $b_{\mu\nu}$ in (\ref{Fmod}).}
\bea
 \delta \tilde{B}_{\mu\nu}{}^{mn}  &=& \frac12\,\Upsilon_{\mu\nu}{}^{[mn]} \;,\qquad
 \delta  b_{\mu\nu}{}^m ~=~ -\partial_n \Upsilon_{\mu\nu}{}^{[nm]}
 \;,
 \label{yetsym}
\eea
we can then completely eliminate the field $b_{\mu\nu}{}^m$.
The field equations following from variation of ${\cal A}_{\mu\nu\rho}$ in (\ref{Ldual}) imply that
\bea
2\,\partial_{[\mu} a_{\nu]} ~=~ - \partial_m b_{\mu\nu}{}^m~=~0
\;.
\label{abpure}
\eea
Thus, $a_\mu$ is also pure gauge and can be set to zero with the local symmetry 
(\ref{deltaa}). As a result, all auxiliary fields 
$a_\mu$, $b_{\mu\nu}{}^m$, and ${\cal A}_{\mu\nu\rho}$
disappear from the equations of motion and we are back to the theory obtained from 
the E$_{6(6)}$-covariant formulation.

Alternatively, we may integrate out the auxiliary gauge fields $a_\mu$, $b_{\mu\nu}$ upon using 
their algebraic field equations. 
The local symmetries (\ref{shifts}), (\ref{newgauge}), (\ref{yetsym}) which formally remain present
in this procedure, show that after integrating out $a_\mu$ and $b_{\mu\nu}$, the resulting Lagrangian
no longer depends on the fields $\varphi$, $A_\mu{}^{\bar{m}}$, and $\tilde B_{\mu\nu}{}^{mn}$.
Instead, the fields ${\cal A}_{\mu\nu\rho}$ and $\tilde{B}_{\mu\nu}{}^{\bar{m}}$ are promoted
to propagating fields with proper kinetic terms.
We thus obtain a dual version of the original Lagrangian with precisely the field content of $D=11$ supergravity.
To conclude this discussion, we will now show in detail that the result indeed 
coincides with the $D=11$ supergravity Lagrangian after Kaluza-Klein decomposition.

With the kinetic terms from (\ref{finalaction}) evaluated according to (\ref{M27}), (\ref{kinfromE6}), 
and covariantized according to (\ref{covnew}), (\ref{Fmod}),
the equations of motion for the auxiliary fields
$a_\mu$, $\tilde{b}_{\mu\nu}{}^m$ read
\bea
a_\mu &=&
{\cal D}_\mu\varphi 
+\frac1{72} \,\epsilon^{klmnpq}\, c_{klm} {\cal D}_{\mu} c_{npq}
+2\,\varepsilon^{\mu\nu\rho\sigma\tau} \,\phi^2
\left(
{D}_\nu\tilde {\cal A}_{\rho\sigma\tau} -\tilde{B}_{\nu\rho\,\bar{m}} F_{\sigma\tau}{}^m
\right) 
\;,
\nonumber\\
\tilde{b}_{\mu\nu}{}^m
&=& 
-({\cal M}_{\bar{m}\bar{n}})^{-1}\,  {\cal M}_{\bar{n}\,M}\,{\cal F}^{\mu\nu\,M}
-\frac23\,\varepsilon^{\mu\nu\rho\sigma\tau} \,({\cal M}_{\bar{m}\bar{n}})^{-1}\,
\left(
2\,\tilde{\cal H}_{\rho\sigma\tau\,\bar{n}}+3\partial_n \tilde{\cal A}_{\rho\sigma\tau} \right)
\;.
\label{EOMaux}
\eea
Inserting this into the Lagrangian produces the new kinetic terms
\bea
e^{-1}\,{\cal L}_{\rm kin,2+3} &=&
-\frac1{24}\,\phi^{4/3}\,\phi^{mn}
\left(
2\,\tilde{\cal H}_{\mu\nu\rho\,\bar{m}}+3\,\partial_m \tilde{\cal A}_{\mu\nu\rho} \right)
\left(
2\,\tilde{\cal H}^{\mu\nu\rho}{}_{\bar{n}}+3\,\partial_n \tilde{\cal A}^{\mu\nu\rho} \right)
\nonumber\\
&&{}
-\frac32\,\phi^{2}
\left(
{D}_{[\mu}\tilde {\cal A}_{\nu\rho\sigma]} -\tilde{B}_{[\mu\nu\,|\bar{m}|} F_{\rho\sigma]}{}^m
\right)
\left(
{D}^{\mu}\tilde {\cal A}^{\nu\rho\sigma} -\tilde{B}^{\mu\nu}{}_{\bar{n}} F^{\rho\sigma}{}^n
\right)
\;,
\label{Lkin23}
\eea
for the two-forms $\tilde{B}_{\mu\nu}{}^{\bar{m}}$ and three-form ${\tilde{\cal A}_{\mu\nu\rho}}$,
while the vector kinetic term turns into
\bea
e^{-1}\,{\cal L}_{\rm kin,1} &=&
-\frac14\, {\cal F}_{\mu\nu}{}^M {\cal F}^{\mu\nu}{}^N \tilde{\cal M}_{MN}
\;,
\label{FFM}
\eea
with the matrix $\tilde{\cal M}_{MN}$ from (\ref{tildeM27}), (\ref{compM2}).
In particular, the form of this matrix shows that the vector fields $A_\mu{}^{\bar{m}}$
have disappeared from the kinetic term (\ref{FFM}) as expected.
In order to calculate the topological term after elimination of the
auxiliary fields, let us first consider the original topological term
(\ref{CSlike}). After explicitly solving the section condition~(\ref{explicit_section})
we can give a fairly compact expression for this term upon integrating up (\ref{vartopo0}) as
\bea
{\cal L}_{\rm top,0}&=& 
\frac1{4\sqrt{2}}\,\varepsilon^{\mu\nu\rho\sigma\tau} \varepsilon^{mnklpq} \,
\Big(\,
\frac12 \,A_{\mu\,mn} {\cal F}_{\nu\rho\,kl}\,\partial_p\tilde{B}_{\sigma\tau\,q}
+\frac13\,D_\mu A_{\nu\,mn}D_\rho A_{\sigma\,kl}\,A_{\tau\,pq}
\nonumber\\
&&{}
\qquad\qquad\qquad\qquad
+\frac13\,\partial_m A_{\mu\,pq} A_{\nu\,kl}A_{\rho\,nr} F_{\sigma\tau}{}^r
+ {\cal O}(A_\mu{}^{\bar{m}}) + {\cal O}(B_{\mu\nu}{}^{mn})
\,\Big).
\label{topdual0}
\eea
Eventually, we are only interested in this term at vanishing $A_\mu{}^{\bar{m}}$, $B_{\mu\nu}{}^{mn}$,
since we know from the general symmetry argument above that these fields will no longer
enter the Lagrangian after elimination of the auxiliary fields.
Moreover, plugging (\ref{EOMaux}) into the original Lagrangian
gives the following additional contributions to the topological term
\bea
{\cal L}_{\rm top, dual} &=&
\frac1{4\sqrt{2}}\,\varepsilon^{\mu\nu\rho\sigma\tau} \varepsilon^{mnklpq} \,
\Big[\,
\frac1{12} c_{mnk} \left(\sqrt{2}\,\partial_l A_{\mu\,pq}+\frac13\,{\cal D}_\mu c_{lpq}\right)\, 
\left(
{D}_\nu\tilde {\cal A}_{\rho\sigma\tau} -\tilde{B}_{\nu\rho\,\bar{r}} F_{\sigma\tau}{}^r
\right)
\nonumber\\
&&{}
\qquad\qquad\qquad\qquad
+\frac1{72}\left(
F_{\mu\nu}{}^r  \,c_{rlp}c_{qmn}
-12\,A_{\mu\,kl}\partial_n A_{\nu\,pq}  
\right)
\left(
2\,\tilde{\cal H}_{\rho\sigma\tau\,\bar{k}}+3\partial_k \tilde{\cal A}_{\rho\sigma\tau} \right)
\nonumber\\
&&{}
\qquad\qquad\qquad\qquad
-\frac1{18\sqrt{2}} \,  c_{pqn}\,{\cal F}_{\mu\nu\,kl} \left(
2\,\tilde{\cal H}_{\rho\sigma\tau\,\bar{m}}+3\partial_m \tilde{\cal A}_{\rho\sigma\tau} \right)
\Big]
\;.
\label{topdual1}
\eea
Comparing the resulting parts of the Lagrangian (\ref{Lkin23})--(\ref{topdual1})
to the Kaluza-Klein decomposition of eleven-dimensional supergravity
presented in section~\ref{subsec:decomposition},
we are led to the following redefinition of fields
\bea
\tilde{{\cal A}}_{\mu\nu\rho}&\longrightarrow& \frac{2\sqrt{2}}3\,A_{\mu\nu\rho}\;,\qquad
\tilde{B}_{\mu\nu\,\bar{m}} ~\longrightarrow~ \sqrt{2}\,A_{\mu\nu\,m}\;,\nonumber\\
{A}_{\mu\,mn} &\longrightarrow& \sqrt{2} \,A_{\mu\,mn}\;,\qquad
c_{mnk} ~\longrightarrow~ -2 A_{mnk}
\;.
\label{trans}
\eea
With this translation, the above combinations of field strengths become 
\bea
2\,\tilde{\cal H}_{\mu\nu\rho\,\bar{m}} +3\,\partial_m \tilde{\cal A}_{\mu\nu\rho} &\longrightarrow&
2\sqrt{2}\, F_{\mu\nu\rho\,m}
\;,\nonumber\\
{D}_{[\mu}\tilde {\cal A}_{\nu\rho\sigma]} -\tilde{B}_{[\mu\nu\,|\bar{m}|} F_{\rho\sigma]}{}^m &\longrightarrow&
 -\frac{\sqrt{2}}{6}\,F_{\mu\nu\rho\sigma}
\;,\nonumber\\
{\cal F}_{\mu\nu}{\,}_{mn} &\longrightarrow& \sqrt{2}\,{\cal F}_{\mu\nu}{\,}_{mn}
\;,\nonumber\\
{\cal D}_\mu c_{klm} &\longrightarrow&  -2\,F_{\mu\,klm}
\;,
\eea
i.e.\ translated directly into the field strengths
(\ref{invfieldstr}), (\ref{newFKK}) introduced in the discussion of the Kaluza-Klein 
decomposition of eleven-dimensional supergravity.
It is then straightforward to verify that the combination of kinetic terms
(\ref{kinfromE6}), (\ref{Lkin23}), (\ref{FFM}),
indeed precisely coincides with the corresponding terms of
(\ref{finalKK}),
from eleven-dimensional supergravity.
Likewise, the combination of the topological terms (\ref{fulltop}), (\ref{topdual0}), (\ref{topdual1}),
using the dictionary  (\ref{trans}) reproduces the eleven-dimensional result~(\ref{finaltop}) up to total derivatives.
Although this comparison is not straightforward since there is no canonical form in which to give
these non-manifestly gauge covariant terms, they can be systematically matched comparing their
general variation w.r.t.\ the various gauge fields.
Similarly, agreement is found between the potential terms (\ref{potfromE6}) and (\ref{finalpot}).
Finally, the Einstein-Hilbert terms from eleven dimensions and from EFT are based on
the improved Riemann tensors (\ref{improvedRE6}) and (\ref{improvedR}),
that are readily identified since
\bea
{\cal F}_{\mu\nu}{}^M \partial_M &\rightarrow& { F}_{\mu\nu}{}^m \partial_m
\;,
\eea
on the solution of the section constraint (\ref{explicit_section}).
Thus we have shown total agreement between the EFT evaluated for (\ref{explicit_section})
and the full eleven-dimensional supergravity cast into the (5+6)-dimensional 
 Kaluza-Klein form.

\section{Embedding of Type IIB Supergravity}\label{sec5}

In the previous section, we have shown that upon imposing the
explicit ${\rm GL}(6)$ invariant solution (\ref{explicit_section}) of the section condition
and subsequent dualization of some of the fields, the E$_{6(6)}$ covariant EFT
precisely reproduces the full eleven-dimensional supergravity in the
5+6 Kaluza-Klein split. 
In this section, we discuss an inequivalent solution~\cite{Hohm:2013pua} to the section condition
upon which the EFT reproduces the full ten-dimensional IIB 
theory~\cite{Schwarz:1983wa,Howe:1983sra}.\footnote{
An analogous IIB solution of the ${\rm SL}(5)$ covariant section condition, 
corresponding to some three-dimensional truncation of type IIB supergravity,  
has been studied recently \cite{Blair:2013gqa} in the truncation of the theory to its potential term.}

\subsection{${\rm GL}(5)\times {\rm SL}(2)$ invariant solution of the section condition}

The corresponding solution of the section condition
preserves the group ${\rm GL}(5)\times {\rm SL}(2)$ embedded according to
\bea
{\rm GL}(5)\times {\rm SL}(2) &\subset& {\rm SL}(6) \times {\rm SL}(2) ~\subset~ {\rm E}_{6(6)}
\;,
\eea
into E$_{6(6)}$.
In this case, the fundamental  and the adjoint representation of ${\rm E}_{6(6)}$ break as
\bea
\bar{{\bf  27}} &\rightarrow& (5,1)_{+4}+(5',2)_{+1}+(10,1)_{-2}+(1,2)_{-5}
\;,
\label{split27B}
\\
{\bf 78} &\rightarrow&  (5,1)_{-6} + (10',2)_{-3} + \left(1+15+20\right)_{0} + (10,2)_{+3} + (5',1)_{+6}
\;,
\label{split78B}
\eea
with the subscripts referring to the charges under ${\rm GL}(1)\subset {\rm GL}(5)$.
An explicit solution to the section condition (\ref{sectioncondition}) is given by 
restricting the $Y^M$ dependence of all fields to the five coordinates in the $(5,1)_{+4}$. 
Explicitly, splitting the coordinates $Y^M$ and the fundamental indices according to (\ref{split27B}) into
\bea
\left\{Y^M\right\} &\rightarrow&\left\{ \,y^m\,,\; y_{m\,\alpha}\,,\; y^{mn}\,,\;  y_\alpha \,\right\}
\;,
\label{YbreakB}
\eea
with internal indices $m, n = 1, \dots, 5$ and SL$(2)$ indices  $\alpha=1, 2$, 
the non-vanishing components of the $d$-symbol are given by
 \bea
d^{MNK} &:&
 d^{m}{}_{n\alpha,\beta} = \frac1{\sqrt{10}}\, \delta^m_n \epsilon_{\alpha\beta}\;,\quad
d^{mn}{}_{k\alpha,l\beta} = \frac1{\sqrt{5}}\, \delta^{mn}_{kl}\,\epsilon_{\alpha\beta}\;,\quad
d^{mn,kl,p} = \frac1{\sqrt{40}}\,\epsilon^{mnklp}\;, \nonumber\\
d_{MNK} &:&   d_{m}{}^{n\alpha,\beta} = \frac1{\sqrt{10}}\, \delta^n_m \epsilon^{\alpha\beta}\;,\quad
d_{mn}{}^{k\alpha,l\beta} = \frac1{\sqrt{5}}\, \delta_{mn}^{kl}\,\epsilon^{\alpha\beta}\;,\quad
d_{mn,kl,p} = \frac1{\sqrt{40}}\,\epsilon_{mnklp}\;, 
 \label{dbreakB}
 \eea
and all those related by symmetry, $d^{MNK}=d^{(MNK)}$\,.
In particular, the ${\rm GL}(1)$ grading guarantees that all components $d^{m\,n\,k}$ vanish,
such that the section condition (\ref{sectioncondition}) indeed is solved by restricting the coordinate dependence
of all fields according to
\bea
\left\{\partial^{m\,\alpha}A =0\;,\; \partial_{mn} A = 0\;,\; \partial^\alpha A = 0\right\}\qquad
\Longleftrightarrow\qquad
A(x^\mu,Y^M) &\longrightarrow& A(x^\mu,y^m)
\;.
\label{explicit_sectionB}
\eea
Moreover, the form of the $d$-symbol (\ref{dbreakB}) shows that
any further coordinate dependence of a field $A$ 
on combinations of the remaining coordinates
violates the section condition. This explicitly shows that (\ref{explicit_sectionB})
is not a subcase of (\ref{explicit_section}), but a different inequivalent solution.

\subsection{${\rm GL}(5)\times {\rm SL}(2)$ invariant reduction of EFT}

In this subsection, we evaluate the EFT Lagrangian (\ref{finalaction}) upon splitting fields and tensors according to
(\ref{split27B})--(\ref{dbreakB}) and
assuming the explicit solution (\ref{explicit_sectionB}) of the section condition.
Having gone through this analysis in great detail for the case of $D=11$ supergravity in section~\ref{sec4},
we will keep the discussion much shorter here, and restrict it to the essential new ingredients.
In particular, in this case, due to the presence of the self-dual four-form in IIB,
there is no known ten-dimensional Lagrangian to which the result can immediately be compared. 
Rather, the procedure will produce an action, in which only an ${\rm SO}(1,4)\times {\rm SO}(5)$ subgroup
of the ten-dimensional Lorentz group is realized, much in the spirit of \cite{Henneaux:1988gg,Schwarz:1993vs}~in 
which Lorentz symmetry appears broken to ${\rm SO}(9)$ but is recovered on the level of the equations of motion.\footnote{
Covariant PST type formulations of IIB supergravity
have been constructed in~\cite{DallAgata:1997ju,DallAgata:1998va}.}

In analogy to the discussion in section~\ref{subsec:gl6} above, let us first 
revisit the resulting field content of the model. 
With the split (\ref{split27B}), (\ref{split78B}),
the full $p$-form field content of the ${\rm E}_{6(6)}$ Lagrangian in this basis
is thus given by
\bea
\left\{ A_\mu{}^m, A_{\mu\,m\,\alpha}, A_{\mu\,kmn},  A_{\mu\,\alpha} \right\}\;,\qquad
\left\{B_{\mu\nu}{}^\alpha ,  B_{\mu\nu\,mn}, B_{\mu\nu}{}^{m\,\alpha}  \right\}
\;,
\label{AB_B}
\eea
where we have defined $A_{\mu\,kmn}=\frac12\epsilon_{kmnpq}A_\mu{}^{pq}$\,.
More precisely, the Lagrangian depends on the two-forms only under derivatives, 
\bea
\left\{\,\partial_m B_{\mu\nu}{}^\alpha \,,\;  \partial_{[k}B_{|\mu\nu|\,mn]}\,,\; \partial_m B_{\mu\nu}{}^{m\,\alpha}\,  \right\}
\;.
\label{onlyBB}
\eea
Similar to the case of $D=11$ supergravity,
the vector fields $A_\mu{}^m$ will be identified with the IIB Kaluza-Klein vector fields.
Indeed, 
they transform under general gauge transformations (\ref{GaugeVar}) according to
\bea
\delta_\Lambda A_\mu{}^m & = & \partial_\mu \Lambda^m - A_\mu{}^n \partial_n \Lambda^m + 
\Lambda^n \partial_n A_\mu{}^m 
\;,
\eea
with the associated
gauge transformations closing into the algebra
 \bea
{}\big[ \delta_{\Lambda_1},\delta_{\Lambda_2}\big] \ = \ \delta_{\Lambda_{12}}
\;,
\qquad
\Lambda_{12}^m ~\equiv~ \Lambda_{2}^k \partial_k \Lambda_{1}^m-\Lambda_{1}^k \partial_k \Lambda_{2}^m
\;,
\eea
of five-dimensional diffeomorphisms, embedded into the E-bracket~(\ref{Ebracket}).
Comparing the remaining fields of (\ref{AB_B}) to the field content of the Kaluza-Klein reduction 
of IIB supergravity suggests to relate the fields $\{ A_{\mu\,m\,\alpha}, B_{\mu\nu}{}^\alpha\}$ in (\ref{AB_B}) 
to the different components of the doublet of ten-dimensional two-forms, and the fields $A_{\mu\,kmn}, B_{\mu\nu\,mn}$
with the components of the (self-dual) IIB four-form. The remaining fields $A_{\mu\,\alpha}, B_{\mu\nu}{}^{m\,\alpha}$
descend from components of the doublet of dual six-forms.
Again, the two-form tensors $B_{\mu\nu\,m}$ that do not figure in the ${\rm E}_{6(6)}$ covariant Lagrangian 
represent the degrees of freedom on-shell dual to the Kaluza-Klein vector fields, i.e.\ descending from the ten-dimensional
dual graviton. 
We recall that in the EFT formulation, all vector fields appear with a Yang-Mills kinetic term 
whereas the two-forms couple via a topological term and are on-shell dual to the vector fields. 
In order to match the structure of IIB supergravity, we will thus have to trade the Yang-Mills vector fields $A_{\mu\,\alpha}$
for a propagating two-form $B_{\mu\nu}{}^\alpha$.

The details of this identification can be worked out 
by evaluating the general formulas of the ${\rm E}_{6(6)}$-covariant formulation
with (\ref{dbreakB}) and imposing the explicit solution of the section condition (\ref{explicit_sectionB}) on all fields.
Without repeating the details of the derivation which goes in close analogy
to the analysis of section~\ref{subsec:gl6}, we summarize the covariant field strengths
for the different vector fields from (\ref{AB_B})
\bea
{\cal F}_{\mu\nu}{}^m &=&  2\partial_{[\mu} A_{\nu]}{}^m 
- A_{\mu}{}^n\partial_n A_{\nu}{}^m
+A_{\nu}{}^n\partial_n A_{\mu}{}^m
\;,
\nonumber\\
{\cal F}_{\mu\nu\,m\alpha} &=&
2D_{[\mu} A_{\nu]\,m\alpha}
+\epsilon_{\alpha\beta} \,  \partial_m \tilde{B}_{\mu\nu}{}^\beta
\;,
\nonumber\\
{\cal F}_{\mu\nu\,kmn} &=&
2D_{[\mu} A_{\nu]\,kmn}
-3\,\sqrt{2} \, \epsilon^{\alpha\beta}  A_{[\mu\,[k|\alpha|} \partial_m A_{\nu]}{}_{n]\beta} 
+3\,\partial_{[k} \tilde{B}_{|\mu\nu|\,mn]} 
\;,
\nonumber\\
{\cal F}_{\mu\nu\,\alpha} &=&
2 D_{[\mu} A_{\nu]\,\alpha}
-2(\partial_k A_{[\mu}{}^k) \, A_{\nu]}{}_\alpha 
-\sqrt{2}\,A_{[\mu}{}^{mn} \partial_n A_{\nu]\,m\alpha}
-\sqrt{2}\,A_{[\mu|m\alpha|}{} \partial_n A_{\nu]}{}^{mn}
\nonumber\\
&&{}
-\epsilon_{\alpha\beta} \,\partial_k \tilde{B}_{\mu\nu}{}^{k\beta}
\;,
\eea
with the modified two-forms
\bea
\tilde{B}_{\mu\nu}{}^\alpha &\equiv&
\sqrt{10}\, B_{\mu\nu}{}^\alpha-\epsilon^{\alpha\beta}\,A_{[\mu}{}^n A_{\nu]}{}_{n\beta}
\;,
\quad\nonumber\\
\tilde B_{\mu\nu\,mn} &\equiv& \sqrt{10}\, B_{\mu\nu\,mn} + A_{[\mu}{}^{k}  A_{\nu]\,kmn}
\;,\nonumber\\
\tilde{B}_{\mu\nu}{}^{k\alpha} &\equiv& 
\sqrt{10}\, B_{\mu\nu}{}^{k\alpha}+\epsilon^{\alpha\beta}\,A_{[\mu}{}^k  A_{\nu]}{}_\beta
\;.
\eea
All covariant derivatives $D_\mu \equiv \partial_\mu - {\cal L}_{A_\mu}$
correspond to the action of five-dimensional internal diffeomorphisms.
The corresponding vector gauge transformations are given by
\bea
\delta A_\mu{}^m &=& D_\mu\Lambda^m\;,
\nonumber\\
\delta A_{\mu\,m\alpha} &=&
D_\mu \Lambda_{m\alpha}
+{\cal L}_\Lambda A_{\mu\,m\alpha}
-  \epsilon_{\alpha\beta} \, \partial_m 
\tilde\Xi_\mu{}^\beta 
\;,
\nonumber\\
 \delta A_{\mu\,kmn} &=&
D_\mu \Lambda_{kmn}
+{\cal L}_\Lambda A_{\mu\,kmn}
-3\sqrt{2} \, \epsilon^{\alpha\beta} \, \partial_{[k} A_{|\mu|}{}_{m|\alpha|} \Lambda_{n]\beta}
-3\,  \partial_{[k}\tilde\Xi_{|\mu|\,mn]} 
\;,
\eea
with
\bea
\tilde{\Xi}_\mu{}^\alpha &\equiv&
\sqrt{10}\, \Xi_\mu{}^\alpha -\epsilon^{\alpha\beta}\,\Lambda^n A_{\mu\,n\beta}
\;,\qquad
\tilde{\Xi}_{\mu\,mn} ~\equiv~ \sqrt{10}\,\Xi_{\mu\,mn} + \Lambda^{k} A_{\mu\,kmn}
\;. 
\eea
As for the vector fields $A_{\mu\,\alpha}$, it will be sufficient to observe that
its gauge variation is given by
\bea
\delta A_{\mu\,\alpha} &=& \dots + \epsilon_{\alpha\beta}\,\partial_k \tilde{\Xi}_{\mu}{}^{k\beta}
\;,
\label{dAa}
\eea
implying that it can entirely be gauged away by the tensor gauge symmetry 
associated with the two-forms ${B}_{\mu\nu}{}^{k\beta}$.
Consequently, it will automatically disappear from the Lagrangian upon integrating out 
$\partial_k {B}_{\mu\nu}{}^{k\beta}$.
The remaining two-form field strengths in turn come with gauge transformations
\bea
\delta \tilde{B}_{\mu\nu}{}^\alpha &=&
2 D_{[\mu} \tilde \Xi_{\nu]}{}^\alpha 
+{\cal L}_\Lambda \tilde{B}_{\mu\nu}{}^\alpha
-\epsilon^{\alpha\beta}\, \Lambda_{n\beta} {F}_{\mu\nu}{}^n
\;,\nonumber\\
\delta \tilde{B}_{\mu\nu\,mn} &=& 
2  D_\mu \left(\tilde{\Xi}_{\nu\,mn} +\frac1{\sqrt{2}}\,\epsilon^{\alpha\beta} \, A_{\nu\,m\alpha} \,\Lambda_{n\beta}\right)
+\sqrt{2}  \,\partial_m A_\mu{}_{n\alpha} \, \tilde{\Xi}_\nu{}^\alpha 
\nonumber\\
&&{}
+{\cal L}_\Lambda \tilde{B}_{\mu\nu\,mn} 
-\frac1{\sqrt{2}} \,  \Lambda_{[m|\alpha|} \,\partial_{n]} \tilde{B}_{\mu\nu}{}^\alpha
+\Lambda_{mnk} \,{F}_{\mu\nu}{}^k
\;,
\label{deltaBB}
\eea
and field strengths
\bea
\tilde{\cal H}_{\mu\nu\rho}{}^\alpha &\equiv&
\sqrt{10}\,{\cal H}_{\mu\nu\rho}{}^\alpha 
~=~
3\,D_{[\mu} \tilde{B}_{\nu\rho]}{}^\alpha 
+3\,\epsilon^{\alpha\beta}\,F_{[\mu\nu}{}^n A_{\rho]\,n\beta}
\;,
\nonumber\\
\tilde{\cal H}_{\mu\nu\rho\,mn} &\equiv&
\sqrt{10}\,{\cal H}_{\mu\nu\rho\,mn}
\nonumber\\
&=&
3\, D_\mu \tilde B_{\nu\rho\,mn}
-3\,F_{\mu\nu}{}^k A_{\rho\,kmn}
-3\sqrt{2}\, \epsilon^{\alpha\beta}\, A_\mu{}_{m\alpha} D_\nu A_\rho{}_{n\beta}
+3\sqrt{2}\, A_\mu{}_{m\alpha} \partial_n \tilde{B}_{\nu\rho}{}^\alpha
\;,\nonumber\\
\label{HB}
\eea
up to terms that are projected out from the Lagrangian under $y$-derivatives.
The expressions on the r.h.s.\ in (\ref{deltaBB}) and (\ref{HB}) are understood to
be projected onto the corresponding antisymmetrizations in their parameters, i.e. $[mn]$,
$[\mu\nu]$, $[\mu\nu\rho]$, etc.

Finally, we note that the topological term (\ref{CSlike}) in this parametrization is given by
\bea
  {\cal L}_{\rm top} &=&
\frac1{8}\,\varepsilon^{\mu\nu\rho\sigma\tau} \epsilon^{klmnp}\,\Big(
\frac{\sqrt{2}}{6}\,
\epsilon^{\alpha\beta} \,
{\cal F}_{\mu\nu\,m\alpha} {\cal F}_{\rho\sigma\,n\beta}\,A_{\tau\,pkl}
+\frac16\,{\cal F}_{\mu\nu}{}_{mnq} {F}_{\rho\sigma}{}^{q}\, A_\tau{}_{klp}
\nonumber\\
&&{}
\qquad
-\frac{\sqrt{2}}{2}\, \epsilon^{\alpha\beta}\, A_{\mu\,m\alpha}\partial_n A_{\nu\,p\beta}  {F}_{\rho\sigma}{}^{q}\,   A_\tau{}_{klq}
+\frac12\,
 \partial_p \tilde{B}_{\mu\nu\,mn}   {F}_{\rho\sigma}{}^{q}\,    A_\tau{}_{klq}
\nonumber\\
&&{}
\qquad
+\sqrt{2}\,  \epsilon^{\alpha\beta}\, 
A_\mu{}_{m\alpha} D_\nu A_\rho{}_{n\beta} \,\partial_p \tilde{B}_{\sigma\tau\,kl}
-\sqrt{2}\, 
A_\mu{}_{m\alpha} \partial_n \tilde{B}_{\nu\rho}{}^\alpha\,\partial_p \tilde{B}_{\sigma\tau\,kl}
\nonumber\\
&&{}
\qquad
 +\frac23\, \epsilon^{\alpha\beta}\, 
A_{\mu\,m\alpha} \partial_n A_{\nu\,k\beta} A_{\rho\,l\gamma} \partial_p \tilde{B}_{\sigma\tau}{}^\gamma
-
 \epsilon^{\alpha\beta}\, \epsilon^{\gamma\delta}
A_{\mu\,m\alpha} \partial_n A_{\nu\,k\beta} A_{\rho\,l\gamma} D_\sigma A_{\tau\,p\delta}
\nonumber\\
&&{}
\qquad
 +\frac{\sqrt{2}}9\,
 \,\partial_m \tilde{\cal H}_{\mu\nu\rho}{}^\alpha
\,A_{\sigma\, n\alpha}  A_{\tau}{}_{klp}
-  D_\mu \tilde B_{\nu\rho\,mn} \partial_p \tilde{B}_{\sigma\tau\,kl}
-\frac{2}3\,
\epsilon_{\alpha\beta} \,\tilde{\cal H}_{\mu\nu\rho}{}^\beta \partial_k \tilde{B}_{\sigma\tau}{}^{k\alpha}
\nonumber\\
&&{}
\qquad
+{\cal O}(A_{\mu\,\alpha})
\Big)
\;.
\label{LtopexpB}
\eea

Let us now move to the scalar field content of the theory. In the EFT formulation,
they parametrize the symmetric matrix ${\cal M}_{MN}$\,. To relate to IIB supergravity,
we need to choose a parametrization of this matrix in accordance with the decomposition (\ref{split78B}).
In standard fashion, we build the matrix as ${\cal M}={\cal V}{\cal V}^T$ from a `vielbein' ${\cal V}$
in triangular gauge
\bea
{\cal V}^T &\equiv& {\rm exp} \left[\Phi\, t_{(0)}\right]\,{\cal V}_2\,{\cal V}_5\,
{\rm exp}\left[b_{mn}{}^\alpha\,t_{(+3)}{}_{\alpha}^{mn}\right]\,{\rm exp}\left[\epsilon^{klmnp}\,c_{klmn} \, t_{(+6)\,p}\right]
\;.
\label{V27B}
\eea
Here, $t_{(0)}$ is the E$_{6(6)}$ generator associated to the GL(1) grading of (\ref{split78B}), ${\cal V}_2$, ${\cal V}_5$ 
denotes matrices in the SL(2) and SL(5) subgroup, respectively,
parametrized by vielbeins $\nu_2$, $\nu_5$ in analogy to (\ref{sl6}).
The $t_{(+n)}$ refer to the E$_{6(6)}$ generators of positive grading in (\ref{split78B}), with
non-trivial commutator
\bea
{}
\left[t_{(+3)}{}_\alpha^{kl}, t_{(+3)}{}_\beta^{mn}\right] &=& \epsilon_{\alpha\beta}\,\epsilon^{klmnp}\,t_{(+6)\,p}
\;.
\eea
All generators are evaluated in the
fundamental ${\bf 27}$ representation (\ref{split27B}), such that the symmetric matrix ${\cal M}_{MN}$ takes the block form
\bea
{\cal M}_{KM} &=& \left(
\begin{array}{cccc}
{\cal M}_{km}&{\cal M}_k{}^{m\beta}&{\cal M}_{k,mn}&{\cal M}_{k}{}^{\beta}\\
{\cal M}^{k\alpha}{}_{m}&{\cal M}^{k\alpha,}{}^{m\beta}&{\cal M}^{k\alpha}{}_{mn}&{\cal M}^{k\alpha,}{}^{\beta}\\
{\cal M}_{kl,m}&{\cal M}_{kl}{}^{m\beta}&{\cal M}_{kl,mn}&{\cal M}_{kl}{}^{\beta}\\
{\cal M}^\alpha{}_{m}&{\cal M}^{\alpha,m\beta}&{\cal M}^\alpha{}_{mn}&{\cal M}^\alpha{}^{\beta}
\end{array}
\right)
\;.
\label{M2B}
\eea
Explicit evaluation of (\ref{V27B}) determines the various blocks in (\ref{M2B}). For instance, its last line is given
by
\bea
{\cal M}^{\alpha\beta} &=& e^{5\Phi/3}\,m^{\alpha\beta}\;,\qquad
{\cal M}^{\alpha}{}_{mn} = \sqrt{2} \,e^{5\Phi/3}\,m^{\alpha\beta} \varepsilon_{\beta\gamma} \,b_{mn}{}^\gamma
\;,
\nonumber\\
{\cal M}^{\alpha,m\beta} &=& 
\frac12\,e^{5\Phi/3}\,m^{\alpha\gamma}\varepsilon_{\gamma\delta}\,\varepsilon^{mklpq}\,b_{kl}{}^\beta b_{pq}{}^\delta
-\frac1{24}\,e^{5\Phi/3}\,m^{\alpha\beta}\,\varepsilon^{mklpq}\,c_{klpq}
\;,
\nonumber\\
{\cal M}^\alpha{}_m &=& 
\frac23\,e^{5\Phi/3}\, m_{\beta\gamma}  \,\varepsilon^{kpqrs}
 \left(
b_{mk}{}^{[\alpha} b_{pq}{}^{\beta]} b_{rs}{}^{\gamma}
+\frac1{8}\,\varepsilon^{\alpha\beta}\,b_{mk}{}^\gamma\,c_{pqrs} \right)
\;,
\eea
with the symmetric matrix $m^{\alpha\beta}=(\nu_2)^\alpha{}_u (\nu_2)^{\beta\,u}$
build from the ${\rm SL}(2)$ vielbein from (\ref{V27B}).
Later, after integrating out some of the fields, we will need the components of (c.f.~the discussion 
in the previous section) 
\bea
\tilde{\cal M}_{MN} &\equiv& {\cal M}_{MN}- {\cal M}_{M}{}^{\alpha} ({\cal M}^{\alpha\beta})^{-1} {\cal M}_{N}{}^\beta
\;,
\eea
for which we find
\bea
\tilde{\cal M}_{mn,kl} &=& e^{2\Phi/3}\,m_{m[k}m_{l]n}
\;,\nonumber\\
\tilde{\cal M}_{mn}{}^{k\alpha} &=& \frac1{\sqrt{2}}\,e^{2\Phi/3}\,
\varepsilon_{mnpqr} m^{kp} m^{qu}m^{rv} b_{uv}{}^\alpha
\;,\nonumber\\
\tilde{\cal M}_{mn,k} &=&
-\frac1{6\sqrt{2}}\,e^{2\Phi/3}\,\varepsilon^{uvpqr}\,m_{mu}m_{nv} \left( c_{kpqr}-6\varepsilon_{\alpha\beta}\, b_{kp}{}^\alpha b_{qr}{}^\beta\right)\;, 
\nonumber\\
\tilde{\cal M}^{m\alpha,n\beta} &=&
e^{-\Phi/3}\,m^{mn}m^{\alpha\beta}
+2\, e^{2\Phi/3}\,\,m^{kp}
\left(
m^{mn} m^{lq} -2\, m^{ml} m^{nq} 
\right) 
b_{kl}{}^\alpha b_{pq}{}^\beta 
\;,
\label{tildeMB}
\eea
etc., with $m_{mn}=(\nu_5)_m{}^a (\nu_5)_n{}^a$.
From the inverse matrix ${\cal M}^{MN}$ we will in particular
need the components
\bea
{\cal M}^{mn} = e^{4\Phi/3}\,m^{mn} \;.
\eea
With (\ref{dbreakB}) we find for the covariant derivatives of the matrix parameters from (\ref{M2B}) 
\bea
{\cal D}_\mu \Phi &=& D_\mu\phi + \frac45\,\partial_k A_\mu{}^k \;,\nonumber\\
{\cal D}_\mu m_{mn} &=& D_\mu m_{mn}+ \frac{2}{5}\,\partial_k A_\mu{}^k\, m_{mn}\;,\nonumber\\
{\cal D}_\mu b_{mn}{}^\alpha &=& D_\mu b_{mn}{}^\alpha -\epsilon^{\alpha\beta} \partial_{[m} A_{n]\beta\,\mu}\;,\nonumber\\
{\cal D}_\mu c_{klmn} &=& D_\mu c_{klmn} 
+4 \sqrt{2}\, \partial_{[k} A_{lmn]\mu}
+12\,b_{[kl}{}^\alpha\,\partial_{m} A_{n]\,\alpha}
\;,
\eea
which will build the kinetic term of the Lagrangian.

As discussed above and similar to the analysis for the embedding of $D=11$ supergravity, 
the precise map with type IIB supergravity requires some dualizations of the fields. To this end, we observe
that in the Lagrangian the two-form tensors $\tilde{B}_{\mu\nu}{}^{k\beta}$
appear only under a divergence, i.e. contracted with $\partial_k$, c.f.\ (\ref{onlyBB}), 
and with algebraic field equations
\bea
\epsilon_{\alpha\beta}\,\partial_k\tilde{B}_{\mu\nu}{}^{k\beta}
&=& 
({\cal M}^{\alpha\beta})^{-1}\,  {\cal M}^\beta{}_{M}\,{\cal F}^{\mu\nu\,M}
-\frac16\,\varepsilon^{\mu\nu\rho\sigma\tau} \,({\cal M}^{\alpha\beta})^{-1}\,
\,\tilde{\cal H}_{\rho\sigma\tau}{}^\beta
\;.
\label{EOMauxB}
\eea
By means of these equations, the fields $\tilde{B}_{\mu\nu}{}^{k\beta}$ can be eliminated
from the Lagrangian. The gauge symmetry (\ref{dAa}) shows that 
in the process, the vector fields $A_{\mu\,\alpha}$ also disappear.
We infer from (\ref{EOMauxB}) that the kinetic term for the remaining vector fields
changes into the form (\ref{FFM}) with $\tilde{\cal M}_{MN}$ from (\ref{tildeMB}).
Moreover, the two-forms $\tilde{B}_{\mu\nu}{}^\alpha$ are promoted into propagating fields
with kinetic term 
\bea
-e^{-5/3\,\Phi}\,m_{\alpha\beta}\,
\tilde{\cal H}_{\mu\nu\rho}{}^\alpha\,\tilde{\cal H}^{\mu\nu\rho\,\beta}\;, 
\eea
and we note that  the cross terms from (\ref{EOMauxB}) give rise to additional contributions
to the topological term in (\ref{LtopexpB}).

Let us conclude by commenting on some of the properties of the resulting Lagrangian. 
At first sight, it may appear surprising 
that we can obtain a ten-dimensional Lagrangian describing the field equations of the full IIB supergravity,
whereas it is known that the presence of a self-dual four-form poses a severe obstruction to the
construction of a Lorentz-covariant Lagrangian. It is the latter property which justifies the existence
of our Lagrangian: what we have constructed is a ten-dimensional Lagrangian in
which however only an ${\rm SO}(1,4)\times {\rm SO}(5)$ subgroup of the ${\rm SO}(1,9)$ 
Lorentz symmetry is realized.
In this respect, its existence is no more surprising than the corresponding 
constructions of \cite{Henneaux:1988gg,Schwarz:1993vs} in which Lorentz symmetry appears broken to ${\rm SO}(9)$
but recovered on the level of the equations of motion.
The self-dual four form is described by propagating degrees of freedom $c_{klmn}$ and $A_{\mu\,kmn}$,
yet the final Lagrangian also carries some of the dual degrees of freedom in the two-forms $\tilde{B}_{\mu\nu\,mn}$.
These do not appear with a kinetic term but couple by a topological term (\ref{LtopexpB}) such that their field equations 
precisely give rise to the first-order duality equations that relate their field strength to the field strength of the $A_{\mu\,kmn}$,
thereby reproducing part of the ten-dimensional self-duality equations.

\section{Summary and Outlook}\label{sec6}

In this paper, we have presented the detailed construction of the E$_{6(6)}$ exceptional field theory 
recently announced in  \cite{Hohm:2013pua}.
This theory is formally defined in $5+27$ dimensions, with $27$ coordinates 
transforming in the fundamental representation of E$_{6(6)}$, 
subject to a covariant section constraint. 
This constraint, which implies that only a subset of the coordinates is physical, 
is the M-theory analogue of the strong constraint in double field theory, 
which in turn is a stronger version of the level-matching constraint in string theory. 
The constraint allows for different solutions, two of which we have discussed in detail.
The first reduces the 27 coordinates to six, thereby breaking E$_{6(6)}$ 
to GL$(6)$, leading to a $(5+6)$-dimensional formulation of the full (untruncated) 
11-dimensional supergravity. The second solution of the constraint
reduces the 27  coordinates to five, breaking  E$_{6(6)}$ to GL$(5)\times {\rm SL}(2)$, 
leading to a $(5+5)$-dimensional formulation of type IIB supergravity with manifest 
SL$(2)$ S-duality. In this sense, the exceptional field theory (\ref{finalaction}) 
unifies M-theory and type IIB in that both are obtained on different `slices' of the generalized 
spacetime. This generalizes type II double field theory,
in which type IIA and type IIB arise on different slices of the doubled spacetime  
 \cite{Hohm:2011zr,Hohm:2011dv}. 
 As a by-product, we have obtained an off-shell action for 
type IIB supergravity, at the cost of sacrificing 10-dimensional Lorentz invariance. 
 
In this paper we have restricted ourselves to the purely bosonic theory, 
but we are confident that the extension to include fermions and the construction of 
a supersymmetric action is straightforward along the lines of the 
supersymmetric $D=5$ gauged supergravity~\cite{deWit:2004nw}. The fermions will be E$_{6(6)}$ singlets 
transforming under the local generalized Lorentz group of the corresponding coset, i.e., in this case 
${\rm H}={\rm USp}(8)$, which will require a notion of generalized Lorentz connection. 
This should also clarify the relation of our construction to that of 
de~Wit and Nicolai \cite{deWit:1986mz,Nicolai:1986jk}, who cast the eleven-dimensional 
supersymmetry transformations into an H-covariant from.  
At first sight it may appear surprising that such a supersymmetric covariant
construction is feasible at all. First we know that conventional supersymmetric theories are restricted to 
dimensions $D\leq 11$. Second, the resulting theory would encode both type IIA and
type IIB, despite the crucial difference of their fermion chiralities.   
The first obstacle is circumvented by virtue of the section constraint, which implies
that the additional coordinates are not physical in the same sense as the usual spacetime 
coordinates. In fact, 
in double field theory supersymmetric extensions are possible and beautifully 
simplify the usually rather involved ${\cal N}=1$ supergravities 
in $D=10$, with the supersymmetry transformations closing 
into the generalized diffeomorphisms \cite{Hohm:2011nu}. 
The second obstacle is circumvented since the EFT formulation does not 
preserve the $D=10$ Lorentz invariance, so that the EFT fermions can 
consistently encode the fermions of type IIA and type IIB. This possibility is then 
no more surprising than the observation 
that both type IIA and type IIB give rise to the same supersymmetric 
theory in $D=5$ upon dimensional reduction. 

A novel feature of the supersymmetric EFT is that usually it is supersymmetry 
which fixes the detailed form of even some of the purely bosonic couplings, most notably
the presence and shape of the scalar potential. 
In contrast,  in (\ref{finalaction}) all bosonic couplings are already uniquely determined 
by the bosonic gauge and duality symmetries. This points to a deep 
connection between the duality covariant geometries of double and exceptional 
field theories on the one hand and supersymmetry on the other, as for instance 
illustrated by the striking economy of the supersymmetric double field theory.  
We leave a discussion of these matters and the detailed 
construction of supersymmetric EFT to a separate publication.

There are many open questions and possible generalizations.  
An obvious question is about the physical significance of the 27 coordinates. 
Beyond the six coordinates of $D=11$ supergravity, are they a purely formal device, 
or do they have a deeper role to play? A comparison with string theory is illuminating. 
Here the doubled coordinates, at least on toroidal backgrounds, 
are undoubtedly physical and real, as made explicit by closed string field theory, 
subject only to the weaker level-matching 
constraint that allows for solutions depending locally both on coordinates and 
their duals \cite{Hull:2009mi}. Thus, although the currently understood double field theory is  
subject to the strong constraint, the latter constraint is well motivated from 
string theory, implementing the level-matching constraint in stronger form. 
The section constraint of exceptional field theory has been 
postulated by analogy to the strong constraint, but since there is no analogue 
to string field theory in M-theory, it cannot be `derived' in a similar fashion. 
However, we may consider a partial solution of the E$_{6(6)}$ covariant 
section constraint that breaks the symmetry to the T-duality group of string theory. 
Specifically, we can embed the SO$(5,5)$ T-duality group that is appropriate 
for a $(5+5)$-dimensional decomposition of type II string theory into 
E$_{6(6)}$.  The fundamental representation then decomposes as 
 \be
  {\rm SO}(5,5)\, \subset \, {\rm E}_{6(6)}\; : \;\qquad
  {\bf 27}\;\;\rightarrow\;\; 10\,\oplus\, 16\,\oplus\, 1\;,
 \ee
where $10$ and $16$ are the vector and spinor representation of  SO$(5,5)$, 
respectively. 
Thus, under this decomposition we obtain the NS-NS fields transforming 
as a vector (or rather, in the generalized metric formulation, as a 2-tensor) 
but also the RR fields transforming as a spinor. The resulting theory will be 
a Kaluza-Klein-type decomposition of the original type II 
double field theory of \cite{Hohm:2011zr,Hohm:2011dv}, 
in the sense of  \cite{Hohm:2013nja}. 
The decomposition of the $d$-symbol  is then such that the section constraint 
implies independence of all fields on the $1+16$ variables, and further restricts 
the field dependence on the remaining 10 variables in the fundamental vector representation
of SO$(5,5)$ as 
 \be
  d^{MNK}\partial_M \partial_N \ = \ 0\qquad \Longrightarrow \qquad 
  \eta^{\check{M}\check{N}}\partial_{\check{M}}  \partial_{\check{N}} \ = \ 0\;, 
 \ee
with the SO$(5,5)$ vector indices denoted by $\check{M}, \check{N}$, 
see e.g.~eqs.~(3.27), (3.29) in \cite{LeDiffon:2008sh}. 
Thus, the section constraint reduces precisely 
to the strong constraint in double field theory. Since 
in string theory the strong constraint is relaxed so that 
the doubled coordinates are physical and real, 
U-duality covariance strongly suggests the same for the 27 coordinates 
of the E$_{6(6)}$ EFT, and similarly for the extended coordinates of 
the higher EFTs w.r.t.~E$_{7(7)}$
and  E$_{8(8)}$.

A related question is about the most 
general solutions of the section constraint (\ref{sectioncondition}), in particular 
whether there are solutions beyond the known $D=10$ and $D=11$ supergravity. 
While we do not have a proof that there are no solutions with $D>11$, this 
appears unlikely. However, it is certainly important to classify all solutions, in particular 
in order to see whether or not there may be any `non-geometric' solutions, for any $D>5$. 
For instance, one may imagine that the gauged diffeomorphisms (\ref{skewD}) 
and the generalized internal diffeomorphisms 
do not organize into conventional diffeomorphisms of a $D$-dimensional theory, 
thereby escaping the conventional classifications. We leave this for future work. 
Even if such more general solutions of the section constraint 
may be excluded, it is still likely that there are non-geometric 
solutions of the EFT field equations that locally depend on the subset of coordinates corresponding to 
one solution of the constraint, but that patch together inequivalent 
solutions in a globally non-trivial manner, as happens in double field theory \cite{Hohm:2013bwa}. 
Perhaps the most intriguing, but also most involved question
is about a genuine relaxation of the section constraint, which would truly transcend 
the framework of supergravity.

Another fascinating prospect is to generalize the presently known EFT to  
include higher-derivative M-theory corrections along the lines of the recent results on 
double field theory \cite{Hohm:2013jaa}. This would entail a deformation
of the E$_{n(n)}$ generalized Lie derivatives and other structures. 
If possible, this would give a scheme to compute the $\alpha'$ corrections of 
type II string theories and the higher-derivative M-theory corrections in a unified manner. 

Let us finally note that the details for the remaining finite-dimensional groups E$_{7(7)}$ and E$_{8(8)}$
will be presented in a separate publication. The general construction proceeds along the 
same lines as the one presented here, with a $4+56$ and $3+248$ dimensional formulation, respectively. 
One novel feature of these cases is that additional field components need to be introduced 
which, from a 11-dimensional perspective, play the role of the dual graviton, a field 
for which a local field theory formulation is usually considered impossible on the 
grounds of the no-go theorems in \cite{Bekaert:2002uh,Bekaert:2004dz}. 
We have shown in \cite{Hohm:2013jma}  how to handle this problem in the covariant approach via introducing 
constrained compensator fields, extending the approach of \cite{Boulanger:2008nd}.  
In three dimensions, the components of the higher-dimensional dual graviton figure among the
coordinates of the scalar target space. The Lagrangian of~\cite{Hohm:2013jma} carries these fields
in a duality covariant way and yields the first-order duality equations which relate them to the corresponding
components of the higher-dimensional metric, all while retaining full higher-dimensional coordinate dependence.
The construction hinges on the introduction of the covariantly constrained compensator fields,
which can be viewed as extra gauge potentials, however, satisfying the analogue of the section constraint, but for 
the field components, so that effectively only a subset of fields survives, c.f.\ equation (2.34) of~\cite{Hohm:2013jma}. 
In fact, these additional gauge fields appear among the $(D-2)$-forms in the covariant formulation in all dimensions 
and neatly fit in the 
structure of the tensor hierarchy. For instance, although such fields are not visible in 
the E$_{6(6)}$ action (\ref{finalaction}) presented in this paper, they would show up 
when extending the tensor hierarchy on-shell to the full set of two-forms $B_{\mu\nu M}$
in the form of compensating gauge fields $C_{\mu\nu\rho\,M}$ among the three-forms.
For our action, they are irrelevant thanks to the extra gauge redundancy corresponding 
to ${\cal O}_{\mu\nu M}$, see (\ref{deltaAB}), whose 3-form gauge potential 
does not enter the action. 
For the $D=4$ decomposition, however, the compensating gauge field 
is a two-form and thus enters explicitly the gauge-covariant field strength of 
the gauge vectors $A_{\mu}{}^{M}$. Finally, in the $D=3$ decomposition the 
compensating gauge fields are among the vectors entering the covariant derivatives, 
as discussed for the Ehlers SL$(2,\mathbb{R})$ subgroup in \cite{Hohm:2013jma}. 
This mechanism also circumvents the seeming 
problem of non-closure of the E$_{8(8)}$ generalized Lie derivatives~\cite{Berman:2012vc}.  
Summarizing, we have arrived at a satisfying homogeneous picture of 
the  exceptional field theory formulations for E$_{n(n)}$, $n=6,7,8$. 
It is a fascinating question whether and if so how these constructions 
can be extended to even larger groups, possibly starting with the infinite-dimensional 
E$_{9(9)}$ and lifting the action functional of \cite{Samtleben:2007an}, but here we can only speculate.

\section*{Acknowledgments}
The work of O.H. is supported by the 
U.S. Department of Energy (DoE) under the cooperative 
research agreement DE-FG02-05ER41360 and a DFG Heisenberg fellowship. 
We would like to thank Hong Liu, Hermann Nicolai, Washington Taylor, and Barton Zwiebach 
for useful comments and discussions.

\appendix

\section{Truncations of Exceptional Field Theory}
\setcounter{equation}{0}
In this appendix we discuss possible truncations of the EFT action (\ref{finalaction}) 
in order to relate it to results in the literature on duality-covariant formulations 
of subsectors of 11-dimensional 
supergravity~\cite{Hillmann:2009pp,Berman:2010is,Berman:2011jh,Coimbra:2011ky,Berman:2011pe,Park:2013gaj}. 
In particular, 
in these formulations all off-diagonal field components and the 
external components of the 3-form are set to zero, 
and it is assumed that all fields depend only on internal 
coordinates. In terms of the fields and coordinates of the E$_{6(6)}$ EFT presented here 
this truncation therefore assumes 
 \be
  A_{\mu}{}^{M} \ = \ 0\;, \qquad B_{\mu\nu\, M} \ = \ 0\;, \qquad \partial_{\mu} \ = \ 0\;. 
  \label{truncder}
 \ee
For the action (\ref{finalaction}) this truncation implies 
\bea
\widehat{R} &\rightarrow& 0\;,\nonumber\\
g^{\mu\nu}{\cal D}_{\mu}{\cal M}^{MN}\,{\cal D}_{\nu}{\cal M}_{MN}&\rightarrow& 0\;,\nonumber\\
{\cal M}_{MN}{\cal F}^{\mu\nu M}{\cal F}_{\mu\nu}{}^N&\rightarrow& 0\;,\nonumber\\
{\cal L}_{\rm top}&\rightarrow& 0\;,
\eea
such that the only surviving term is a truncation of the potential term $V({\cal M}_{MN},g_{\mu\nu})$\,.
The available formulations in the literature differ in the treatment 
of the remaining fields, i.e., the external metric $g_{\mu\nu}$ and the generalized
metric ${\cal M}_{MN}$ encoding the internal field components. 
The original work by Hillmann on E$_{7(7)}$ covariance~\cite{Hillmann:2009pp} sets the external
metric to the flat Minkowski metric, 
 \be\label{truncation}
  g_{\mu\nu} \ = \  \eta_{\mu\nu}\qquad\Rightarrow  \qquad  \sqrt{-g}\ = \ e \ = \ 1\,,
 \ee
so that the volume factor becomes unity.    
In the analogous truncation of the E$_{6(6)}$ EFT,  the action (\ref{finalaction}) 
reduces to the `potential term' only, 
 \be\label{Vaction}
  S_{{\rm EFT}} \ \longrightarrow \ -\int d^{27}Y \,V({\cal M})\;,  
 \ee 
with $V({\cal M})$ obtained from (\ref{fullpotentialIntro}) by setting 
$g_{\mu\nu}=\eta_{\mu\nu}$. It is useful to investigate what are the 
residual gauge symmetries after this truncation. Of course, the 
$(4+1)$-dimensional diffeomorphisms are broken, 
but also the `internal' generalized diffeomorphisms are not completely 
preserved, for the presence of $g$-dependent terms in the potential 
was crucial for gauge invariance, as discussed in sec.~3.2. In particular, 
the volume factor $e$ with the appropriate weight is needed.   
Requiring that the condition $e=1$ be preserved under gauge transformations we obtain  
 \be\label{residualgauge}
  \delta_{\Lambda}e \ = \ \Lambda^N\partial_{N}e+\frac{5}{3}\,e\, \partial_N\Lambda^N 
  \overset{!}{ \ = \ } 0 \quad \Longrightarrow \quad 
  \partial_N\Lambda^N \ = \ 0\;. 
 \ee 
In fact, Hillmann found that his formulation matches the considered 
truncation of $D=11$ supergravity only in `uni-modular gauge' 
of the internal metric \cite{Hillmann:2009ci},  
for which the residual gauge transformations are indeed compatible with  
(\ref{residualgauge}). 

For a proper duality-covariant truncation of (\ref{finalaction}),
the volume factor of the internal metric has to be kept as a separate degree of freedom,
as already noted in \cite{Hillmann:2009ci}.
Specifically, (\ref{truncation}) is relaxed to 
 \be\label{warpgansatz}
  g_{\mu\nu} \ = \ e^{2\Delta}\, \eta_{\mu\nu}\;, 
 \ee
with a warp-factor that in accordance with (\ref{truncder})
is a function of $Y$ only and transforms as a scalar-density of weight $\lambda=\frac23$ under
$\Lambda$ gauge transformations (\ref{genLie}). For this truncation, the EFT action (\ref{finalaction}) 
again reduces to its potential term, now with extra contributions in $\Delta$
\bea
S_{{\rm EFT}} &\longrightarrow & \int d^{27}Y \,
   e^{5\Delta}\;
\Big(
\frac{1}{24}{\cal M}^{MN}\partial_M{\cal M}^{KL}\,\partial_N{\cal M}_{KL}
-\frac{1}{2} {\cal M}^{MN}\partial_M{\cal M}^{KL}\partial_L{\cal M}_{NK}
\nonumber\\
&&{}
\qquad\qquad\qquad    
-5\,\partial_M \Delta\,\partial_N{\cal M}^{MN}
-20\, {\cal M}^{MN}\partial_M \Delta\partial_N \Delta
\Big)
\;.
\label{truncationpot}
\eea
This truncated action is duality and $\Lambda^M$ gauge invariant. Note that $\Delta$
is a separate degree of freedom that transforms independently of the 42 scalars parametrizing 
the ${\rm E}_{6(6)}$ matrix ${\cal M}_{MN}$.
It may be convenient to combine ${\cal M}_{MN}$ and $\Delta$ into a single object
 \be\label{rescaledM}
  \widehat{\cal M}_{MN} \ = \ e^{\gamma\,\Delta}\,{\cal M}_{MN}\;, 
 \ee
with some factor $\gamma$, and rewrite the potential in terms of $\widehat{\cal M}$ only. 
This rescaled matrix is no longer an 
element of the duality group E$_{n(n)}$, but can rather be thought 
of as taking values in ${\rm E}_{n(n)}\times \mathbb{R}^+$, which is the 
starting point in the approach of~\cite{Coimbra:2011ky}. 
The formulations of \cite{Berman:2010is,Berman:2011pe,Berman:2011jh}
employ the object (\ref{rescaledM}) (with different choices for $\gamma$),
but identify $\Delta$ with one of the internal components of ${\cal M}_{MN}$,
which breaks the ${\rm E}_{n(n)}$ covariance of (\ref{truncationpot}) down to
the subgroup commuting with that parameter, as pointed out in~\cite{Berman:2010is,Park:2013gaj}. 
The resulting truncation for the ${\rm E}_{6(6)}$ case~\cite{Berman:2011jh} coincides with 
(\ref{truncationpot}), (\ref{rescaledM}) (choosing $\gamma=-5$).

We close by pointing out that, in principle, one may also 
separate the $\mathbb{R}^+$ factor in the full, un-truncated EFT in (\ref{finalaction}), 
by re-defining $g_{\mu\nu}=e^{2\Delta}\widehat{g}_{\mu\nu}$, with uni-modular
metric $\widehat{g}$, and then rescaling the generalized metric ${\cal M}_{MN}$
as in (\ref{rescaledM}). This has various technical disadvantages, however, as
for instance the Einstein-Hilbert and scalar-kinetic terms start mixing 
in an intricate fashion, thereby obscuring the manifest 
E$_{6(6)}$ covariance of the current formulation.



\begin{thebibliography}{10}

\bibitem{Cremmer:1979up}
E.~Cremmer and B.~Julia, { The ${SO}(8)$ supergravity},  { Nucl. Phys.} { B159}
  (1979)
141.

\bibitem{Cremmer:1978km}
E.~Cremmer, B.~Julia, and J.~Scherk, { Supergravity theory in 11 dimensions},
  { Phys. Lett.} { B76} (1978)
409--412.

\bibitem{Hull:1994ys}
C.~Hull and P.~Townsend, { {Unity of superstring dualities}},  { Nucl.Phys.} {
  B438} (1995) 109--137,
[\href{http://xxx.lanl.gov/abs/hep-th/9410167}{{\tt hep-th/9410167}}].

\bibitem{Julia:1982gx}
B.~Julia, { Kac-{M}oody symmetry of gravitation and supergravity theories},  in
  { Lectures in Applied Mathematics AMS-SIAM, Vol. 21}, p.~335, 1985.

\bibitem{deWit:1986mz}
B.~de~Wit and H.~Nicolai, { $d = 11$ supergravity with local {$SU(8)$}
  invariance},  { Nucl.Phys.} { B274} (1986)
363.

\bibitem{Nicolai:1986jk}
H.~Nicolai, { ${D} = 11$ supergravity with local ${SO}(16)$ invariance},  {
  Phys. Lett.} { B187} (1987)
316.

\bibitem{Obers:1998fb}
N.~Obers and B.~Pioline, { {U} duality and {M} theory},  { Phys.Rept.} { 318}
  (1999) 113--225,
[\href{http://xxx.lanl.gov/abs/hep-th/9809039}{{\tt hep-th/9809039}}].

\bibitem{Koepsell:2000xg}
K.~Koepsell, H.~Nicolai, and H.~Samtleben, { An exceptional geometry for {$D =
  11$} supergravity?},  { Class.Quant.Grav.} { 17} (2000) 3689--3702,
[\href{http://xxx.lanl.gov/abs/hep-th/0006034}{{\tt hep-th/0006034}}].

\bibitem{deWit:2000wu}
B.~de~Wit and H.~Nicolai, { {Hidden symmetries, central charges and all that}},
   { Class.Quant.Grav.} { 18} (2001) 3095--3112,
[\href{http://xxx.lanl.gov/abs/hep-th/0011239}{{\tt hep-th/0011239}}].

\bibitem{West:2001as}
P.~C. West, { {${{E}}_{11}$ and {M} theory}},  { Class. Quant. Grav.} { 18}
  (2001) 4443--4460,
[\href{http://xxx.lanl.gov/abs/hep-th/0104081}{{\tt hep-th/0104081}}].

\bibitem{HenryLabordere:2002dk}
P.~Henry-Labord\`ere, B.~Julia, and L.~Paulot, { Borcherds symmetries in {M}
  theory},  { JHEP} { 0204} (2002) 049,
[\href{http://xxx.lanl.gov/abs/hep-th/0203070}{{\tt hep-th/0203070}}].

\bibitem{Damour:2002cu}
T.~Damour, M.~Henneaux, and H.~Nicolai, { {${E}_{10}$} and a `small tension
  expansion' of {M} theory},  { Phys. Rev. Lett.} { 89} (2002) 221601,
[\href{http://xxx.lanl.gov/abs/hep-th/0207267}{{\tt hep-th/0207267}}].

\bibitem{Damour:2002et}
T.~Damour, M.~Henneaux, and H.~Nicolai, { {Cosmological billiards}},  {
  Class.Quant.Grav.} { 20} (2003) R145--R200,
[\href{http://xxx.lanl.gov/abs/hep-th/0212256}{{\tt hep-th/0212256}}].

\bibitem{West:2003fc}
P.~C. West, { {$E_{11}$}, {$SL(32)$} and central charges},  { Phys.Lett.} {
  B575} (2003) 333--342,
[\href{http://xxx.lanl.gov/abs/hep-th/0307098}{{\tt hep-th/0307098}}].

\bibitem{West:2004iz}
P.~C. West, { Brane dynamics, central charges and {$E_{11}$}},  { JHEP} { 0503}
  (2005) 077,
[\href{http://xxx.lanl.gov/abs/hep-th/0412336}{{\tt hep-th/0412336}}].

\bibitem{Hull:2007zu}
C.~Hull, { Generalised geometry for {M}-theory},  { JHEP} { 0707} (2007) 079,
[\href{http://xxx.lanl.gov/abs/hep-th/0701203}{{\tt hep-th/0701203}}].

\bibitem{DallAgata:2007sr}
G.~Dall'Agata, N.~Prezas, H.~Samtleben, and M.~Trigiante, { Gauged
  supergravities from twisted doubled tori and non-geometric string
  backgrounds},  { Nucl. Phys.} { B799} (2008) 80--109,
[\href{http://xxx.lanl.gov/abs/arXiv:0712.1026 [hep-th]}{{\tt arXiv:0712.1026
  [hep-th]}}].

\bibitem{Pacheco:2008ps}
P.~P. Pacheco and D.~Waldram, { {M}-theory, exceptional generalised geometry
  and superpotentials},  { JHEP} { 0809} (2008) 123,
[\href{http://xxx.lanl.gov/abs/0804.1362}{{\tt 0804.1362}}].

\bibitem{Hillmann:2009ci}
C.~Hillmann, { Generalized ${E}_{7(7)}$ coset dynamics and {$D=11$}
  supergravity},  { JHEP} { 0903} (2009) 135,
[\href{http://xxx.lanl.gov/abs/0901.1581}{{\tt 0901.1581}}].

\bibitem{Hillmann:2009pp}
C.~Hillmann, { {$E_{7(7)}$ and d=11 supergravity}},
  \href{http://xxx.lanl.gov/abs/0902.1509}{{\tt 0902.1509}}.
PhD thesis, Humboldt-Universit\"at zu Berlin, 2008,

\bibitem{Gunaydin:2009zz}
M.~G\"unaydin and O.~Pavlyk, { Quasiconformal realizations of ${E}_{6(6)},
  {E}_{7(7)}, {E}_{8(8)}$ and ${SO}(n+3,m+3)$ , ${N}\ge4$ supergravity and
  spherical vectors},  { Adv. Theor. Math. Phys.} { 13} (2009)
[\href{http://xxx.lanl.gov/abs/0904.0784}{{\tt 0904.0784}}].

\bibitem{Aldazabal:2010ef}
G.~Aldazabal, E.~Andres, P.~G. Camara, and M.~Grana, { {U}-dual fluxes and
  generalized geometry},  { JHEP} { 11} (2010) 083,
[\href{http://xxx.lanl.gov/abs/1007.5509}{{\tt 1007.5509}}].

\bibitem{Berman:2010is}
D.~S. Berman and M.~J. Perry, { Generalized geometry and {M} theory},  { JHEP}
  { 1106} (2011) 074,
[\href{http://xxx.lanl.gov/abs/1008.1763}{{\tt 1008.1763}}].

\bibitem{Berman:2011pe}
D.~S. Berman, H.~Godazgar, and M.~J. Perry, { {$SO(5,5)$} duality in {M}-theory
  and generalized geometry},  { Phys.Lett.} { B700} (2011) 65--67,
[\href{http://xxx.lanl.gov/abs/1103.5733}{{\tt 1103.5733}}].

\bibitem{Berman:2011jh}
D.~S. Berman, H.~Godazgar, M.~J. Perry, and P.~West, { Duality invariant
  actions and generalised geometry},  { JHEP} { 1202} (2012) 108,
[\href{http://xxx.lanl.gov/abs/1111.0459}{{\tt 1111.0459}}].

\bibitem{Coimbra:2011ky}
A.~Coimbra, C.~Strickland-Constable, and D.~Waldram, { {$E_{d(d)} \times
  \mathbb{R}^+$ generalised geometry, connections and M theory}},
\href{http://xxx.lanl.gov/abs/1112.3989}{{\tt 1112.3989}}.

\bibitem{Coimbra:2012af}
A.~Coimbra, C.~Strickland-Constable, and D.~Waldram, { Supergravity as
  generalised geometry {II}: {$E_{d(d)} \times \mathbb{R}^+$} and {M} theory},
\href{http://xxx.lanl.gov/abs/1212.1586}{{\tt 1212.1586}}.

\bibitem{Berman:2012uy}
D.~S. Berman, E.~T. Musaev, and D.~C. Thompson, { Duality invariant {M}-theory:
  {G}auged supergravities and {S}cherk-{S}chwarz reductions},  { JHEP} { 1210}
  (2012) 174,
[\href{http://xxx.lanl.gov/abs/1208.0020}{{\tt 1208.0020}}].

\bibitem{Berman:2012vc}
D.~S. Berman, M.~Cederwall, A.~Kleinschmidt, and D.~C. Thompson, { {The gauge
  structure of generalised diffeomorphisms}},  { JHEP} { 1301} (2013) 064,
[\href{http://xxx.lanl.gov/abs/1208.5884}{{\tt 1208.5884}}].

\bibitem{Park:2013gaj}
J.-H. Park and Y.~Suh, { {U}-geometry : {SL(5)}},  { JHEP} { 04} (2013) 147,
[\href{http://xxx.lanl.gov/abs/1302.1652}{{\tt 1302.1652}}].

\bibitem{Aldazabal:2013mya}
G.~Aldazabal, M.~Gra{\~n}a, D.~Marqu{\'e}s, and J.~Rosabal, { {Extended
  geometry and gauged maximal supergravity}},  { JHEP} { 1306} (2013) 046,
[\href{http://xxx.lanl.gov/abs/1302.5419}{{\tt 1302.5419}}].

\bibitem{Godazgar:2013dma}
H.~Godazgar, M.~Godazgar, and H.~Nicolai, { {Generalised geometry from the
  ground up}},
\href{http://xxx.lanl.gov/abs/1307.8295}{{\tt 1307.8295}}.

\bibitem{Hohm:2013pua}
O.~Hohm and H.~Samtleben, { Exceptional form of ${D}=11$ supergravity},  {
  Phys.Rev.Lett.} { 111} (2013) 231601,
[\href{http://xxx.lanl.gov/abs/1308.1673}{{\tt 1308.1673}}].

\bibitem{Schwarz:1983wa}
J.~H. Schwarz and P.~C. West, { {Symmetries and transformations of chiral
  ${N}=2$ ${D}=10$ supergravity}},  { Phys. Lett.} { B126} (1983)
301.

\bibitem{Howe:1983sra}
P.~S. Howe and P.~C. West, { {The complete ${N}=2$, ${D}=10$ supergravity}},  {
  Nucl. Phys.} { B238} (1984)
181.

\bibitem{HSEFT7}
O.~Hohm and H.~Samtleben, { Exceptional Field Theory {II}}: E$_{7(7)}$, 
 \href{http://xxx.lanl.gov/abs/1312.4542}{{\tt 1312.4542}}.

\bibitem{Siegel:1993th}
W.~Siegel, { {Superspace duality in low-energy superstrings}},  { Phys.Rev.} {
  D48} (1993) 2826--2837,
[\href{http://xxx.lanl.gov/abs/hep-th/9305073}{{\tt hep-th/9305073}}].

\bibitem{Hull:2009mi}
C.~Hull and B.~Zwiebach, { Double field theory},  { JHEP} { 0909} (2009) 099,
[\href{http://xxx.lanl.gov/abs/0904.4664}{{\tt 0904.4664}}].

\bibitem{Hull:2009zb}
C.~Hull and B.~Zwiebach, { The gauge algebra of double field theory and
  {C}ourant brackets},  { JHEP} { 0909} (2009) 090,
[\href{http://xxx.lanl.gov/abs/0908.1792}{{\tt 0908.1792}}].

\bibitem{Hohm:2010jy}
O.~Hohm, C.~Hull, and B.~Zwiebach, { {Background independent action for double
  field theory}},  { JHEP} { 1007} (2010) 016,
[\href{http://xxx.lanl.gov/abs/1003.5027}{{\tt 1003.5027}}].

\bibitem{Hohm:2010pp}
O.~Hohm, C.~Hull, and B.~Zwiebach, { {Generalized metric formulation of double
  field theory}},  { JHEP} { 1008} (2010) 008,
[\href{http://xxx.lanl.gov/abs/1006.4823}{{\tt 1006.4823}}].

\bibitem{Tseytlin:1990va}
A.~A. Tseytlin, { {Duality symmetric closed string theory and interacting
  chiral scalars}},  { Nucl.Phys.} { B350} (1991)
395--440.

\bibitem{Duff:1989tf}
M.~Duff, { Duality rotations in string theory},  { Nucl.Phys.} { B335} (1990)
610.

\bibitem{Siegel:1993xq}
W.~Siegel, { {Two vierbein formalism for string inspired axionic gravity}},  {
  Phys.Rev.} { D47} (1993) 5453--5459,
[\href{http://xxx.lanl.gov/abs/hep-th/9302036}{{\tt hep-th/9302036}}].

\bibitem{Kugo:1992md}
T.~Kugo and B.~Zwiebach, { {Target space duality as a symmetry of string field
  theory}},  { Prog.Theor.Phys.} { 87} (1992) 801--860,
[\href{http://xxx.lanl.gov/abs/hep-th/9201040}{{\tt hep-th/9201040}}].

\bibitem{Hohm:2011ex}
O.~Hohm and S.~K. Kwak, { Double field theory formulation of heterotic
  strings},  { JHEP} { 1106} (2011) 096,
[\href{http://xxx.lanl.gov/abs/1103.2136}{{\tt 1103.2136}}].

\bibitem{Hohm:2011nu}
O.~Hohm and S.~K. Kwak, { {$N=1$} supersymmetric double field theory},  { JHEP}
  { 1203} (2012) 080,
[\href{http://xxx.lanl.gov/abs/1111.7293}{{\tt 1111.7293}}].

\bibitem{Coimbra:2011nw}
A.~Coimbra, C.~Strickland-Constable, and D.~Waldram, { Supergravity as
  generalised geometry {I}: type {II} theories},  { JHEP} { 1111} (2011) 091,
[\href{http://xxx.lanl.gov/abs/1107.1733}{{\tt 1107.1733}}].

\bibitem{Jeon:2011sq}
I.~Jeon, K.~Lee, and J.-H. Park, { Supersymmetric double field theory:
  {S}tringy reformulation of supergravity},  { Phys.Rev.} { D85} (2012) 081501,
[\href{http://xxx.lanl.gov/abs/1112.0069}{{\tt 1112.0069}}].

\bibitem{Hohm:2011zr}
O.~Hohm, S.~K. Kwak, and B.~Zwiebach, { Unification of type {II} strings and
  {T}-duality},  { Phys.Rev.Lett.} { 107} (2011) 171603,
[\href{http://xxx.lanl.gov/abs/1106.5452}{{\tt 1106.5452}}].

\bibitem{Hohm:2011dv}
O.~Hohm, S.~K. Kwak, and B.~Zwiebach, { Double field theory of type {II}
  strings},  { JHEP} { 1109} (2011) 013,
[\href{http://xxx.lanl.gov/abs/1107.0008}{{\tt 1107.0008}}].

\bibitem{Hohm:2011cp}
O.~Hohm and S.~K. Kwak, { Massive type {II} in double field theory},  { JHEP} {
  1111} (2011) 086,
[\href{http://xxx.lanl.gov/abs/1108.4937}{{\tt 1108.4937}}].


\bibitem{Jeon:2012kd}
  I.~Jeon, K.~Lee and J.~-H.~Park,
  { Ramond-Ramond Cohomology and O(D,D) T-duality},
  {JHEP} { 1209} (2012) 079
  [\href{http://xxx.lanl.gov/abs/1206.3478}{{\tt 1108.4937}}].

\bibitem{Hohm:2010xe}
O.~Hohm and S.~K. Kwak, { Frame-like geometry of double field theory},  {
  J.Phys.} { A44} (2011) 085404,
[\href{http://xxx.lanl.gov/abs/1011.4101}{{\tt 1011.4101}}].

\bibitem{Hohm:2011si}
O.~Hohm and B.~Zwiebach, { On the {R}iemann tensor in double field theory},  {
  JHEP} { 1205} (2012) 126,
[\href{http://xxx.lanl.gov/abs/1112.5296}{{\tt 1112.5296}}].

\bibitem{Hohm:2012gk}
O.~Hohm and B.~Zwiebach, { Large gauge transformations in double field theory},
   { JHEP} { 1302} (2013) 075,
[\href{http://xxx.lanl.gov/abs/1207.4198}{{\tt 1207.4198}}].

\bibitem{Hohm:2012mf}
O.~Hohm and B.~Zwiebach, { {Towards an invariant geometry of double field
  theory}},  { J. Math. Phys.} { 54} (2013) 032303,
[\href{http://xxx.lanl.gov/abs/1212.1736}{{\tt 1212.1736}}].

\bibitem{Jeon:2010rw}
I.~Jeon, K.~Lee, and J.-H. Park, { {Differential geometry with a projection:
  Application to double field theory}},  { JHEP} { 1104} (2011) 014,
[\href{http://xxx.lanl.gov/abs/1011.1324}{{\tt 1011.1324}}].

\bibitem{Jeon:2011cn} 
  I.~Jeon, K.~Lee and J.~-H.~Park,
   { {Stringy differential geometry, beyond Riemann}}, 
    {Phys.\ Rev.\ D} { \bf 84} (2011) 044022,
[\href{http://xxx.lanl.gov/abs/1105.6294}{{\tt 1105.6294}}].

\bibitem{Hitchin:2004ut}
N.~Hitchin, { {Generalized Calabi-Yau manifolds}},  { Quart.J.Math.Oxford Ser.}
  { 54} (2003) 281--308,
[\href{http://xxx.lanl.gov/abs/math/0209099}{{\tt math/0209099}}].

\bibitem{Gualtieri:2003dx}
M.~Gualtieri, { Generalized complex geometry},  { Ann. of Math. (2)} { 174}
  (2011), no.~1 75--123, [\href{http://xxx.lanl.gov/abs/math/0401221}{{\tt
  math/0401221}}].

\bibitem{Gualtieri:2007bq}
M.~Gualtieri, { Branes on {P}oisson varieties},  in { The many facets of
  geometry}, pp.~368--394.
\newblock Oxford Univ. Press, Oxford, 2010.

\bibitem{Aldazabal:2011nj}
G.~Aldazabal, W.~Baron, D.~Marques, and C.~Nunez, { The effective action of
  double field theory},  { JHEP} { 1111} (2011) 052,
[\href{http://xxx.lanl.gov/abs/1109.0290}{{\tt 1109.0290}}].

\bibitem{Geissbuhler:2011mx}
D.~Geissb\"uhler, { Double field theory and {$N=4$} gauged supergravity},  {
  JHEP} { 1111} (2011) 116,
[\href{http://xxx.lanl.gov/abs/1109.4280}{{\tt 1109.4280}}].

\bibitem{Andriot:2012wx}
D.~Andriot, O.~Hohm, M.~Larfors, D.~L\"ust, and P.~Patalong, { {A geometric
  action for non-geometric fluxes}},  { Phys.Rev.Lett.} { 108} (2012) 261602,
[\href{http://xxx.lanl.gov/abs/1202.3060}{{\tt 1202.3060}}].

\bibitem{Andriot:2012an}
D.~Andriot, O.~Hohm, M.~Larfors, D.~L\"ust, and P.~Patalong, { Non-geometric
  fluxes in supergravity and double field theory},  { Fortsch.Phys.} { 60}
  (2012) 1150--1186,
[\href{http://xxx.lanl.gov/abs/1204.1979}{{\tt 1204.1979}}].

\bibitem{Geissbuhler:2013uka}
D.~Geissb\"uhler, D.~Marques, C.~Nunez, and V.~Penas, { Exploring double field
  theory},  { JHEP} { 1306} (2013) 101,
[\href{http://xxx.lanl.gov/abs/1304.1472}{{\tt 1304.1472}}].

\bibitem{Hohm:2013jaa}
O.~Hohm, W.~Siegel, and B.~Zwiebach, { Doubled $\alpha'$-geometry},
\href{http://xxx.lanl.gov/abs/1306.2970}{{\tt 1306.2970}}.

\bibitem{Hohm:2013bwa}
O.~Hohm, D.~L\"ust, and B.~Zwiebach, { {The Spacetime of Double Field Theory:
  Review, Remarks, and Outlook}},
  { Fortsch.Phys.} { 61} (2013) 926--966, 
\href{http://xxx.lanl.gov/abs/1309.2977}{{\tt 1309.2977}}.

\bibitem{Aldazabal:2013sca}
G.~Aldazabal, D.~Marques, and C.~Nunez, { Double field theory: {A} pedagogical
  review},  { Class.Quant.Grav.} { 30} (2013) 163001,
[\href{http://xxx.lanl.gov/abs/1305.1907}{{\tt 1305.1907}}].

\bibitem{Berman:2013eva}
D.~S. Berman and D.~C. Thompson, { Duality symmetric string and {M}-theory},
\href{http://xxx.lanl.gov/abs/1306.2643}{{\tt 1306.2643}}.

\bibitem{Hohm:2013jma}
O.~Hohm and H.~Samtleben, { {U-duality covariant gravity}},  { JHEP} { 1309}
  (2013) 080,
[\href{http://xxx.lanl.gov/abs/1307.0509}{{\tt 1307.0509}}].

\bibitem{Hohm:2013nja}
O.~Hohm and H.~Samtleben, { {Gauge theory of Kaluza-Klein and winding modes}},
  { Phys.Rev.} { D88} (2013) 085005,
[\href{http://xxx.lanl.gov/abs/1307.0039}{{\tt 1307.0039}}].

\bibitem{deWit:2005hv}
B.~de~Wit and H.~Samtleben, { Gauged maximal supergravities and hierarchies of
  nonabelian vector-tensor systems},  { Fortschr. Phys.} { 53} (2005) 442--449,
[\href{http://xxx.lanl.gov/abs/hep-th/0501243}{{\tt hep-th/0501243}}].

\bibitem{deWit:2008ta}
B.~de~Wit, H.~Nicolai, and H.~Samtleben, { Gauged supergravities, tensor
  hierarchies, and {M}-theory},  { JHEP} { 0802} (2008) 044,
[\href{http://xxx.lanl.gov/abs/arXiv:0801.1294}{{\tt arXiv:0801.1294}}].

\bibitem{Nicolai:2000sc}
H.~Nicolai and H.~Samtleben, { Maximal gauged supergravity in three
  dimensions},  { Phys. Rev. Lett.} { 86} (2001) 1686--1689,
[\href{http://xxx.lanl.gov/abs/hep-th/0010076}{{\tt hep-th/0010076}}].

\bibitem{deWit:2004nw}
B.~de~Wit, H.~Samtleben, and M.~Trigiante, { The maximal ${D} = 5$
  supergravities},  { Nucl. Phys.} { B716} (2005) 215--247,
[\href{http://xxx.lanl.gov/abs/hep-th/0412173}{{\tt hep-th/0412173}}].

\bibitem{Cremmer:1980gs}
E.~Cremmer, { Supergravities in 5 dimensions},  in { Superspace and
  supergravity : proceedings} (S.~Hawking and M.~Rocek., eds.), Cambridge Univ.
  Press, 1980.
\newblock Nuffield Gravity Workshop, Cambridge.

\bibitem{Cederwall:2013naa}
M.~Cederwall, J.~Edlund, and A.~Karlsson, { Exceptional geometry and tensor
  fields},  { JHEP} { 1307} (2013) 028,
[\href{http://xxx.lanl.gov/abs/1302.6736}{{\tt 1302.6736}}].

\bibitem{Aulakh:1985un}
C.~Aulakh and D.~Sahdev, { The infinite-dimensional gauge structure of
  {K}aluza-{K}lein theories. {$D = 1+4$}},  { Phys.Lett.} { B164} (1985)
293.

\bibitem{Hohm:2005sc}
O.~Hohm, { {On the infinite-dimensional spin-2 symmetries in Kaluza-Klein
  theories}},  { Phys.Rev.} { D73} (2006) 044003,
[\href{http://xxx.lanl.gov/abs/hep-th/0511165}{{\tt hep-th/0511165}}].

\bibitem{Cremmer:1997ct}
E.~Cremmer, B.~Julia, H.~Lu, and C.~N. Pope, { {Dualisation of dualities. I}},
  { Nucl. Phys.} { B523} (1998) 73--144,
[\href{http://xxx.lanl.gov/abs/hep-th/9710119}{{\tt hep-th/9710119}}].

\bibitem{Nicolai:2003bp}
H.~Nicolai and H.~Samtleben, { Chern-{S}imons vs. {Y}ang-{M}ills gaugings in
  three dimensions},  { Nucl. Phys.} { B668} (2003) 167--178,
[\href{http://xxx.lanl.gov/abs/hep-th/0303213}{{\tt hep-th/0303213}}].

\bibitem{deWit:2003ja}
B.~de~Wit, I.~Herger, and H.~Samtleben, { Gauged locally supersymmetric ${D} =
  3$ nonlinear sigma models},  { Nucl. Phys.} { B671} (2003) 175--216,
[\href{http://xxx.lanl.gov/abs/hep-th/0307006}{{\tt hep-th/0307006}}].

\bibitem{Blair:2013gqa}
C.~D.~A. Blair, E.~Malek, and J.-H. Park, { {M}-theory and {F}-theory from a
  duality manifest action},
\href{http://xxx.lanl.gov/abs/1311.5109}{{\tt 1311.5109}}.

\bibitem{Henneaux:1988gg}
M.~Henneaux and C.~Teitelboim, { Dynamics of chiral (self-dual) $p$-forms},  {
  Phys. Lett.} { B206} (1988)
650.

\bibitem{Schwarz:1993vs}
J.~H. Schwarz and A.~Sen, { Duality symmetric actions},  { Nucl. Phys.} { B411}
  (1994) 35--63,
[\href{http://xxx.lanl.gov/abs/hep-th/9304154}{{\tt hep-th/9304154}}].

\bibitem{DallAgata:1997ju}
G.~Dall'Agata, K.~Lechner and D.~P.~Sorokin,
Covariant actions for the bosonic sector of $d = 10$ IIB supergravity,
 Class. Quant. Grav.  {\bf 14} (1997) L195
 [{\tt hep-th/9707044}].

\bibitem{DallAgata:1998va}
G.~Dall'Agata, K.~Lechner and M.~Tonin,
$D = 10$, $N = IIB$ supergravity: Lorentz invariant actions and duality,
 JHEP {9807} (1998) 017
 [{\tt hep-th/9806140}].


\bibitem{LeDiffon:2008sh}
A.~Le~Diffon and H.~Samtleben, { Supergravities without an action: {G}auging
  the trombone},  { Nucl. Phys.} { B811} (2009) 1--35,
[\href{http://xxx.lanl.gov/abs/0809.5180}{{\tt 0809.5180}}].

\bibitem{Bekaert:2002uh}
X.~Bekaert, N.~Boulanger, and M.~Henneaux, { {Consistent deformations of dual
  formulations of linearized gravity: A no-go result}},  { Phys.Rev.} { D67}
  (2003) 044010,
[\href{http://xxx.lanl.gov/abs/hep-th/0210278}{{\tt hep-th/0210278}}].

\bibitem{Bekaert:2004dz}
X.~Bekaert, N.~Boulanger, and S.~Cnockaert, { {No self-interaction for two-column massless fields}},  
{ J.Math.Phys.} { 46}
  (2005) 012303,
[\href{http://xxx.lanl.gov/abs/hep-th/0407102}{{\tt hep-th/0407102}}].


\bibitem{Boulanger:2008nd}
N.~Boulanger and O.~Hohm, { {Non-linear parent action and dual gravity}},  {
  Phys.Rev.} { D78} (2008) 064027,
[\href{http://xxx.lanl.gov/abs/0806.2775}{{\tt 0806.2775}}].

\bibitem{Samtleben:2007an}
H.~Samtleben and M.~Weidner, { {Gauging hidden symmetries in two dimensions}},
  { JHEP} { 08} (2007) 076,
[\href{http://xxx.lanl.gov/abs/arXiv:0705.2606 [hep-th]}{{\tt arXiv:0705.2606
  [hep-th]}}].

\end{thebibliography}


\providecommand{\href}[2]{#2}\begingroup\raggedright\endgroup

\end{document}